\documentclass[final,3p,times]{elsarticle}

\usepackage{amssymb}
\usepackage{amsmath}

\usepackage{booktabs}
\usepackage{algorithm}
\usepackage{algpseudocode}
\usepackage{longtable}
\usepackage{array}
\usepackage{xurl}      
\usepackage{hyperref}  
\usepackage{pifont} 
\usepackage{tablefootnote}
\usepackage{pifont}   
\usepackage{amssymb}  
\usepackage{xcolor}

\definecolor{darkgreen}{RGB}{0,110,0}
\definecolor{darkred}{RGB}{190,0,0}

\definecolor{cust_purple}{RGB}{76,0,153}
\definecolor{cust_pink}{RGB}{177,22,255}
\definecolor{cust_red}{RGB}{173,19,19}
\definecolor{cust_cyan}{RGB}{16,115,158}

\journal{Expert Systems with Applications}

\begin{document}

\begin{frontmatter}



\title{Graph-Based Re-ranking in Information Retrieval and Beyond: A Survey}

\author[label1]{Md Shahir Zaoad}
\author[label1]{Niamat Zawad}
\author[label2]{Priyanka Ranade}
\author[label1]{Latifur Khan}

\affiliation[label1]{organization={Department of Computer Science, University of Texas at Dallas},
             city={Richardson},
             postcode={75080},
             state={Texas},
             country={USA}}

\affiliation[label2]{organization={The Laboratory for Physical Sciences},
             city={Baltimore},
             postcode={20740},
             state={Maryland},
             country={USA}}

\address{
    {\small \textsuperscript{a}\{mdshahir.zaoad, nxz190009, lkhan\}@utdallas.edu, \textsuperscript{b}\ psranad@lps.umd.edu}
}



\begin{abstract}
The two-stage information retrieval (IR) framework, also known as the retrieve-then-re-rank pipeline, has empowered numerous AI paradigms, including retrieval-augmented generation (RAG) and question-answering (QA) systems that leverage ad hoc knowledge to address the rapidly expanding information landscape. In this setting, graph structures have emerged as a promising mechanism for context augmentation, further enhancing re-ranking frameworks through structured relational modeling, semantic dependency representation, and adaptive retrieval strategies. Consequently, graph representation learning techniques have been actively explored alongside leading IR paradigms. Despite increased research interest in graph-based re-ranking methods, a comprehensive study that connects existing approaches and provides a clear overview of this paradigm remains absent. In this survey, we provide an in-depth review of graph-based re-ranking models, tracing their history, evolution, and state-of-the-art development. To facilitate a clear understanding of this paradigm, we introduce an intuitive taxonomy that organizes graph-based re-ranking models by application domain and methodological characteristics. We also present a chronological timeline that illustrates the evolution of graph-based re-ranking methods. In addition, we analyze the experimental setups of representative studies and provide detailed insight into their findings. We conclude by providing recommendations on future research based on community-wide challenges and opportunities.
\end{abstract}



\begin{keyword}
Re-ranking \sep Information Retrieval \sep Search \sep Graph Re-ranking \sep Two-stage retrieval \sep Multi-stage retrieval \sep RAG \sep QA \sep KGQA \sep LLM \sep Survey


\end{keyword}

\end{frontmatter}



\section{Introduction}
\label{sec:introduction}

Information retrieval (IR), a fundamental task in computing and information science, aims to meet the user's information needs by retrieving relevant information from extensive collections. Its utility extends beyond traditional search systems, serving as a pivotal component in numerous AI applications, such as question answering (QA), retrieval-augmented generation (RAG), answer sentence selection (AS2), and recommender systems. Since its inception, text-based IR models have evolved substantially, from early term-based, statistical, and learning-to-rank (LTR) approaches \cite{robertson1995okapi, burges2005learning, burges2006learning} to neural ranking methods \cite{huang2013learning, guo2016deep, xiong2017end}, and more recently to transformer-based architectures \cite{macavaney2019cedr, li2023parade, nogueira2019passage, nogueira2020document, khattab2020colbert}. As IR models grow larger and more sophisticated, the need for multi-stage architectures, particularly the two-stage retrieve-then-re-rank pipeline, has become increasingly apparent. Lightweight lexical retrieval models are computationally efficient, but suffer from limited precision. In contrast, dense neural models often achieve superior accuracy, though at the expense of substantial computational overhead. To balance this efficiency-effectiveness trade-off, the widely adopted retrieve-then-re-rank pipeline first retrieves a manageable list of candidate documents for a given query from a large corpus using an efficient retriever, and subsequently applies an advanced re-ranker to produce the final ranking. 

A substantial body of IR research approaches re-ranking from a pointwise perspective, independently scoring each query-document pair while overlooking structural and relational information within the candidate list. Recently, advances in large language models (LLMs) have inspired listwise, zero-shot re-ranking approaches that directly generate a ranking permutation over a list of candidate documents; i.e., the documents to be re-ranked. These models implicitly capture inter-document relational signals during re-ranking. However, such relational modeling remains limited and often weakly grounded. Furthermore, LLM-based approaches are constrained by context length limitations and are vulnerable to the lost-in-the-middle problem \cite{tang2024found}. In addition, the continuously expanding information space poses significant challenges for IR methods that primarily rely on the Euclidean representation of text. Such approaches often struggle to adequately model complex, heterogeneous, and semantically diverse search corpora. This further underscores the importance of explicit inter-document relationships within non-Euclidean graph spaces, where structural information can be naturally represented and propagated. Compounding these challenges, natural language queries often fail to adequately represent users' information needs. As a result, incorporating external knowledge is often necessary to enrich query representation and improve ranking effectiveness. 

Graph-based approaches offer a principled solution to address these challenges. Numerous studies explicitly model inter-document relations through graph structures to enhance query-document relevance estimation \cite{luo2024prp, dong2024don, yu2022kg, zhang2021answering}. Many of these methods construct heterogeneous graphs to capture intricate structural and semantic dependencies that play a crucial role in re-ranking performance \cite{christmann2024rag, liu2025knowledge}. The integration of external knowledge through knowledge graphs (KGs) further refines query and document representations. KGs are particularly valuable  in domain-specific tasks, such as biomedical information retrieval, where high variability in synonymous expressions and abbreviated medical terminologies demands structured semantic grounding \cite{gupta2024empowering}. More recently, graph-based methods have been proposed to adaptively retrieve and re-rank candidates to mitigate the infamous recall limitations of two-stage IR pipelines \cite{macavaney2022adaptive, frayling2024effective, macavaney2022adaptiveagent, jaenich2024fairness, rathee2025guiding, rathee2025quam}. Traditional pipelines are inherently constrained by the recall of the initial retrieval stage; any relevant document that fails to make the initial candidate list remains inaccessible to the re-ranker. Adaptive retrieval methods address this limitation by pre-computing a corpus graph that enables exploration beyond the initial candidate list. By traversing the neighbors of the top-ranked documents in the corpus graph, these methods can extract additional relevant documents, thereby improving recall and ranking quality. 

These examples highlight only a fraction of the potential of graph-based modeling in re-ranking. Graph representations and associated algorithms have the potential to elevate IR to a new dimension. The emergence of graph transformers \cite{ying2021transformers, kreuzer2021rethinking, rampavsek2022recipe, shehzad2026graph} and graph foundation models (GFMs) \cite{zhang2023graph, fatemi2023talk} further strengthens this opportunity. Despite growing interest in integrating graph-based techniques into IR, their potential remains underexplored. A primary cause is the fragmented and scattered nature of the literature. Beyond general IR, graph-based re-ranking has been explored in QA, RAG, and various specialized tasks. However, these contributions are scattered across a rapidly expanding research landscape, making it challenging to develop a coherent understanding of the field. As a result, identifying relevant work and synthesizing insights often becomes analogous to searching for a needle in a haystack. This fragmentation underscores a need for a unified survey that presents graph-based re-ranking from a comprehensive, bird's-eye perspective, tracing its evolution, current state-of-the-art (SOTA) methodologies, limitations, and future directions. Although existing surveys address related areas, such as LLMs for information retrieval \cite{zhu2025large} and neural ranking models for information retrieval \cite{ guo2020deep}, they seldom focus on graph-based aspects. Similarly, while other domains have benefited from comprehensive overviews of graph-based methods, such as surveys on graph neural networks (GNNs) for recommender systems \cite{ gao2023survey}, a dedicated and comprehensive review of graph-based re-ranking approaches remains absent. 

To address this gap, this survey aims to unify graph-based re-ranking methodologies within a single, cohesive study. We examine their history, development, and modern techniques, identify prevailing challenges and limitations, and highlight promising future directions to guide subsequent research in this emerging area. We examine graph-based re-ranking from four major perspectives: (1) General Information Retrieval, (2) Question Answering, (3) Retrieval Augmented Generation, and (4) Auxiliary \& Specialized Tasks, as illustrated in our proposed taxonomy (Figure \ref{fig:taxonomy}). The taxonomy further organizes existing work into fine-grained categories to facilitate intuitive comprehension and highlight relationships among different approaches. Additionally, we introduce a timeline (\ref{fig:timeline}) to depict the chronological evolution of graph-based re-ranking methods. 

The methodological discussion follows the structure of the taxonomy, presenting each approach with a balanced combination of intuitive explanation and technical rigor. Our goal is to provide readers with a comprehensive understanding of graph-based re-ranking approaches without requiring them to consult every original paper individually. For experimental settings, we provide detailed insights into datasets, evaluation metrics, learning paradigms, training objectives, and the availability of open-source implementations. Our survey considers graph-based re-ranking at varying levels of granularity. Any substantial incorporation of graph structures within the re-ranking pipeline qualifies a study for inclusion. Some approaches leverage external information from the knowledge graph to complement candidate documents. Other constructs graphs over candidate documents to explicitly model structural and relational dependencies during re-ranking. Still others apply graph-based algorithms to directly produce ranking scores. In certain cases, models integrate multiple graph-based strategies within a unified re-ranking framework. Through an extensive review of the literature, we consolidate these diverse approaches under a common analytical perspective. The GitHub repository associated with this work is publicly available\footnote{\url{https://github.com/Shahir47/awesome-graph-reranking-ir}}. 

In summary, our contributions are as follows:
\begin{itemize}
    \item To the best of our knowledge, we present the first comprehensive survey dedicated to graph-based re-ranking approaches, consolidating a fragmented body of literature. 
    \item We propose an intuitive and structured taxonomy that unifies graph-based re-ranking models across diverse research paradigms, enabling a clear understanding of their methodological relationships.
    \item We introduce a chronological timeline that captures the evolution of graph-based re-ranking research. 
    \item We analyze the role of graphs at different levels of granularity within re-ranking pipelines across various tasks, including IR, QA, RAG, and specialized tasks. 
\end{itemize}

The remainder of the survey is structured as follows: In Section \ref{sec:background}, we discuss relevant background topics, including information retrieval, different re-ranking paradigms, graph fundamentals, and an overview of GNNs. The core contribution of this paper lies in Section \ref{sec:graph-reranking}, where we present a detailed discussion of graph-based re-ranking methods. In Section \ref{sec:experimental_settings}, we discuss the experimental settings associated with the reviewed literature. In this section, we provide an organized overview of different datasets, evaluation metrics, learning paradigms, and training objectives, followed by a detailed discussion. Following this, in Section \ref{sec:findings}, we present findings of our comprehensive review, including limitations and potential future directions of graph-based re-ranking approaches. Finally, we conclude the survey by presenting the overview of this study in Section \ref{sec:conclusion}.

\section{Background}
\label{sec:background}

\subsection{Information Retrieval \& Re-ranking}
Early IR systems predominantly relied on handcrafted features, heuristics, and probabilistic models to satisfy users' information demands \cite{salton1965smart, robertson1995okapi}. These pioneering approaches laid the foundation for the architecture of modern IR systems. Today, nearly every IR pipeline begins with a first-stage retriever built upon these classical models, typically within a multi-stage re-ranking pipeline, most commonly, a two-stage re-ranking pipeline. Re-ranking can be formulated as a task of producing a new ordering of a finite list of documents $\mathcal{D} = \{d_i\}_{i \in [n]}$, given a query $q$ and a candidate list retrieved $\textit{a priori}$ by an initial retriever. This is typically achieved through a scoring function $r_q: \mathcal{D} \rightarrow (0,1)$, where $r_q(d_i)$ denotes the relevance score between query $q$ and candidate document $d_i \in \mathcal{D}$. The re-ranked document list $\mathcal{D}'$ is obtained by ordering the documents in $\mathcal{D}$ in descending order according to their relevance scores $r_q(d_i)$. In line with the standard IR conventions, the term document here broadly refers to any textual unit at varying levels of granularity, including full documents, passages, or individual sentences \cite{nogueira2019multi}.

In the early 2000s, with the rapid growth of the web, manual tuning became difficult, and the need for automatic ranking models emerged, leading to the development of LTR models. Instead of relying solely on handcrafted features and probabilistic methods, LTR applies machine learning (ML) to learn a scoring function in a data-driven manner. Subsequently, three core LTR paradigms emerged: pointwise, pairwise, and listwise. Pointwise approaches treat each document independently, estimating its relevance to a given query without considering its relationships to other candidate documents \cite{crammer2001pranking, li2007mcrank}. In contrast, pairwise methods determine ranking order by comparing the relevance of each document pair to the query \cite{burges2005learning, freund2003efficient}. Listwise approaches, on the other hand, optimize over the entire list of candidate documents for a given query \cite{cao2007learning, xu2007adarank}. Traditional LTR models remain constrained by manual feature engineering and limited semantic representation capabilities. Neural ranking models address these limitations by modeling texts as distributed representations \cite{huang2013learning, guo2016deep, xiong2017end}. Within neural re-ranking, two broad categories emerged: representation-based and interaction-based models. The former computes query-document relevance by measuring similarity between their corresponding embeddings. Notable examples include DSSM \cite{huang2013learning} and its convolutional variant, CLSM (also known as CDSSM) \cite{shen2014latent}. These models are computationally efficient and effective at capturing global semantics, but often fail to capture fine-grained semantics. In contrast, interaction-based models capture query-document relevance at the token level, enabling more precise matching signals. Among the most notable models are DRMM \cite{guo2016deep}, KNRM \cite{xiong2017end}, and PACRR \cite{hui2017pacrr}.  

The advent of transformer architecture \cite{vaswani2017attention}, together with the subsequent rise of pre-trained language models (PLMs) and LLMs, marked a new era in information retrieval. Transformer-based contextual embeddings facilitate accurate modeling of query-document relevance, thereby improving ranking performance \cite{macavaney2019cedr, li2023parade}. These advances surpass traditional neural ranking models and paved the way for more sophisticated transformer-based ranking architectures, particularly cross-encoder re-rankers built upon PLMs such as BERT \cite{devlin2019bert}, ELECTRA \cite{clark2020electra}, and T5 \cite{raffel2020exploring}. Cross-encoder re-rankers model query-document relevance by capturing fine-grained token-level interactions \cite{nogueira2019multi, nogueira2020document}. PLM-based cross-encoders have become representative supervised ranking architectures in contemporary research, whereas LLM-based methods have come to dominate the unsupervised paradigm. Prompt engineering empowers zero-shot LLM-based models to perform effective ranking, while overcoming common limitations of supervised methods, including poor transferability and the need for an expensive labeled dataset. Similar to LTR paradigms, LLM-based models are categorized into pointwise \cite{sachan2022improving, chen2024attention, sun2023instruction, zhuang2024beyond}, pairwise \cite{qin2024large, luo2024prp, yan2024consolidating}, and listwise \cite{sun2023chatgpt, yoon2024listt5, ma2024fine, pradeep2023rankvicuna} approaches. A fourth paradigm, setwise ranking \cite{zhuang2024setwise, zhuang2025rank, podolak2025beyond}, has recently gained attention as a solution to the limitations of pairwise and listwise LLM-based ranking models. The pairwise method incurs quadratic time complexity with respect to the number of candidate documents. In contrast, listwise approaches are limited by the LLMs' fixed context window and are susceptible to the lost-in-the-middle effect. Setwise ranking offers a middle ground between these two paradigms by prompting an LLM to generate the correct order of a small, manageable set of candidate documents at a time, thereby balancing efficiency and effectiveness. This section presents a concise overview of information retrieval and re-ranking methods; for a more comprehensive discussion, we refer the reader to \cite{zhu2025large, guo2020deep}.

\subsection{Graph Fundamentals}
A graph is a non-linear data structure defined mathematically as $G=(V, E)$, where $V$ denotes the set of nodes (or vertices) and $E$ represents the set of edges. Each node $v \in V$ corresponds to an element of interest, while an edge $e_{ij} \in E$ specifies a connection between two nodes $v_i$ and $v_j$. In computational settings, a graph is typically represented by an adjacency matrix $\mathbf{A}\in R^{N \times N}$, where $N$ denotes the total number of nodes in the graph. Depending on whether an edge $e_{ij} \in E$ exists, the corresponding entry in the adjacency matrix is defined as $\mathbf{A}_{ij} = \{0, 1\}$. In addition, a graph might contain a node feature matrix $X \in R^{N \times d}$, where $d$ denotes the feature dimension associated with each node. Similarly, graphs may include an edge feature matrix that encodes attributes of edges. 

Knowledge graphs are a specialized form of graph that play an important role in information retrieval and re-ranking. They represent semantic relationships among real-world entities such as people, places, facts, and concepts. The fundamental unit of a knowledge graph is a triplet of the form  $<e_h, r, e_t>$, which connects a head entity $e_h$ to a tail entity $e_t$ through a relation $r$, where both the head and tail entities are represented as nodes in the knowledge graph. For example, $<Paris, isCapitalOf, France>$ represents a fact where the relation $ isCapitalOf$ links the entity $Paris$ to the entity $France$. In Section \ref{sec:findings}, we present detailed statistics of popular knowledge graphs used in the re-ranking task.

\subsection{Graph Neural Network}
\label{subsec:GNN}
Graph neural networks aim to learn node- or graph-level representations that capture the structural relationships and semantic information present in graph-structured data. The term graph neural network was first coined in \cite{gori2005new}, which extends recursive neural networks to process graphs. Recurrent GNNs \cite{gori2005new, scarselli2008graph} are designed to iteratively update node representations until they reach convergence. Despite their success in processing graphs, these methods are computationally intensive and lack scalability. To address this limitation, subsequent research turned to spectral graph theory, introducing convolution on graphs that generalize the success of convolutional neural networks (CNNs) on one-dimensional (1D) sequences and two-dimensional (2D) images to graph domains. The pioneering spectral-based convolutional GNN utilizes the eigendecomposition of the graph Laplacian to define the graph convolution \cite{bruna2013spectral}. However, the cubic complexity $O(N^3)$ associated with eigendecomposition limits its scalability and practical applicability. To avoid explicit eigendecomposition, subsequent methods approximated spectral filters using efficient techniques \cite{defferrard2016convolutional, kipf2016semi}. A prominent example is the graph convolutional network (GCN) \cite{kipf2016semi}, which approximates spectral graph convolution by a localized first-order approximation, greatly improving computational efficiency and scalability. By constraining the layer-wise convolution operation to $k=1$, it bridges the gap between spectral-based and spatial-based approaches. Spatial-based methods, another prominent framework for convolutional GNNs, update node representations by aggregating features from neighboring nodes. This mechanism, commonly referred to as message passing \cite{gilmer2017neural, shehzad2026graph}, forms the foundation of many widely used and influential GNNs \cite{albarede2022passage, schlichtkrull2018modeling, hamilton2017inductive}. The message-passing mechanism is formulated as:

\begin{equation}
\label{eq:GNN_background_1}
h_v^{(l)} = U_l \left( 
h_v^{(l-1)}, 
\sum_{u \in N(v)} M_l\left(h_u^{(l-1)}, h_v^{(l-1)}, e_{uv}\right)
\right)
\end{equation}

where, $h_v^{(l)}$ refers to the hidden representation of node $v$ at layer $l$, $U_l$ represents the update function, $M_l$ is the message function, and $e_{uv}$ denotes the feature associated with the edge connecting nodes $u$ and $v$. Spectral approaches are often restricted to transductive settings and struggle to generalize to unseen nodes during inference. Consequently, most contemporary methods have shifted toward spatial methods that perform message passing within a node's local neighborhood, enabling inductive learning and better generalization to previously unseen nodes. GraphSAGE \cite{hamilton2017inductive}, a pioneering inductive learning framework for GNN, updates node representations by aggregating information from a fixed-size set of sampled neighbors. Graph attention network (GAT) \cite{velivckovic2017graph} further enhances this framework by introducing the attention mechanism to GNNs. Instead of assigning equal importance to neighboring nodes, GAT employs masked self-attention to dynamically compute attention weights for each neighbor during message passing. Recently, graph transformers \cite{ying2021transformers, kreuzer2021rethinking, rampavsek2022recipe, shehzad2026graph} have gained significant attention. Based on transformer architectures \cite{vaswani2017attention}, these models can capture both local and global structural dependencies in graph data. They mitigate key limitations of traditional message passing GNNs, such as information over-squashing and over-smoothing, while enabling more effective modeling of long-range dependencies in graphs. We refer readers to \cite{wu2020comprehensive, zhou2020graph, shehzad2026graph} for a detailed discussion of GNNs. 

\section{Graph-based Re-ranking}
\label{sec:graph-reranking}

\begin{figure}[htbp]
	\includegraphics[width=\columnwidth]{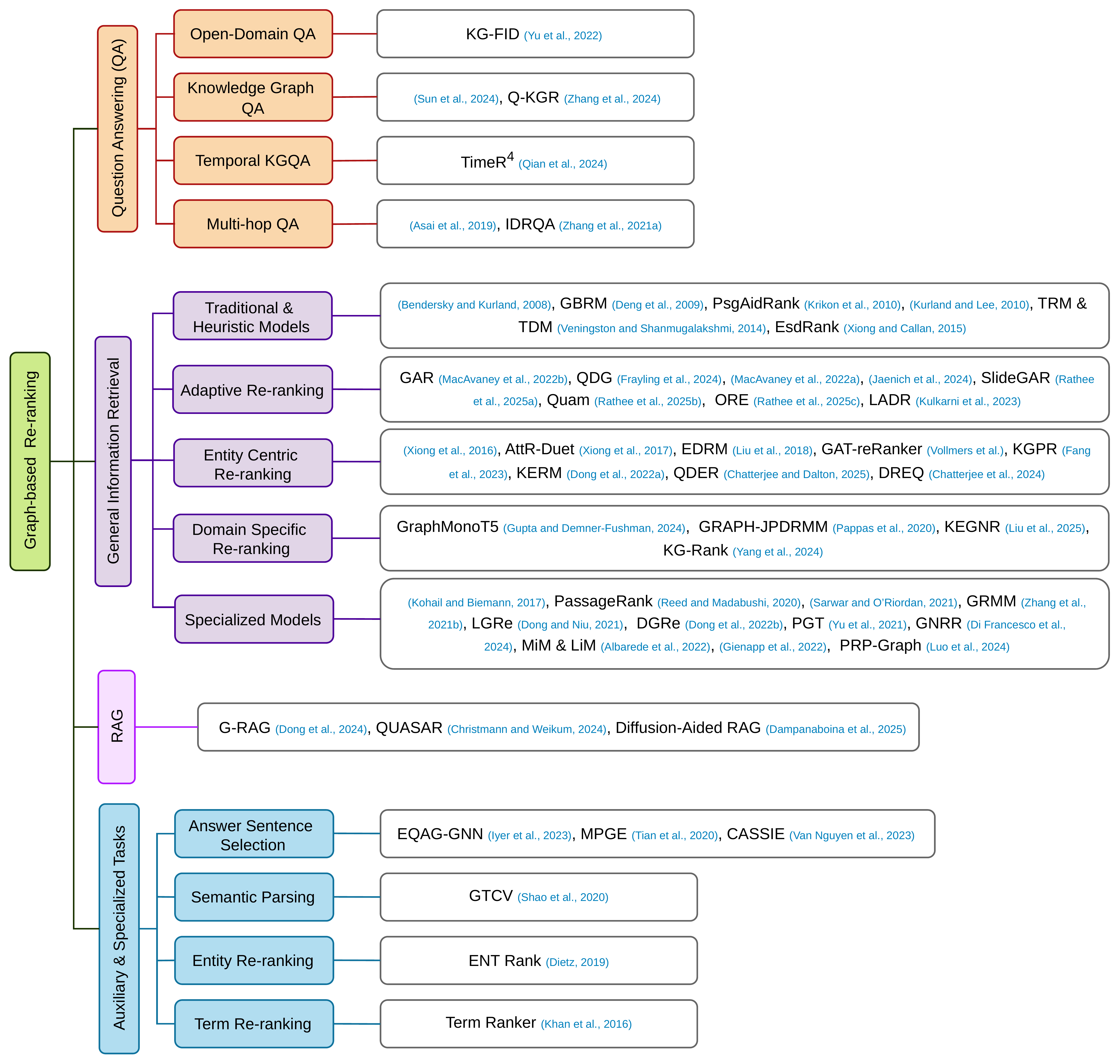}
	\caption{The taxonomy of graph-based re-ranking methods}
	\label{fig:taxonomy}
\end{figure}

In this section, we present a comprehensive overview of graph-based re-ranking approaches. These methods are not limited to general information retrieval and also facilitate numerous other tasks, including QA and RAG, which are scattered across the literature space. To consolidate this fragmented body of work, we propose a unified taxonomy in Figure \ref{fig:taxonomy}, that organizes graph-based re-ranking methods into four primary categories based on their application domain: General Information Retrieval, QA, RAG, and Auxiliary and Specialized Tasks. 

The General Information Retrieval category is further divided into five subcategories based on methodological distinctions: Traditional and Heuristic Models, Adaptive Re-ranking methods, Entity-Centric Re-ranking methods, Domain-Specific Re-ranking methods, and Specialized models. The QA category is also structured to reflect graph-based re-ranking in different QA tasks, including Open-Domain Question Answering (ODQA), Knowledge Graph Question Answering (KGQA), Temporal Knowledge Graph Question Answering (TKGQA), and Multi-hop Question Answering. The Auxiliary and Specialized Tasks are similarly classified into Answer Sentence Selection, Semantic Parsing, Entity Re-ranking, and Term Re-ranking. Relevant studies are listed alongside each category and sub-category within the taxonomy. Studies that do not introduce official model names are only represented by author information and publication years. Throughout our discussion of graph-based re-ranking techniques, we adhere to the original notations used in respective studies.

\begin{figure}[htbp]
	\includegraphics[width=\columnwidth]{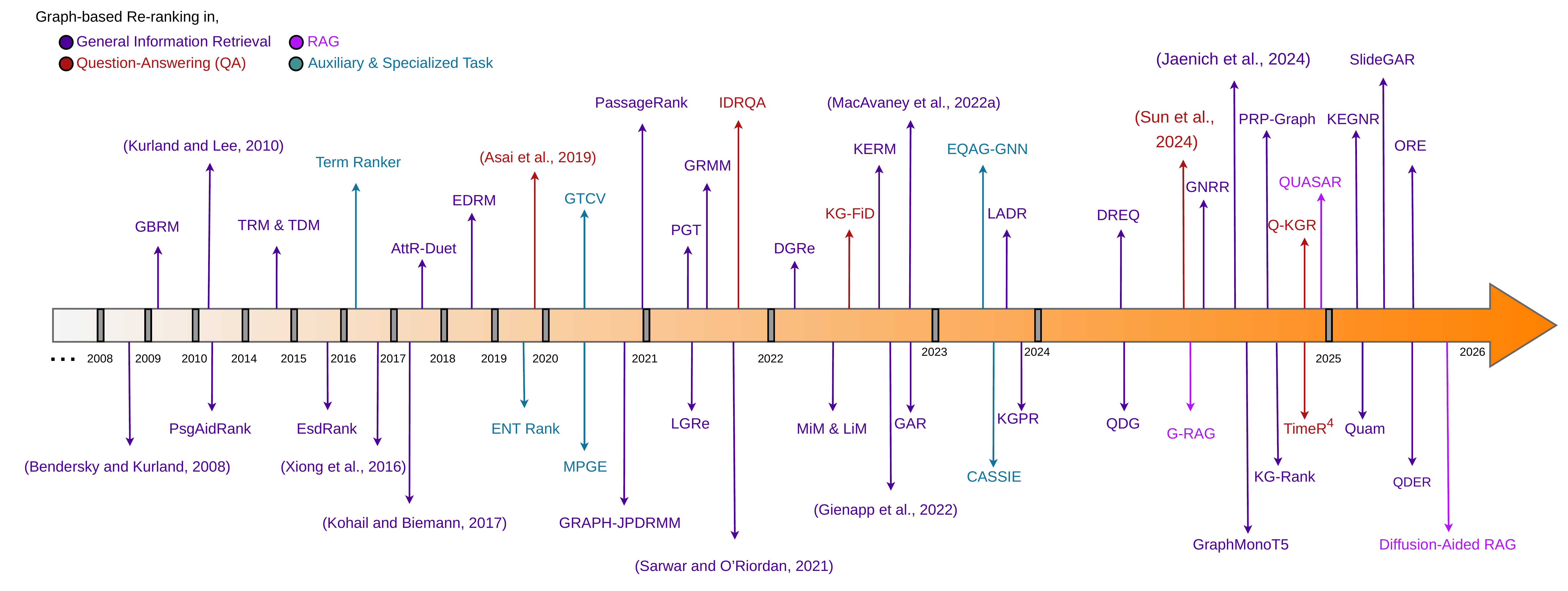}
	\caption{Chronological timeline of the evolution of graph-based re-ranking methods}
	\label{fig:timeline}
\end{figure}

To illustrate the evolution of graph-based re-ranking methods, we also introduce a chronological timeline in Figure \ref{fig:timeline} covering the period from 2008 to 2025. Although the primary focus of this survey is graph-based re-ranking in neural and deep learning paradigms, we trace its evolution from traditional and heuristic models to contemporary SOTA methods. Accordingly, we include several notable traditional graph-based re-rankers, though this selection is not exhaustive. The timeline does not allocate space proportionally across years; instead, it allocates space based on the number of studies published within each period. However, the month-level proportionality has been preserved, to a reasonable approximation, within each year span. 

\subsection{General Information Retrieval}
This category focuses on graph-based re-ranking methods dedicated to satisfying users' information needs by identifying relevant content. Historically, such models have relied on algorithms like PageRank \cite{page1999pagerank, brin1998anatomy} and other graph-based centrality measures \cite{kleinberg1999authoritative}. A prominent contemporary direction is adaptive re-ranking methods, which address the recall limitation  of the initial retriever by dynamically extracting additional relevant candidates from a corpus graph during the re-ranking stage. These approaches strengthen modern re-ranking architectures built on PLMs and LLMs, thereby improving ranking performance. In a similar vein, entity-centric methods assist neural, PLM, and LLM-centric re-rankers by incorporating external information from knowledge graphs. Graph-based methods also play a pivotal role in domain-specific IR tasks, where PLM-based re-rankers alone may struggle. Beyond these directions, there is a wide range of specialized models that incorporate graph-based techniques, including inter-document relationships and structural information in the re-ranking process.

\subsubsection{Traditional \& Heuristic Models}
One of the earliest works, \cite{bendersky2008re}, introduces passage centrality to complement the query-document relevance score. The proposed approach assumes that passages similar to many candidate documents are likely to be query-relevant. Starting from a query and an initial candidate list, the proposed method splits the documents into overlapping passages and constructs a bipartite graph between documents and top-relevant passages. Passage importance is subsequently estimated using graph-based centrality measures, either by aggregating incoming edge weights (influx) or by applying the HITS authority algorithm \cite{kleinberg1999authoritative}. The final query-document relevance is estimated by combining the original similarity score with the centrality of the top-ranked passage.

To improve upon methods that rely solely on textual content, many studies incorporate link-based features, such as document relations, using algorithms like PageRank. However, these approaches often treat content and link information as independent signals and combine their scores only at a later stage via linear fusion, which may lead to suboptimal performance. To address this limitation, \cite{deng2009effective} proposes the graph-based re-ranking model (GBRM), which refines the initial ranking scores over a graph structure constrained on the initial content-based ranking. The model introduces a joint regularization framework to enforce global consistency across the graph, such that documents sharing edges exhibit similar ranking scores. Furthermore, GBRM integrates content and link information into a unified-latent space graph, enabling re-ranking by jointly exploiting content relevance and structural dependencies. The proposed framework includes two main stages: an offline stage and an online stage. In the first phase, the document representations $X$ are generated using latent semantic analysis (LSA) \cite{deerwester1990indexing}, which jointly models content and link signals. In the subsequent stage, a latent space graph $G = (V, E)$ is constructed over the K nearest neighbors (KNNs) of the candidate documents, where nodes correspond to document representations $X$ and the edge-set $E$ defines the document relations weighted via heat kernel as represented in Eq. \ref{eq:M18_3}.

\begin{equation}
w_{ij} = \exp\!\left(-\frac{\lVert x_i - x_j \rVert^{2}}{2\sigma^{2}}\right)
\label{eq:M18_3}
\end{equation}






Here, $\sigma$ is the parameter associated with the heat kernel. A cost function $R(\cdot)$ within a joint regularization framework \cite{zhou2003learning} facilitates the final relevance estimation by unifying the refined ranking score with the initial ranking score, as formulated in Eq. \ref{eq:M18_4}.

\begin{equation}
R(F, q, G) = \frac{1}{2}\sum_{i,j=1}^{n}w_{ij}\left\lVert\frac{f(d_i, q)}{\sqrt{D_{ii}}}-\frac{f(d_j, q)}{\sqrt{D_{jj}}}\right\rVert^{2}+\mu\sum_{i=1}^{n}\left\lVert
f(d_i, q) - f^{0}(d_i, q)\right\rVert^{2}
\label{eq:M18_4}
\end{equation}

Here, $\mu > 0$ denotes  a regularization parameter, while $f^0(d_i, q)$ and $f(d_i, q)$ represent the initial and refined ranking scores, respectively. The first term in Eq. \ref{eq:M18_4} ensures global consistency of the relevance score within the graph structure, whereas the second term constrains the refined ranking scores to remain close to the initial score. Subsequently, a closed form version of Eq. \ref{eq:M18_4} can be formulated as follows: 

\begin{equation}
\begin{gathered}
\hat{\mathbf{F}}^{*} = \left(\mathbf{I}-\mu_{\alpha} \hat{\mathbf{S}}\right)^{-1}\hat{\mathbf{F}}^{0} \\
\mathbf{S} = \mathbf{D}^{-\frac{1}{2}}\mathbf{W}\mathbf{D}^{-\frac{1}{2}} \\
\mu_{\alpha} = \frac{1}{1 + \mu}
\end{gathered}
\label{eq:M18_5}
\end{equation}

This closed-form solution is approximated on the top-$n$ initial candidates. The parameter $\mu_\alpha$ controls the trade-off between the original ranking scores and global consistency. Here, $F^0$ represents the initial ranking matrix, while $\hat{F^*}$ denotes the refined ranking matrix. $I$ stands for the identity matrix, and $\hat{S}$ denotes a positive semi-definite matrix corresponding to the subgraph induced by the top-$n$ documents.

Another study \cite{krikon2010utilizing} introduces a language-model-based re-ranking framework that enhances search precision by jointly modeling document- and passage-level evidence alongside centrality information. Operating on an initially retrieved list of documents, the proposed approach estimates query-document relevance by integrating centrality scores with query-based similarity. Centrality is estimated by constructing two separate similarity graphs over documents and their constituent passages and applying PageRank to derive an importance score. By interpolating document-level and passage-level evidence, the model effectively captures the relevance signals even for long or heterogeneous documents.

Inspired by the PageRank algorithm, \cite{kurland2010pagerank} proposes a structural re-ranking method to refine the initial retrieval results by exploiting asymmetric implicit relationships without relying on hyperlinks. Following the initial retrieval, candidate documents are modeled within a generation graph using generation links, where two documents ($d_i$, $d_j$) share a directed edge $e_{i \rightarrow j}$ if the language model associated with $d_j$ assigns a high generation probability to $d_i$. Next, variants of the PageRank algorithm are applied to the generation graph to compute document centrality and identify authoritative documents relevant to the information need. The final ranking is obtained by combining the centrality estimates with the initial query-document relevance scores.

\cite{veningston2014information} introduces a document re-ranking strategy that utilizes term graphs to model word association across documents. The term graph is constructed from frequent item-sets, where each document is represented by its most frequent terms that exceed a predefined threshold. Each unique term in the frequent item-sets is represented as a node in the term graph, and an edge is established between two nodes if and only if both terms co-occur within the same frequent item-set. The weight assigned to the edge is the highest support value among all candidate frequent item-sets that contain the corresponding pair of nodes, where a document's $d$ support is computed as the normalized aggregated term frequency of an item-set, as shown in Eq. \ref{eq:M3_1},

\begin{equation}
\begin{gathered}
\mathrm{Support}_{d} = \frac{\sum_{i=1}^{n} f_{d}(t_i)} {\sum_{j=1}^{N} \sum_{i=1}^{n} f_{d_j}(t_i)},\; \text{where } f_d(t_i) = \frac{\mathrm{term\_frequency}_d(t_i)} {\max\limits_i \left( \mathrm{term\_frequency}_d(t_i) \right)}
\end{gathered}
\label{eq:M3_1}
\end{equation}

Here, $f_d(t_i)$ represents the support of term $t_i$, $n$ indicates the number of terms contained in an item-set, and $N$ refers to the total number of item-sets for each query. The resulting term graph then facilitates document re-ranking using either of the two methods: term-rank-based document re-ranking or term-distance-matrix-based document re-ranking. The first method relies on the PageRank algorithm that computes a term's $t_a$ ranking using Eq. \ref{eq:M3_2}, where $T_b$ is the set of neighboring terms that $t_a$ points to, $T_a$ denotes the set of neighboring terms directed toward $t_a$, and $c$ serves as a normalization factor.

\begin{equation}
\mathrm{Rank}(t_a) = c \sum_{t_b \in T_a} \frac{\mathrm{Rank}(t_b)}{N_{t_b}}
\label{eq:M3_2}
\end{equation}

Term ranks produced through PageRank are used to re-rank candidate documents by either using a heuristic-based approach or the Spearman correlation coefficient (SCC). In the heuristic method, documents are ranked using the TermRank scores of the top-$k$ terms linked to the query. In the other method, document relevance is estimated by identifying sets of terms linearly related to the query, computing SCC between their TermRank-ordered lists, and promoting documents associated with term sets that show strong positive correlation with the highest-ranked terms. The term-distance-matrix-based re-ranking method, by contrast, uses a term distance matrix, which is an adjacency matrix that captures the shortest distance between pairs of terms as the least number of hops. A document receives a higher rank when its terms are located closer to the query terms.

\cite{xiong2015esdrank} posits that query and corpus signals alone are insufficient for effective document re-ranking. To address this, the authors introduce EsdRank, a framework that integrates external semi-structured data, such as controlled vocabularies and knowledge bases, into the re-ranking process. The rich external information, including synonyms, entities, and relational structures, enables search engines to model query-document relationships from multiple perspectives, thereby improving ranking performance. EsdRank represents semi-structured external data as intermediate “objects” between queries and documents and constructs two types of evidence features: query-object features that measure how well an object represents the query, and object-document features that capture object-document similarity. Based on these representations, a latent-listwise ranking model, Latent-ListMLE \cite{liu2009learning}, is employed to learn a ranking function by jointly modeling queries, objects, and documents. 

\subsubsection{Adaptive Re-ranking}
In traditional two-phase IR systems, the initial retriever aims to maximize recall while the re-ranker aims to maximize precision. It introduces an efficiency-effectiveness tradeoff in the number of documents retrieved by the initial retriever. To address this issue, \cite{macavaney2022adaptive} introduces GAR, an adaptive retrieval and re-ranking approach grounded in the clustering hypothesis, which iteratively updates the initial candidate list to overcome the recall limitations of traditional two-phase approaches. The clustering hypothesis posits that documents that are closely related are also likely to be relevant to the same query \cite{jardine1971use}. Consequently, GAR expands the re-ranking scope by incorporating new documents related to previously scored ones. To this end, it leverages a corpus graph $G = (V, E)$, where nodes in $V$ represent documents and edges in $E$ model their relationships. GAR operates on an initial pool of candidate documents $R_0$, together with a batch size $b$, a re-ranking budget $c$ indicating how many documents will be re-ranked, and a corpus graph $G$, and returns a re-ranked document pool, $R_1$. The re-ranking procedure also relies on a dynamically updated re-ranking pool $P$ and a graph frontier $F$. Initially, $F$ starts as empty, while $P$ is initialized with $R_0$. During each iteration, the top-$b$ documents $B$ from the $P$ are re-ranked, and both $R_0$ and  $R_1$ are updated accordingly. The documents in $B$ are also removed from the frontier $F$, if present. Subsequently, the frontier $F$ is updated with the neighboring documents of $B$ from $G$. Rather than scoring only the top-$b$ documents from the original pool $R_0$, GAR alternates scoring documents from $R_0$ and from the frontier $F$ by updating $P$ after every iteration, as depicted in Eq. \ref{eq:NC_1}. As a result, GAR ensures that documents beyond the initial pool are considered and scored. This process continues as long as the re-ranking budget $c$ permits. 

\begin{equation}
P \leftarrow
\begin{cases}
R_0, & \text{if } P = F \\
F, & \text{if } P = R_0
\end{cases}
\label{eq:NC_1}
\end{equation}







\cite{macavaney2022adaptiveagent} replaces GAR's naive strategy of alternating between the candidate set and frontier set by an agentic approach, while all other components, including the corpus graph configuration, remain unchanged. The proposed approach introduces four agent types: TwoPhase, Threshold, Greedy, and Oracle. TwoPhase initially re-ranks $K < c$ documents until $|R_1| < K$, where $K$ is the size of the initial  phase. In the subsequent phase, it iteratively re-ranks candidates from the frontier, which contains the neighbors of $R1$, until the re-ranking budget $c$ is exhausted. The Threshold agent maintains a single pool $R_0$, to which neighbors of top-$b$ documents are added if their relevance scores exceed a predefined threshold $r$. Unlike the original and the TwoPhase approaches, this method uses absolute relevance scores of documents. In the Greedy approach, the decision between the initial ranking pool and the frontier is made by selecting the source with the largest maximum score from the latest batch. The Oracle agent, though impractical, ranks both the candidate set and the frontier at each step and chooses one based on the evaluation result against the ground truth.

\cite{jaenich2024fairness} builds on GAR \cite{macavaney2022adaptive} by introducing policies intended to ensure fair exposure for different document groups within the candidate pool. The proposed policies are divided into two categories: pre-retrieval policies, which include two strategies for modifying the graph associated with GAR, and in-process policies, which comprise four strategies that revise the GAR pipeline itself. The first policy, \textit{Diverse Group Linking}, dictates that only documents from different groups share an edge in the corpus graph $G$. The second policy, \textit{Balanced Neighbor Quota}, extends this idea by defining an upper limit on the number of documents each group can represent in $G$. The third policy, \textit{Removing Neighbors from the same Group}, modifies the neighbor set of each re-ranked document $d_i$ in a batch prior to adding them to the frontier $F$. Specifically, any document belonging to the same group as $d_i$ is discarded from the neighbors set. The fourth Policy, \textit{Equal Proportions in the Neighbors}, extends on this by defining a quota that limits the number of documents each group can contribute to the frontier. Rather than selecting only the top-$b$ documents in each batch, the fifth policy, \textit{Select Highest Scoring Documents from Each Group}, requires selecting a specific number of the highest-scoring documents from each group. Therefore, further ensuring variances among the neighbors in the corpus graph. In addition to selecting a batch based on the highest-scoring document from each group, the sixth policy, \textit{Priorities Neighbors from the First Iterations}, also considers the order in which documents were inserted into the frontier, thereby prioritizing documents scored in the first iteration.

Another study, \cite{rathee2025guiding}, extends GAR \cite{macavaney2022adaptive} by incorporating an LLM-based listwise re-ranker. While the original work relies on the probability ranking principle and computes relevance scores independently, it does not account for relationships among documents. To address this limitation, the proposed SlideGAR transforms GAR from a pointwise LTR paradigm to a listwise one, enabling explicit modeling of document relationships. SlideGAR handles LLM context constraints by applying a sliding-window strategy defined by window size $w$ and step size $b$. Similar to GAR, the re-ranking pool $P$ in SlideGAR is initialized with $R_0$. Initially, the window $L$ is populated with top-$w$ documents from $P$, which are then re-ranked using a listwise re-ranker, RankZephyr\cite{pradeep2023rankzephyr}. The top-$b$ documents from the re-ranked list $B$, denoted as $L_1$, are carried to the next window while the rest are added to the final ranked list, $R_1$. The $B$ documents are also removed from $R_0$. Following this, the frontier $F$ is updated with the neighboring documents of $B$ (that are absent in both $R_1$ and $L_1$), prioritized according to their ranking in $B$. Following \cite{macavaney2022adaptive}, the alternating strategy constructs $P$ from either $R_0$ or $F$. Accordingly, $L$ is prepared with $L_1$ together with top-$b$ documents from $P$, which are then used in the subsequent round. The iterative process continues until the size of $R_1$ reaches $c - b$; at which point the documents in $L_1$ are appended to the top of $R_1$ to complete the ranking.

The existing adaptive retrieval approaches suffer from two major drawbacks. Firstly, the corpus graph $G_c$ is built on heuristic choices; i.e., it models lexical or semantic similarities without considering query-based relevance, which might introduce non-relevant content in the frontier $F$. Another limitation is the lack of a degree of similarity between documents in adaptive re-ranking algorithms such as GAR \cite{macavaney2022adaptive}. As a result, during the expansion process, GAR will weigh each neighbor of a candidate equally rather than selecting the truly relevant ones. \cite{rathee2025quam} address both of these problems by improving the corpus graph and introducing a novel adaptive retrieval strategy, QUAM, that selects neighboring documents during the expansion process on the basis of  learnt-affinity L\textsc{aff} scores. Here, affinity refers to co-relevance; i.e., the degree of similarity between documents as L\textsc{aff} scores from a learnt-affinity model $f$. Leveraging the L\textsc{aff} scores as edge weights, the corpus graph $G_c$ is updated and pruned to construct the affinity graph $G_a$, in which the affinity between neighboring documents $d_i$ and $d_j$ is represented as $f(d_i, d_j)$. The affinity model is implemented using a BERT-based architecture trained with the binary cross-entropy loss, as shown in Eq. \ref{eq:M19_1}. In this equation, $D$ denotes the training set and $d$ is the target document. The variable $x$ represents either a positive or negative document with respect to (w.r.t.) $d$, and $y$ is a relevance label, either 0 or 1.

\begin{equation}
L(\mathcal{D}) = -\frac{1}{|\mathcal{D}|}
\sum_{(x,d,y)\in\mathcal{D}}
\left[
y \log\bigl(f(x,d)\bigr)
+ (1 - y)\log\bigl(1 - f(x,d)\bigr)
\right]
\label{eq:M19_1}
\end{equation}

Like other adaptive retrieval models, QUAM begins with an initial retrieved set $R_0$. At each iteration, the top-ranked newly added documents are used to explore neighboring nodes in the affinity graph, which is then inserted into the frontier. The documents in the frontier are then scored using a set affinity (SetAff) score, defined in Eq. \ref{eq:M19_2}, where $S$ denotes the top-$s$ documents from the re-ranked pool at a given iteration and $d$ is a document in the frontier. Since $d'$ has already been re-ranked, its relevance estimate with respect to the query is used to compute $P(Rel(d'))$, as shown in Eq. \ref{eq:M19_3}. Here, $\phi(.)$ denotes a relevance model; specifically, MonoT5 re-ranker in their implementation. 

\begin{equation}
\label{eq:M19_2}
\mathrm{SetAff}(d, S) = \sum_{d' \in S} P(\mathrm{Rel}(d')) \cdot f(d, d')
\end{equation}

\begin{equation}
\label{eq:M19_3}
P(\mathrm{Rel}(d')) = \frac{e^{\phi(q, d')}}{\sum_{d' \in S} e^{\phi(q, d')}}
\end{equation}


To address the limitation of GAR's query-agnostic nature, \cite{frayling2024effective} proposes Query-Document Graph (QDG), a weighted bipartite graph comprising query and document nodes, which explicitly accounts for user queries while leveraging inter-document relationships. QDG is motivated by the converse of the clustering hypothesis – two documents are similar if they are related to the same query. The QDG nodes are divided into two disjoint vertex sets: one for unique documents and another for unique queries. A generative model, Doc2Query \cite{nogueira2019document}, is employed to generate a collection of augmented queries for each document, since it is infeasible to rely on available datasets for such queries. Each edge encodes the relevance score between a document and its associated query. The nearest neighbor documents of a source document are determined by ranking all the documents two hops away from it, as shown in Eq. \ref{eq:M7_1}.

\begin{equation}
    Sim(d_1, d_2 | q_{org}) = Sim(q_{org}, q’) * Rel(q’, d_2)
\label{eq:M7_1}
\end{equation}

Here, $d_1$ denotes the source document, $d_2$ a document in the two-hop neighborhood of $d_1$, $q_{org}$ the original query, and $q’$ the augmented query shared between $d_1$ and $d_2$. QDG uses two re-ranking strategies, Reverted Adaptive Retrieval (RAR) and Resource Selection. Each approach replaces the corpus graph $G$ of the GAR \cite{macavaney2022adaptive} with the QDG. RAR defines neighbors in terms of augmented queries. The augmented queries relevant to each highly ranked document in a GAR step are used to retrieve additional documents to add to the frontier $F$. However, it does not measure the relevance of the neighbors associated with an augmented query, leading to resource waste. Hence, the latter approach ranks each neighborhood by averaging the relevance score of a set of sampled documents. Therefore, it only adds documents from the relevant neighborhood to the frontier.

In the same vein as adaptive re-ranking-based approaches \cite{macavaney2022adaptive, frayling2024effective, macavaney2022adaptiveagent, jaenich2024fairness, rathee2025guiding, rathee2025quam}, \cite{rathee2025breaking} addresses two notorious limitations of the multi-stage information retrieval pipeline: the recall limitation because of the initial retriever and the fixed re-ranking order dictated by the initial retrieval set. To this end, the authors propose Online Relevance Estimation (ORE), a dynamic re-ranking method inspired by the stochastic linear bandit problem \cite{chaudhuri2019pac, kalyanakrishnan2012pac} that iteratively updates the relevance estimates of the entire candidate pool. Here, candidate documents from an initially retrieved set are analogous to arms, and the relevance score from an expensive cross-encoder re-ranker is analogous to rewards. ORE estimates relevance scores over a large candidate pool while minimizing the relative error between estimated relevance score EstRel$(.)$ and that of an expensive cross-encoder re-ranker $\phi(.)$ (e.g., MonoT5). It employs $\phi(.)$ to score a smaller document subset constrained by budget $m$, which serves as ground-truth for learning the parameters $\vec{\alpha}$ of EstRel$(.)$, as defined in Eq. \ref{eq:M29_1}. The lightweight nature of EstRel$(.)$ enables efficient estimation of a large number of documents, thereby addressing the recall limitation. It is optimized using Eq. \ref{eq:M29_2} to minimize the discrepancy from the cross-encoder $\phi(.)$.

\begin{equation}
\label{eq:M29_1}
\mathrm{EstRel}(\vec{\alpha}, \vec{x}_d) = \vec{\alpha} \cdot \vec{x}_d^{\,T}
\end{equation}

\begin{equation}
\label{eq:M29_2}
E(\vec{\alpha}; q, d, \vec{x}_d)
= \frac{1}{2} \left| \phi(q, d) - \mathrm{EstRel}(\vec{\alpha}, \vec{x}_d) \right|^2
\end{equation}

During re-ranking, the relevance estimator is trained over a limited number of batches, which are then used to estimate candidate relevance for the remaining batches. In each iteration, ORE extracts the top-$b$ documents $B$ from the candidate list $D_q$ based on their EstRel($\cdot$) scores, which are then re-ranked using the expensive cross-encoder re-ranker within a limited budget $m$. Once the budget $m$ is exhausted, in subsequent batches, the trained EstRel($\cdot$) is used to re-rank the remaining candidates. The re-ranked documents are then used to update the initial candidate pool $D_q$ following the adaptive retrieval approach. In subsequent iterations, the estimated relevance is computed as follows: 

\begin{equation}
\alpha_1 \cdot \mathrm{Q2DAFF}(q, d) + \alpha_2 \cdot \mathrm{D2SetAFF}(q, d) + \alpha_3 \cdot x_7
\label{eq:M29_3}
\end{equation}

\begin{equation}
\mathrm{Score}(q, d') =
\begin{cases}
\phi(q, d') + \psi(q, d') & \text{if } d' \text{ is scored using } \phi \\
\mathrm{EstRel}(\vec{\alpha}, \vec{x}_{d'}) & \text{otherwise}
\end{cases}
\label{eq:M29_4}
\end{equation}

where $\psi$ denotes a dual-encoder model. The query-document affinity Q2DAff$(q,d)$ is computed using BM25$(q,d)$, while D2SetAff$(d,S)$ represents the affinity of a target document $d$ to the set $S$ of highly relevant documents at a given stage, as defined in Eq. \ref{eq:M29_5}. Here, D2DAff(d,d’) denotes the edge weight between documents $(d, d’)$, and $N_d$ refers to the neighborhood of document $d$ in the learned affinity graph $G_a$. Furthermore, the authors incorporate an additional signal, $x_7$, to ground the relevance estimation of document $d$ by averaging the relevance of its neighboring documents in the set $S$. Mapping these variables to Eq. \ref{eq:M29_1}, EstRel($\cdot$) is determined by the coefficients $\vec{\alpha} = [\alpha_1, \alpha_2, \alpha_3]$ and the corresponding input features $\vec{x}_d = [x_1, x_6, x_7]$. This process continues until the re-ranking budget $c$ is fully consumed. 

\begin{equation}
\mathrm{D2SetAff}(d, S) =
\frac{\sum_{d' \in S \cap N_d} \mathrm{D2DAff}(d, d')}
{\lvert S \cap N_d \rvert}
\label{eq:M29_5}
\end{equation}

\begin{equation}
x_7 =
\frac{\sum_{d' \in S \cap N_d} \mathrm{Score}(q, d')}
{\lvert S \cap N_d \rvert}
\label{eq:M29_6}
\end{equation}

\cite{kulkarni2023lexically} proposes LADR, Lexically-Accelerated Dense Retrieval, to bridge the efficiency-effectiveness trade-off between lexical and dense retrieval models. Analogous to a two-stage re-ranking pipeline, LADR initially retrieves a small set of seed documents $R_0$ using a lexical model and subsequently expands them by traversing a pre-computed document proximity graph in a dense vector space. The graph is constructed by linking every document with its 128 closest neighbors using TCT-ColBERT-HNP \cite{lin2021batch}. The authors propose two alternative strategies for the dense retrieval and re-ranking stage: proactive and reactive approaches. The former approach expands the candidate list by traversing the $k$ nearest neighbors of documents in $R_0$, followed by re-ranking with a dense retriever model. Adaptive LADR is an iterative approach designed to achieve higher recall. Starting with an initial set of seed documents $R_0$ scored via dense retrieval, it iteratively expands the candidate set by extracting and scoring neighboring documents of the top-$c$ results from $R_0$ until convergence.

\subsubsection{Entity Centric Re-ranking}
\cite{xiong2016bag} introduces knowledge graph information in document re-ranking by substituting the conventional bag-of-words (BoW) representation with a bag-of-entities (BoE) approach. To construct semantic representations, entities are extracted for queries and documents using three automatic entity linking systems, FACC1 \cite{gabrilovich2013facc1}, TagMe \cite{ferragina2011fast}, and CMNS \cite{hasibi2015entity}, applied to Freebase. Each query and document is then transformed into a BoE vector, where each dimension corresponds to a unique Freebase entity and is weighted by its frequency of occurrence in the text. Finally, the model employs two simple matching models, Coordinate Match and Entity Frequency, to compute the query-document relevance score. 

\cite{xiong2017word} proposes AttR-Duet, an attention-based re-ranking architecture that extends traditional word-based textual representation in IR by introducing a word-entity duet representation. It enriches query and document embeddings by integrating features from three distinct interaction layers: the word-space, the entity space, and the cross-space, the latter derived from knowledge graphs. Initially, dual representations are constructed for both queries and documents. A standard BoW model is used to obtain word-based representation ($Qw$ and $Dw$), while the entity-based representations ($Qe$ and $De$) are derived using a BoE approach through automatic entity linking \cite{xiong2016bag}. Based on these representations, the word-entity duet generates four types of ranking features for AttR-Duet: query words to document words ($Qw$-$Dw$), query entities to document words ($Qe$-$Dw$), query words to document entities ($Qw$-$De$), and query entities to document entities ($Qe$-$De$). The $Qw$-$Dw$ interaction is modeled using standard IR similarity functions, such as BM25 and Language Models. The $Qe$-$Dw$ interaction aligns textual attributes of query entities, namely, their names and descriptions, with the document text. The $Qw$-$De$ interaction captures the relationship between query terms and the textual attributes of document-linked entities as represented in Eq. \ref{eq:M20_1}.

\begin{equation}
\phi_{Qw\text{-}De}\supset\mathrm{max}\text{-}k\bigl(\{\, \mathrm{score}(q, e) \mid \forall e \in \mathrm{De} \,\}\bigr)
\label{eq:M20_1}
\end{equation}

The $Qe$-$De$ interaction leverages Explicit Semantic Ranking (ESR) \cite{xiong2017explicit} to capture two complementary matching signals in the entity space: exact matching and soft matching based on semantic relevance. At first, an entity translation matrix $T(\cdot)$ is constructed from the query and document entity embeddings generated by the TransE \cite{bordes2013translating} model. The translation matrix $T(e_i, e_j)$ encodes the similarity between entities $e_i$ and $e_j$ using L1 distance. Histogram pooling is then applied over the translation matrix to generate ranking features, which are subsequently used by AttR-Duet to compute the final relevance score. The model is jointly optimized using pairwise hinge loss to suppress noisy entities while effectively modeling query-document relevance.

Motivated by the importance of entity-level semantics in neural-IR, \cite{liu2018entity} introduces the Entity-Duet Neural Ranking model (EDRM). EDRM models query-document similarity by integrating knowledge graph semantics \cite{xiong2017word} into an interaction-based neural network \cite{dai2018convolutional}. The model learns to integrate KG semantics into neural ranking models end-to-end using user feedback from commercial search logs. Specifically, EDRM unifies three types of embeddings, entity embeddings, description embeddings, and type embeddings, to capture comprehensive semantic representations of entities, computed as follows:

\begin{equation}
\vec{v}_{e}^{\mathrm{sem}} = \vec{v}_{e}^{\mathrm{emb}} + W_{e}\!\left( \vec{v}_{e}^{\mathrm{des}} \oplus \vec{v}_{e}^{\mathrm{type}} \right)^{\top} + \vec{b}_{e}
\label{eq:M51_1}
\end{equation}

where the entity embedding $\vec{v}_{e}^{\mathrm{emb}}$ is a standard learned embedding vector. The description embedding $\vec{v}_{e}^{\mathrm{des}}$ is obtained by applying a CNN to the textual description associated with the entity, followed by max pooling, as defined in Eq. \ref{eq:M51_2}. To capture multiple entity types, the type embedding $\vec{v}_{e}^{\mathrm{type}}$ employs an attention mechanism that assigns weights to entity types conditioned on the query or document context, as shown in Eq. \ref{eq:M51_3}. In this formulation, $a_j$ denotes the attention matrix, computed from a projected BoW representation of the query or document through a learned matrix $W_{bow}$. Building on these representations, EDRM adopts the neural entity-duet framework \cite{xiong2017word} that matches query and document representation across four interaction spaces, word-to-word, word-to-entity, entity-to-word, and entity-to-entity, using cosine similarity. The resulting interaction matrices from the entity-duet framework are then integrated into existing ranking architectures, such as K-NRM \cite{xiong2017end}, and Conv-KNRM \cite{dai2018convolutional}, to estimate query-document relevance scores.

\begin{equation}
\vec{v}_{e}^{\mathrm{des}} = \max\!\left( \vec{g}_{e}^{\,1}, \ldots, \vec{g}_{e}^{\,j}, \ldots, \vec{g}_{e}^{\,m} \right),\; \text{where } \vec{g}_{e}^{\,j} = \mathrm{ReLU}\!\left( \mathbf{W}_{\mathrm{CNN}} \cdot \vec{v}_{w}^{\,j:j+h} + \vec{b}_{\mathrm{CNN}} \right)
\label{eq:M51_2}
\end{equation}

\begin{equation}
\begin{gathered}
a_j = \frac{\exp(P_j)}{\sum_{l=1}^{n} \exp(P_l)},\; \text{where } P_j = \left( \sum_i \mathbf{W}_{\mathrm{bow}} \vec{v}_{t_i} \right) \cdot \vec{v}_{f_j} \\
\vec{v}_{e}^{\mathrm{type}} = \sum_{j=1}^{n} a_j \vec{v}_{f_j},\; \text{where } \vec{v}_{f_j}^{\mathrm{emb}} = \mathrm{Emb}_{\mathrm{tp}}(e)
\end{gathered}
\label{eq:M51_3}
\end{equation}

Motivated by the importance of inter-document relationships, GAT-reRanker \cite{vollmersdocument} introduces a GAT as a cross-encoder over a query-document sub-graph $G = (V, E)$, in which nodes $v \in V$ represent entities extracted from query-document pairs and edges $e \in E$ capture their semantic relationships. First, the model employs FLAIR\footnote{\url{https://github.com/flairNLP/flair}} to extract entities from documents, while Llama3 is used for query-side entity disambiguation and extraction due to its superior performance on shorter text. GENRE\footnote{\url{https://github.com/facebookresearch/GENRE}}, an auto-regressive entity-linking model, then maps these extracted entities to their corresponding URIs within a knowledge graph. The linking process involves generating a candidate list of potential KG matches, which are then ranked by relevance to determine the final candidate. The Subgraph Retrieval Toolkit (SRTK)\footnote{\url{https://github.com/happen2me/subgraph-retrieval-toolkit}} is then utilized to extract RDF triples from the target knowledge graph. A GAT cross-encoder is used to refine node embeddings, as shown in Eq. \ref{eq:NE_1}, where $\alpha_{ij}^{(l)}$ refers to the attention coefficient. Finally, the query-document relevance is estimated using Eq. \ref{eq:NE_2}, where $h$ denotes refined entity embeddings.

\begin{equation}
h_i^{(l+1)} = \sigma\!\left( \sum_{j \in \mathcal{N}(i)} \alpha_{ij}^{(l)} \mathbf{W}^{(l)} h_{j}^{(l)} \right)
\label{eq:NE_1}
\end{equation}

\begin{equation}
r(q,d) = \sum_{i \in q} \sum_{j \in d} h_{i}^{\top} h_{j}
\label{eq:NE_2}
\end{equation}

Cross-encoders serve as a core component of many SOTA re-ranking systems. However, these approaches do not explicitly incorporate background knowledge, namely, query-related information that can be particularly valuable for passage retrieval. To address this limitation, \cite{fang2023kgpr} introduces KGPR, a KG-enhanced cross-encoder for passage re-ranking. This approach is built on LUKE \cite{yamada2020luke}, an entity-aware pre-trained language model, which is used to introduce external information from a knowledge graph into the re-ranking process. The LUKE-based cross-encoder estimates query-passage relevance using query, passage, and their corresponding entity embedding. A limitation of LUKE is that it does not model the relationship among entities. To bridge this gap, the proposed method uses a knowledge subgraph extracted from Freebase \cite{bollacker2008freebase} to represent the entity relationships associated with each query-passage pair. Subgraph extraction proceeds in two stages: entity linking followed by subgraph retrieval. The first stage identifies entities in both the query and the passage, aligns them with nodes in Freebase. For this purpose, KGPR employs ELQ \cite{li2020efficient}, an entity-linking model designed for questions. The extracted passage entities are then traced to within one hop of the query entities in Freebase to build the subgraph, $G_{q,d}$, while filtering out edges unrelated to passage entities. The authors also introduce a dedicated embedding for relations in a KG triple, $<h, r, t>$ where $h$, $r$, and $t$ correspond to the head entity, relation, and tail entity, respectively. It allows encoding the triples representing the subgraph as an additional input to LUKE for calculating the query-passage relevance score while accounting for background information.

A knowledge graph helps PLM-based re-rankers address the limitation of inadequate domain-specific information by supplementing explicit knowledge with the PLM's parametric (implicit) knowledge. Nevertheless, the noise present in existing KGs can reduce their effectiveness for re-ranking. To address this issue, \cite{dong2022incorporating} proposes the Knowledge Enhanced Re-ranking Model (KERM), which extends cross-encoder re-rankers by distilling a knowledge meta graph from existing KGs. The framework consists of two primary parts: the knowledge graph distillation module and the knowledge aggregation module. In the first stage, KERM removes noisy information from a global knowledge graph $G$ to obtain a refined graph $G'$ by selecting relevant entities using TransE embeddings \cite{bordes2013translating}. During pruning, KERM retains only the top-$\pi$ closest neighbors per entity, where the distance metric dist$(e_h, e_t)$ is computed using Eq. \ref{eq:M2_1} with $r$ representing the relation between the head entity $e_h$ and the tail entity $e_t$ and $\mathbf{E}(\cdot)$ representing the TransE embedding function.

\begin{equation}
\mathrm{Dist}(e_h, e_t) = \frac{1}{\mathbf{E}(e_h) \cdot \mathbf{E}(r) + \mathbf{E}(e_h) \cdot \mathbf{E}(e_t) + \mathbf{E}(r) \cdot \mathbf{E}(e_t)}
\label{eq:M2_1}
\end{equation}

Next, KERM constructs a bipartite meta-graph $G_{q,p}$ for each query-passage pair by selecting a representative sentence $s^*$ using Eq. \ref{eq:M2_2}, where $\mathbf{E}(\cdot)$ denotes the  Word2Vec \cite{mikolov2013linguistic} embedding, extracting entities from both sources and connecting them via $k$-hop Breadth-First Search (BFS) expansion over $G'$.

\begin{equation}
\mathrm{Rel}_{qs}(q, s_i)
=
\frac{\sum_{q=1}^{\lvert q \rvert} \mathbf{E}(w_q)}{\lvert q \rvert}
\cdot
\frac{\sum_{s=1}^{\lvert s_i \rvert} \mathbf{E}(w_s)}{\lvert s_i \rvert}
\label{eq:M2_2}
\end{equation}

While the meta graph provides explicit knowledge crucial for re-ranking tasks in niche domains, integrating this structured data with the implicit parametric knowledge of PLMs remains challenging due to their heterogeneity arising from substantial discrepancies in their sources. To mitigate this limitation, the knowledge aggregation phase leverages a knowledge injector that fuses word embeddings and entity embeddings through a Graph Meta Network (GMN). Initially, entities in $G_{q,p}$ are initialized with TransE embedding, while a cross-encoder initializes the query-passage pair. At each $l$-th layer of the knowledge injector, the query-passage embeddings are first processed through a multi-head attention mechanism, followed by alignment with the entity embeddings from the meta-graph \cite{zhang2019ernie}. The fused representation is subsequently fed into a feed-forward network to produce an updated text representation for the $l$-th layer. Simultaneously, the GMN updates the meta-graph nodes by aggregating information from neighboring nodes, ensuring alignment with the textual representation. The final output $O_M$ of the knowledge injection module is then used to compute the query-passage relevance score, as:

\begin{equation}
f(q, p \mid \mathcal{G}) = \sigma\!\left(\mathbf{O}_{M}^{\text{[CLS]}} \mathbf{W}^{4} + \mathbf{b}^{4} \right)
\label{eq:M2_4}
\end{equation}

\cite{chatterjee2025qder} introduces Query-Specific Document and Entity Representations (QDER), a re-ranking framework that unifies entity-oriented Neural IR \cite{liu2018entity, xiong2017word, tran2022dense} with multi-vector ranking models \cite{khattab2020colbert, luan2021sparse}. QDER addresses two key limitations of existing IR approaches: the reliance on query-agnostic document embeddings and the use of coarse matching signals (e.g., cosine similarity), which lack the granularity to model intricate query-document token interactions. At its core, QDER represents documents as collections of token and entity embeddings and computes query-document relevance using a late interaction mechanism. Specifically, a BERT-based text channel encodes query and document tokens, while an entity channel captures higher-level semantic concepts, as shown in Eq. \ref{eq:M43_1}: 

\begin{equation}
\begin{gathered}
Q^{t} = \mathrm{Encoder}(q) \in \mathbb{R}^{l_q \times d_t},\; D^{t} = \mathrm{Encoder}(d) \in \mathbb{R}^{l_d \times d_t} \\
Q^{e} \in \mathbb{R}^{n_q \times d_e},\; D^{e} \in \mathbb{R}^{n_d \times d_e}
\end{gathered}
\label{eq:M43_1}
\end{equation}

where $l_q$ and $l_d$ denote the sequence lengths, while $n_q$ and $n_d$ represent the respective entity counts for the query and the document. The text and entity embedding dimensions are denoted by $d_t$ and $d_e$, respectively. Document entities are extracted through entity linking, whereas query entities are derived by pooling concepts from candidate documents, following the methodology of \cite{meij2010conceptual}. Departing from traditional query-agnostic representations, QDER employs an attention mechanism to synthesize query-specific document embedding across both the text and entity channels as defined in Eq. \ref{eq:M43_2}. Subsequently, QDER captures two complementary relevance signals: semantic alignment, which models  precise query-document similarity (Eq. \ref{eq:M43_4}) and semantic complementarity, which captures how the document expands upon the query (Eq. \ref{eq:M43_5}). 

\begin{equation}
\begin{gathered}
\tilde{D}^{t} = A^{t} D^{t},\; \text{where } A^{t} = \mathrm{softmax}\!\left( Q^{t} (D^{t})^{\top} \right) \in \mathbb{R}^{l_q \times l_d} \\
\tilde{D}^{e} = A^{e} D^{e},\;  \text{where } A^{e} = \mathrm{softmax}\!\left( Q^{e} (D^{e})^{\top} \right) \in \mathbb{R}^{n_q \times n_d} 
\end{gathered}
\label{eq:M43_2}
\end{equation}

\begin{equation}
M^{t} = Q^{t} \odot \tilde{D}^{t} \;\text{and}\; M^{e} = Q^{e} \odot \tilde{D}^{e}
\label{eq:M43_4}
\end{equation}

\begin{equation}
C^{t} = Q^{t} + \tilde{D}^{t} \;\text{and}\; C^{e} = Q^{e} + \tilde{D}^{e}
\label{eq:M43_5}
\end{equation}

These multi-granular interaction patterns are consolidated into a unified feature vector $h = [h_m^t; h_c^t; h_m^e; h_c^e]$, where $h_*^{t} = \mathrm{MeanPool}(*) \text{ for } * \in \{ C^{t}, M^{t} \}, \quad h_*^{e} = \mathrm{MeanPool}(*) \text{ for } * \in \{ C^{e}, M^{e} \}$. A bilinear interaction over this joint representation is then used to estimate query-document relevance (Eq. \ref{eq:M43_6}), where $M \in \mathbb{R}^{d \times d}$ represents a learnable interaction matrix. This formulation is further extended with a hybrid scoring function to incorporate lexical matching signals for robust retrieval performance, as defined in Eq. \ref{eq:M43_7}, where $\lambda \in [0, 1]$ controls the contributions of each component.

\begin{equation}
\mathrm{score}_{\mathrm{QDER}}(q,d) = h^{\top} M h = \sum_{i=1}^{d} \sum_{j=1}^{d} h_i M_{i,j} h_j
\label{eq:M43_6}
\end{equation}

\begin{equation}
\mathrm{score}_{\mathrm{hybrid}}(q,d) = \lambda \cdot \mathrm{score}_{\mathrm{BM25}}(q,d) + (1 - \lambda) \cdot \mathrm{score}_{\mathrm{QDER}}(q,d)
\label{eq:M43_7}
\end{equation}

Despite significant advances, entity-oriented neural IR models often fail to account for the varying degree of influence that individual entities exert on query-document relevance estimation. To overcome this limitation, \cite{chatterjee2024dreq} proposes Document Re-ranking using Entity-based Query Understanding (DREQ), which amplifies query-relevant entities within the document representation while suppressing those with marginal relevance. Specifically, DREQ first extracts a set of entities $\mathcal{E}$ from candidate documents and generates BERT-based entity embedding $\mathbf{e} \in R^k$ by jointly encoding their textual descriptions from a Knowledge Base (DBpedia \cite{lehmann2015dbpedia}) and the query tokens. Entity relevance is then estimated via a linear scoring function $S(e, Q) = W_{1} \cdot \mathbf{e} + b_{1}$, where $W_1$ represents a weight matrix and $b_1$ denotes scalar bias. Subsequently, for each document $d \in \mathcal{D}$, a query-specific representation $\mathbf{V}_{e_d}^{Q}$ is generated via a weighted sum of its entity embeddings derived from Wikipedia2Vec \cite{yamada2020wikipedia2vec}, as shown in Eq. \ref{eq:M44_1}.

\begin{equation}
\mathbf{V}_{e_d}^{Q} = \sum_{e \in e_d} w_{e} \cdot \mathbf{e},\; \text{where } w_{e} = \frac{S(e, Q)}{\sum_{e' \in d} S(e', Q)}
\label{eq:M44_1}
\end{equation}

Additionally, a text-centric document representation $\mathbf{V}_{td}^Q$ is obtained by splitting each candidate document into passages, encoding them with a BERT model, and averaging their embeddings. Subsequently, the entity-centric and text-centric representations are combined to form a hybrid document representation $\mathbf{d}^{Q}$, as depicted in Eq. \ref{eq:M44_2}. In this formulation, $W_2$ is the weight matrix and $\mathbf{b}$ is the bias term. The resulting hybrid representation $\mathbf{d}^Q$ captures both fine-grained entity-level information and the overall contextual semantics of the document text. The query-document relevance score $S(d, Q)$ is then computed by applying a linear projection over fine-grained interactions between the query and the hybrid document representations, including additive, subtractive, and multiplicative interactions, as defined in Eq. \ref{eq:M44_3}. 

\begin{equation}
\mathbf{d}^{Q} = W_{2} \cdot \left[ \mathbf{V}_{t_d}^{Q},\, \mathbf{V}_{e_d}^{Q} \right] + \mathbf{b}
\label{eq:M44_2}
\end{equation}

\begin{equation}
S(d, Q) = W_{3} \cdot  \mathbf{V} + b,\; \text{where } \mathbf{V} = \left[ \mathbf{Q} \,;\, \mathbf{d}^{Q} \,;\, \mathbf{V}_{\mathrm{add}}^{d,Q} \,;\, \mathbf{V}_{\mathrm{sub}}^{d,Q} \,;\, \mathbf{V}_{\mathrm{mul}}^{d,Q} \right]
\label{eq:M44_3}
\end{equation}

\subsubsection{Domain Specific Re-ranking}
PLM-based re-rankers struggle on domain-specific tasks, particularly in the biomedical domain, where high variability exists among synonymous and abbreviated medical terminologies. In response to this shortcoming, \cite{gupta2024empowering} introduces GraphMonoT5, which incorporates external information from a knowledge graph into a GNN-fused T5 encoder \cite{raffel2020exploring}, enabling effective biomedical document re-ranking. GraphMonoT5 comprises $R$ layers of T5 text encoder, a GNN module, $S$ layers of aggregation module, and a T5 decoder. Inspired by \cite{zhang2022greaselm}, it leverages interaction tokens $t_{int}$ and interaction nodes $n_{int}$ to facilitate information exchange between text and graph representations. To generate node embeddings, entities are first extracted from each query-document pair and are linked to a knowledge graph. A subgraph $\mathcal{G}_{q,d} = (\mathcal{V}_{q,d}, \mathcal{E}_{q,d})$ is then constructed by connecting all the nodes that lie on 2-hop paths between the corresponding entities. The vertices $\mathcal{V}_{q,d} = \mathcal{V}_q \cup \mathcal{V}_d$ and edges $\mathcal{E}_{q,d} = \mathcal{E}_q \cup \mathcal{E}_d$ in $\mathcal{G}_{q,d}$ depict the total nodes and edges corresponding to query $q$ and document $d$, respectively. The T5 encoder takes the token sequence $\mathcal{T} = \{ t_{int}, t_1, t_2, ..., t_N \}$ from query-document pair $(q,d)$ and generates corresponding embedding as $H^l = \{ h_{int}^l, h_1^l, h_2^l, ..., h_N^l \} \in \mathcal{R}^{(N+1) \times d_l}$. Similarly, the GNN module generates node embeddings $U^l = \{ u_{int}^l, u_1^l, u_2^l, ..., u_M^l \} \in \mathcal{R}^{(M+1) \times d_g}$ from nodes $\{ n_{int}, n_1, n_2, ..., n_M \}$ in the subgraph $\mathcal{G}_{q,d}$. Following this, the aggregation module models the interaction between two representations using a feed-forward network as $x^l = f(h_{int}^l \oplus u_{int}^l)$, which is further refined using Mutual Information (MI). The fused representation is finally utilized in the T5 decoder to generate the query-document relevance score.

\cite{pappas2020aueb} addresses biomedical document retrieval by integrating  graph-based document representations with a neural re-ranking model. To support embedding generation, the authors construct a biomedical entity co-occurrence graph, where nodes correspond to entities extracted from PubMed\footnote{\url{http://www.ncbi.nlm.nih.gov/pubmed/}} abstracts, and edges link entities that co-occur within the same abstract. Rare co-occurrences are pruned to reduce noise. Word embeddings are then learned by applying Node2Vec \cite{grover2016node2vec}, a graph-based node embedding method, to the resulting co-occurrence graph.

Neural ranking models struggle to capture query-document semantic similarity in biomedical IR. This stems from two key factors: the diverse representations of biomedical terms and the dispersed nature of query-related fact descriptions. To account for this, \cite{liu2025knowledge} presents KEGNR, which models semantic association between query-document pairs using a graph with heterogeneous edges. To further reduce the query-document semantic gap, it integrates external knowledge into a unified knowledge-query-doc graph. An edge-driven GNN is employed to capture the dispersed matching signal on a heterogeneous graph, and the resulting graph representation is then used in a graph classification framework to estimate query-document relevance. At first, KEGNR employs SciBERT \cite{beltagy2019scibert} to encode the query-document representation $H \in \mathbb{R}^{n \times d_h}$, defined as ${H} = [{h}_1, {h}_2, \ldots, {h}_n] = \mathrm{SciBERT}(\text{[CLS]}~ q ~\text{[SEP]}~ d ~\text{[SEP]})$. Subsequently, an undirected, heterogeneous knowledge-query-doc graph $G = (V, E)$ is constructed, comprising three node types $V$: query term nodes ($Q$), document term nodes ($T$), and sentence nodes ($S$). To capture the dispersed yet relevant information, the graph defines five edge types: Query Term-Query Term (QQ), Query Term-Document Term (QT), Document Term-Document Term (TT), Document Term-Sentence (TS), and Sentence-Sentence (SS). The proposed edge-driven graph model includes $L$ identical relational graph convolution network (R-GCN) layers, an edge-driven self-attention graph pooling layer (E-SAGP), and a readout layer. R-GCN updates node $v_i$ at layer $l$ by aggregating information from its neighbors while explicitly accounting for edge types, as shown in Eq. \ref{eq:M40_1}.

\begin{equation}
\mathbf{n}_i^{\,l} :=
\sigma\!\left(
\sum_{r \in \mathcal{R}}
\sum_{v_j \in \mathcal{X}_{v_i}^{r}}
\frac{1}{\lvert \mathcal{X}_{v_i}^{r} \rvert}
\mathbf{W}_r^{\,l-1}\mathbf{n}_j^{\,l-1}
+ \mathbf{W}_o^{\,l-1}\mathbf{n}_i^{\,l-1}
\right)
\label{eq:M40_1}
\end{equation}

In this formulation, $\mathcal{R}$ denotes the set of edge types, and $\mathcal{X}_{v_i}^r$ represents the neighbors of node $v_i$ connected via edge type $r \in \mathcal{R}$. The matrices $W_r$ and $W_o$ are learned parameters, and $\sigma(.)$ denotes the ReLU activation function. Next, E-SAGP identifies the most salient nodes and edges by concatenating features across all $L$ R-GCN layers, computing node-level self-attention scores via another R-GCN, and selecting the top nodes $X$ based on a pooling ratio $K \in (0, 1]$, as defined in Eq. \ref{eq:M40_2}, Eq. \ref{eq:M40_3}, and Eq. \ref{eq:M40_4}, respectively.

\begin{equation}
\mathbf{N} = \mathrm{Concat}\!\left([ \mathbf{N}^{1}, \mathbf{N}^{2}, \ldots, \mathbf{N}^{L} ]\right)
\label{eq:M40_2}
\end{equation}

\begin{equation}
\mathbf{A} = \mathrm{R\text{-}GCN}(\mathbf{N}, \mathcal{X})
\label{eq:M40_3}
\end{equation}

\begin{equation}
\begin{gathered}
\mathit{idx} = \mathrm{top\_rank}(\mathbf{A}, [k \cdot t]) \\
\mathbf{N}^{L} = \mathbf{N}^{L}_{\mathit{idx}} \\
\mathbf{A}_{\text{mask}} = \mathbf{A}_{\mathit{idx}} \\
\mathbf{X} = \mathbf{N}^{L} \odot \mathbf{A}_{\text{mask}}
\end{gathered}
\label{eq:M40_4}
\end{equation}





A readout layer then aggregates node features into a graph-level representation $\mathbf{s}$, as defined in Eq. \ref{eq:M40_5}. The relevance score is then computed by passing $\mathbf{s}$ through three fully connected layers optimized with cross-entropy loss, as shown in Eq. \ref{eq:M40_6}. Here, $L$ denotes the loss function and rel$(q,d)$ represents the query-document relevance score. 

\begin{equation}
\mathbf{s} =
\frac{1}{Z}
\sum_{i=1}^{Z}
\mathbf{X}_{i}
\;\Vert\;
\max_{1 \le j \le d_n} \mathbf{X}_{:.j}
\label{eq:M40_5}
\end{equation}
 
\begin{equation}
\begin{gathered}
o = \mathrm{ReLU}\!\left(\mathrm{ReLU}( \mathbf{s}\mathbf{W}_1 + \mathbf{b}_1 )\mathbf{W}_2 + \mathbf{b}_2\right)\mathbf{W}_3 + \mathbf{b}_3 \\
\mathrm{rel}(q, d) = \hat{y} = \mathrm{sigmoid}(o) \\
L(q, d) = -\left(y \log(\hat{y}) + (1 - y)\log(1 - \hat{y}) \right)
\end{gathered}
\label{eq:M40_6}
\end{equation}

To improve the factuality in medical QA, KG-Rank \cite{yang2024kg} uses an external knowledge graph in a multi-stage retrieval and filtering pipeline. First, medical terminology is extracted from the query using PMedNER, a prompt-based named-entity recognition (NER) model, and mapped to their corresponding concepts in the Unified Medical Language System (UMLS) \cite{bodenreider2004unified}. The one-hop triplets associated with these entities are then retrieved from UMLS and filtered using three ranking strategies: similarity ranking, answer expansion (AE) ranking, and maximal marginal relevance (MMR) \cite{carbonell1998use}. Similarity ranking uses embedding space proximity to align triplets with the question, while AE ranking utilizes an initial response generated by an LLM to identify contextually relevant triplets. In contrast, the MMR ranking promotes diversity by iteratively selecting triplets that differ from the set of triplets already selected. Filtered triplets are further refined using MedCPT \cite{jin2023medcpt}, a medical cross-encoder re-ranker. The top-ranked triplets are subsequently incorporated into the generator via a KG-enhanced prompt to produce the final, factually grounded answer.

\subsubsection{Specialized Models}
\cite{kohail2017matching} introduces a supervised document re-ranker for Semantic Textual Similarity (STS) that integrates dependency graph-based similarity and coverage features with traditional lexical features. STS is a natural language processing (NLP) task that quantifies the degree of semantic similarity between a pair of texts. The proposed method combines lexical, semantic, and syntactic information for enhanced relevance estimation. Each text pair is modeled using multiple similarity measures derived from diverse representations, including BoW, topic distributions, named entities, dependency graph similarity, and lexical expansion via distributional thesaurus. A novel \textit{approximate dependency subgraph alignment} algorithm aligns the query's dependency structure with a subgraph extracted from the candidate document, accommodating node gaps and syntactic mismatches. Furthermore, to prevent bias towards longer documents, the authors introduce coverage features that model one-to-one correspondence between dependency graphs. The similarity measures and coverage features are then utilized in a Multilayer Perceptron (MLP) to re-rank the top-$N$ candidates from the initial retriever.

The presence of multiple passages within long documents has motivated PLM-based methods \cite{yan2019idst, chen2019ucas} to estimate relevance through passage-level aggregation. This design, however, incurs significant computational overhead due to the quadratic attention complexity of transformers. To bridge this gap, \cite{reed2020faster} samples a subset of representative passages and estimates query-document relevance using the most relevant passage from the sampled set. Therefore, the overall effectiveness of the system depends heavily on the quality of the passage extraction technique. Among the proposed approaches, the most promising method introduced in this work is PassageRank. Inspired by TextRank \cite{mihalcea2004textrank}, PassageRank models each passage as a node in a directed graph $G = (V, E)$, while edges capture similarity relations among nodes. The score of each node $V_i$ is calculated using Eq. \ref{eq:M11_1}, where $d$ is the dampening factor. $In(V_i)$ and $Out(V_i)$ denote the incoming and outgoing neighbor sets of $V_i$, respectively. The process starts with arbitrary node scores and runs until convergence. The most relevant passage then determines the ranking score of the corresponding query-document pair. 

\begin{equation}
\mathrm{WS}(V_i) =(1 - d) + d \sum_{V_j \in \mathrm{In}(V_i)}\frac{w_{ji}}{\sum_{V_k \in \mathrm{Out}(V_j)} w_{jk}} \, \mathrm{WS}(V_j)
\label{eq:M11_1}
\end{equation}

Numerous IR research \cite{macavaney2022adaptive, frayling2024effective, macavaney2022adaptiveagent, jaenich2024fairness, rathee2025guiding, rathee2025quam} built their foundation on the clustering hypothesis \cite{jardine1971use}, which argues that documents within a cluster satisfy the same information need. However, inter-document similarity signals can be distorted by irrelevant passages, thereby adversely affecting document relevance. To address this problem, \cite{sarwar2021graph} proposes a cohesion-driven graph-based re-ranker that mitigates the impact of noisy passages by incorporating a cohesion score into the query-document relevance estimation. Document cohesion is measured by assessing topic shifts across passages; topic-consistent passages yield higher cohesion scores. The methods first segment each document into passages using a fixed-length, half-overlapping  window. The passages are then organized into a weighted directed graph $G = (V, E)$ where each vertex $v_i \in V$ corresponds to a passage and an edge $e_{i,j} \in E$ models the relationship between two passages $p_i$ and $p_j$ weighted by their similarity $sim(p_i, p_j)$ computed using Lucene\footnote{\url{https://lucene.apache.org/core/3 5 0/scoring.html}}. Based on the graph $G$, the model computes a cohesion score $C(d_i)$ for each document $d_i$, as defined below: 

\begin{equation}
C(d_i) = \frac{\sum_{\forall p_{ji}} \mathrm{sim}(p_{ji}, n_{ji})}{N(N-1)} \mid n_{ji} \in d_i
\label{eq:M15_1}
\end{equation}

where $N$ denotes the total number of passages in a document, $p_{ji}$ represents the $j$-th passage of $d_i$, and $n_{ji}$ is the neighbor of $p_{ji}$. Eq. \ref{eq:M15_2} integrates the cohesion score into the final query-document relevance estimation $osim(\cdot)$, thereby amplifying the score of cohesive documents. Empirical results show that weighting the cohesion score at $10\%$ in the relevance estimation leads to the best overall performance.   

\begin{equation}
\mathrm{osim}(d_i, q) = \mathrm{sim}(d_i, q) + \big( C(d_i) \times X \big) \mid X = 0.1
\label{eq:M15_2}
\end{equation}

While many neural IR models prioritize local term-level interaction signals, they often struggle to capture the long-range dependencies essential for understanding document context. To bridge this gap, \cite{zhang2021graph} proposes the graph-based relevance matching model (GRMM). GRMM re-conceptualizes a document as a graph-of-words $\mathcal{G} = (\mathcal{V}, \mathcal{E})$, where nodes correspond to unique words. Node features are derived from the query-document word interaction in embedding space, defined as $S_{ij} = \mathrm{cosine}\!\left( e_{i}^{(d)},\, e_{j}^{(q)} \right)$. An edge is introduced between two nodes if their corresponding document words co-occur within a sliding window, with the edge weight defined by the frequency of such co-occurrences. Node representations are subsequently refined using a GNN through message passing to incorporate intra-document interactions alongside query-document relations, as formulated in Eq. \ref{eq:M58_1}, where $a_{i}^{t} \in \mathbb{R}^M$ denotes the aggregated message. 

\begin{equation}
a_{i}^{t} = \sum_{(w_i, w_j) \in \mathcal{E}} \tilde{A}_{ij}\,\mathbf{W}_{a}\mathbf{h}_{j}^{t}
\label{eq:M58_1}
\end{equation}

To mitigate noise and prevent over-smoothing, GRMM utilizes a gated recurrent unit (GRU) and further refine node states: $\mathbf{h}_{i}^{t+1} = \mathrm{GRU}\!\left( \mathbf{h}_{i}^{t},\, \mathbf{a}_{i}^{t} \right)$, where the reset gate servers as a learnable filter to suppress irrelevant signals. To maintain robustness across document lengths, the model employs $k$-max pooling over the feature matrices for each query term, retaining only the most salient interaction signals. These pooled features are then weighted by a soft gating network and processed by a multi-layer perceptron (MLP) to produce the final relevance score $rel(q, d)$.

Standard PLM-based ranking models compute all pairwise token relations within the attention matrix, whereas traditional IR models primarily focus on query-document relevance. As a result, irrelevant information, such as query-query and document-document token relations, can introduce noise and degrade the performance of PLM-based ranking models \cite{dai2019deeper}, particularly with short queries. To address this issue, \cite{dong2021latent} proposes the latent graph recurrent network (LGRe), which refines word representations produced by PLMs (e.g., BERT) through graph recurrent neural networks (GRNNs). Using a masking strategy, LGRe organizes transformer-derived word representations into a bipartite-core query-document word graph, which is then refined through GRNN-based propagation. The resulting representations are passed through a fully connected layer to compute the final ranking scores. Initially, the model generates BERT-based representations for each query-document pair (q, d), where the input is formatted as a concatenated token sequence in the embedding space $I^{(q,d)}$. At each transformer layer $l$, the contextualized token representation $E_l^{(q,d)}(i) \in \mathbb{R}^{d_k}$ for the $i$-th token is computed using Eq. \ref{eq:M26_1}, where $A^{(q, d)}$ denotes the attention matrix and $E_0^{(q,d)}$ is initialized from $I^{(q,d)}$. 

\begin{equation}
\begin{gathered}
A_{l-1}^{(q,d)} = \mathrm{softmax}\!\left(\frac{\left(\mathbf{W}_{B}\mathbf{E}_{l-1}^{(q,d)}\right)\left(\mathbf{W}_{B}\mathbf{E}_{l-1}^{(q,d)}\right)^{\top}}{\sqrt{d_k}}\right) \\
\mathbf{E}_{l}^{(q,d)}(i) = \mathbf{E}_{l-1}^{(q,d)}(i) + \sum_{j}A_{l-1}^{(q,d)}(i,j)\,\mathbf{E}_{l-1}^{(q,d)}(j)
\end{gathered}
\label{eq:M26_1}
\end{equation}

Subsequently, the $L$ transformer layers produce a set of attention matrices $\left\{ A_{l}^{(q,d)} \right\}_{l=1}^{L}$. However, these matrices include query-query and document-document token interactions, which may introduce noise into the query-document relevance estimation. To reduce these extraneous interactions, a masking matrix $M_l^{(q,d)}$ is applied at each transformer layer, producing a masked word adjacent matrix that characterizes the bipartite-core graph $\hat{A}_{l}^{(q,d)}$, as defined in Eq. \ref{eq:M26_2}, where $\epsilon$ is a small constant.

\begin{equation}
\hat{A}_{l}^{(q,d)} = \mathrm{softmax}\!\left( \frac{ (\mathbf{W}_{A}\mathbf{E}_{l}^{(q,d)})(\mathbf{W}_{A}\mathbf{E}_{l}^{(q,d)})' }{\sqrt{d_k}} + \epsilon \left( 1 - \mathcal{M}_{l}^{(q,d)} \right) \right)
\label{eq:M26_2}
\end{equation}

In both the attention and masking matrices, the first $m$ rows and columns correspond to query tokens, while the subsequent $n$ indices correspond to document tokens. To filter irrelevant interactions, the authors introduce two masking strategies, named according to the resulting graph structures: \textit{the query-document bipartite word graph} and the \textit{query-document bipartite and neighbor word graph}. In the former, only query-document interactions are preserved through the masking matrix defined in Eq. \ref{eq:M26_3}. However, recognizing that word order is critical for search tasks, particularly for long candidate documents, the \textit{query-document bipartite and neighbor word graph} extends Eq. \ref{eq:M26_3} to preserve selected document-document interactions, as specified in Eq. \ref{eq:M26_4}.

\begin{equation}
\mathcal{M}_{l}^{(q,d)}(i,j)
=
\begin{cases}
1, & 1 \le i \le m,\; m+2 \le j \le m+n+2 \\
1, & i = j \\
0, & \text{otherwise}
\end{cases}
\label{eq:M26_3}
\end{equation}

\begin{equation}
\mathcal{M}_{l}^{(q,d)}(i,j)
=
\begin{cases}
1, & 1 \le i \le m,\; m+2 \le j \le m+n+2 \\
1, & m+2 \le i \le m+n+2,\; j = i, \ldots, i+r \\
0, & \text{otherwise}
\end{cases}
\label{eq:M26_4}
\end{equation}

Using the bipartite-core graph $\hat{A}_{l}^{(q,d)}$, the BERT-derived word representations $E_l^{(q,d)}$ are refined using gated graph neural networks (GGNNs), which update each word node by propagating information from its neighboring nodes, as defined in Eq. \ref{eq:M26_5}. Subsequently, a graph-level representation $h_l^{G_{d,q}}$ for each query-document pair is computed using Eq. \ref{eq:M26_6}, where $h_T^{att}$ denotes the word attention matrix. Finally, relevance $f(q,d)$ score is obtained by aggregating layer-wise scores $\mathrm{s_l}(q,d)$, where each $\mathrm{s_l}(q,d)$ is produced by feeding $h_l^{G_{q,d}}$ into a fully connected layer, as depicted in Eq. \ref{eq:M26_7}.

\begin{equation}
\mathbf{h}_t = \mathrm{GRU}\!\left([\mathbf{h}_{t-1},\, \hat{A}_{l}^{(q,d)} \mathbf{h}_{t-1}]\right),\; \text{where } \mathbf{h}_0 = \mathbf{E}_{l}^{(q,d)}
\label{eq:M26_5}
\end{equation}

\begin{equation}
\begin{gathered}
\mathbf{h}_{T}^{\mathrm{att}} = (\mathbf{W}_{a}\mathbf{h}_{T}) \cdot (\mathbf{W}_{h}\mathbf{h}_{T})' \\
\mathbf{h}_{l}^{G_{q,d}} = \bigl[ \mathrm{sum}(\mathbf{h}_{T}^{\mathrm{att}}) + \max(\mathbf{h}_{T}^{\mathrm{att}}),\, \mathbf{E}_{l}^{(q,d)}(0) \bigr]
\end{gathered}
\label{eq:M26_6}
\end{equation}

\begin{equation}
f(q,d) = \mathbf{w}_{f}\,\mathrm{s_l}(q,d)_{1\times L} + b_{f},\; \text{where } \mathrm{s_l}(q,d) = \mathbf{W}_{s}\mathbf{h}_{l}^{G_{q,d}} + \mathbf{b}_{s}
\label{eq:M26_7}
\end{equation}

Building upon \cite{dong2021latent}, the disentangled graph recurrent network (DGRe) \cite{dong2022disentangled} enhances document re-ranking by incorporating refined word representations into transformer-based ranking models. Grounded in the causal inference framework \cite{pearl1995causal}, DGRe leverages causal graphs to explicitly model causal effects in relevance estimation, thereby mitigating the influence of confounding factors such as spurious word relations and irrelevant contextual information. In this setting, the document ranking problem is formulated as a do-operation query $P(Y|do(X))$ which, intuitively, asks: what is the relevance between a query and a document when only genuine word relations and relevant contextual signals are considered? Irrelevant contextual information is removed by blocking the front door path $X \xrightarrow{} Z \xrightarrow{} Y$, as formulated in Eq. \ref{eq:M27_1}, where $X$ denotes the query-document pair embedding, $Z$ represents their contextualized representations, and $Y$ denotes the relevance label. The context representation is consequently divided into query-relevant $Z_r$ and query-irrelevant $Z_n$ components. Furthermore, to eliminate spurious correlations, the back-door path $X \leftarrow B \xrightarrow{} Y$ is blocked by estimating the do-operation query $P(Z|do(X))$, as defined in Eq. \ref{eq:M27_2}. In this formulation, $B$ denotes the self-attention matrix generated from $X$, and $B_+$ and $B_-$ correspond to useful and spurious relations, respectively. By combining these two mechanisms, the final prediction function is computed, as shown in Eq. \ref{eq:M27_3}. 

\begin{equation}
P(Y \mid do(X)) = \sum_{z} P(Y \mid do(Z)) P(Z \mid do(X)) = \sum_{Z_j \in \{Z_r, Z_n\}} P(Y \mid Z_j) P(Z_j \mid do(X))
\label{eq:M27_1}
\end{equation}

\begin{equation}
P(Z_j \mid do(X)) = \sum_{B_i \in \{B_{+}, B_{-}\}} P(Z_j \mid X, B_i) P(B_i)
\label{eq:M27_2}
\end{equation}

\begin{equation}
P(Y \mid do(X)) = \mathbb{E}_{Z}\,\mathbb{E}_{B}\!\left[ P(Y \mid X, Z) \right] \propto \exp\!\left( g\!\left( \mathbb{E}_{Z}[Z],\, \mathbb{E}_{B}[X] \right) \right)
\label{eq:M27_3}
\end{equation}

Here, the expectation $\mathbb{E}_Z[Z]$ corresponds to word representations with irrelevant contextual information removed, while $\mathbb{E}_B[X]$ denotes word representations free from spurious relations. Although the heuristic masking strategy, defined in Eq. \ref{eq:M26_3}, restricts query-document pair interactions using a bipartite word graph, not all document words contribute meaningfully to query-document relevance estimation. To bridge this gap, DGRe employs an adaptive masking strategy that dynamically distinguishes useful from spurious relations, as shown in Eq. \ref{eq:M27_4}. Specifically, the ReLU activation function filters out spurious relations whose word similarities are negative. The resulting attention matrix $G_l^{(q,d)}$ is then normalized using its infinite norm and passed through a modified softmax function to produce the disentangled word graph $\hat{A}_l^{(q,d)}$ in each transformer layer.

\begin{equation}
G_{l}^{(q,d)} = \mathrm{ReLU}\!\left( \frac{ (\mathbf{W}_{A}\mathbf{E}_{l}^{(q,d)})(\mathbf{W}_{A}\mathbf{E}_{l}^{(q,d)})' }{\sqrt{d_k}} + \epsilon \left( 1 - \mathcal{M}_{l}^{(q,d)} \right) \right)
\label{eq:M27_4}
\end{equation}

Following \cite{dong2021latent}, a GGNN is employed to refine term embeddings and approximate the expectation $\mathbb{E}_B[x]$ (Eq. \ref{eq:M27_7}). A conventional attention mechanism then evaluates the significance of document terms relative to the query intent. Specifically, for each document representation $Z_d^l$ derived from the latent space $Z^l$, its importance is computed via a sigmoid function (Eq. \ref{eq:M27_8}). This process facilitates the disentanglement of the embedding into query-relevant component $Z_{d_n}^l$ and query-irrelevant component $Z_{d_r}^l$. By prioritizing the query-related expectation $\mathbb{E}_Z[Z]$, the model synthesizes these refined representations in Eq. \ref{eq:M27_10} to estimate the final query-document relevance. 

\begin{equation}
\mathbb{E}_{B}[X] \propto Z^{l} = \mathrm{softmax}\!\left( (\mathbf{W}_{a}\mathbf{h}_{T}^{l}) \cdot (\mathbf{W}_{h}\mathbf{h}_{T}^{l})' \right)\mathbf{h}_{T}^{l} = \bigl[ Z_{\mathrm{CLS}}^{l}, Z_{q}^{l}, Z_{\mathrm{SEP}}^{l}, Z_{d}^{l}, Z_{\mathrm{SEP}}^{l} \bigr]
\label{eq:M27_7}
\end{equation}

\begin{equation}
\begin{gathered}
Z_{d_r}^{l} = \sigma\!\left( (\mathbf{W}_{q} Z_{q}^{l}) \cdot (\mathbf{W}_{d} Z_{d}^{l})' \right) \cdot Z_{d}^{l} \\
Z_{d_n}^{l} = \left( 1 - \sigma\!\left( (\mathbf{W}_{q} Z_{q}^{l}) \cdot (\mathbf{W}_{d} Z_{d}^{l})' \right) \right) \cdot Z_{d}^{l}
\end{gathered}
\label{eq:M27_8}
\end{equation}

\begin{equation}
P(Y \mid do(X)) \approx f(q,d) = \sigma\!\left( g(q,d) \right),\; \text{where } g(q,d) = \mathbf{w}_{f}\!\left( \mathbf{W}_{s}\!\left[ \mathbf{Z}^{l}, \mathbf{Z}_{d_r}^{l}, \mathbf{E}_{l}^{(q,d)}(0) \right] + \mathbf{b}_{s} \right)_{1 \times L} + b_{f}
\label{eq:M27_10}
\end{equation}

Natural language queries often fail to adequately capture the user's information need. To address this, pseudo relevance feedback (PRF) uses top-ranked documents from the initial retriever to refine the query and improve subsequent retrieval. Historically, PRF has been predominantly applied with vector space \cite{rocchio1971relevance} and probabilistic retrieval models \cite{robertson2009probabilistic}. Extending PRF to the transformer architecture remains challenging because the computational cost of standard self-attention scales quadratically with the input sequence length. To mitigate this limitation,  \cite{yu2021pgt} introduces PGT, a PRF method based on a graph-based Transformer-XH architecture \cite{zhao2020transformer}, which sparsifies attention to enable efficient modeling of a large number of feedback documents. By representing the text sequence as a graph, Transformer-XH applies full self-attention within each node while sparsifying attention across nodes, thereby significantly reducing computational cost. Similarly, PGT represents the query, candidate documents, and feedback documents as a graph, enabling efficient modeling of the multi-sequence text involved in PRF. The hidden representation of the first token, $[CLS]$, in each sequence is used to compute the document-level inter-sequence attention, defined as follows:

\begin{equation}
\hat{h}^{\,l}_{s,0} = \sum_{s' \in \mathcal{N}(s)}\mathrm{softmax}_{s'}\!\left(\frac{\hat{q}_{s,0}^{\top} \cdot\hat{k}_{s',0}}{\sqrt{d_k}}\right)\cdot\hat{v}_{s',0}
\label{eq:M21_1}
\end{equation}

where $s$ denotes a document sequence, $N(.)$ represents the set of neighboring document sequences, and $l$ indicates the encoder layer. The symbols $q$, $k$, and $v$ correspond to the query, key, and value representations used in the attention mechanism. Through the $[CLS]$ token, the neighboring information is propagated to the remaining tokens during the intra-sequence attention in subsequent layers, enabling PGT to effectively encode both local sequence-level information and global graph-level context. The representations from all $[CLS]$ tokens are then aggregated via a weighted sum to produce the final relevance score.

Traditional Neural re-rankers evaluate query-document pairs in isolation, neglecting the underlying document distribution, which can provide valuable contextual signals to improve re-ranking performance. In response to this shortcoming, \cite{di2024graph} proposes Graph Neural Re-ranking (GNRR), which improves re-ranking by introducing a corpus graph to model inter-document relationships. GNRR consists of three main stages: data retrieval, subgraph construction, and feature and score computation. Initially, a semantic corpus graph $\mathcal{G}^C = \{ \mathcal{V}^C, \mathcal{E}^C \}$ is constructed offline from the document corpus $C$ using TCT-ColBERT \cite{lin2020distilling}. During data retrieval, BM25 \cite{robertson1995okapi} is employed to retrieve the top-$1000_q$ documents for a given query $q$. Subsequently, a query-induced subgraph $\mathcal{G}^C_q = \{ \mathcal{V}^q, \mathcal{E}^q \}$ is extracted from $\mathcal{G}^C$, where the vertex set $\mathcal{V}^q$ corresponds to the top-$1000_q$ retrieved documents, and edges are inherited from the pre-constructed corpus graph $\mathcal{G}^C$. As a result, the query-induced subgraph preserves both lexical signals from the sparse retriever and structural information encoded in the corpus graph. Each node in $\mathcal{G}^C_q$ is initialized as $x_i = z_q \odot z_{d_i}$ where, $z_q$ and $z_{d_i}$ denote the embeddings of the query $q$ and document $d_i$, respectively. To incorporate ranking inductive bias, the initial ranking score is also integrated into the node representations. Re-ranking scores are then computed by jointly modeling the local document interactions and individual query-document relevance, where an MLP facilitates the individual relevance calculation $H_{q,loc}$, while the local interaction $H_{q,ind}$ is modeled using a GNN. The final query-document relevance score $s_q \in \mathbb{R}^{|V^q|}$ is calculated by feeding the merged local and induced embedding $CONCAT(H_{q,loc}, H_{q,ind})$ into a scoring model.

A prevalent strategy in passage retrieval and re-ranking is the use of contextual information to improve retrieval precision. To this end, \cite{albarede2022passage} employs a GAT for passage contextualization by constructing a query-specific directed corpus graph that models both inter- and intra-document similarities. The nodes in the graph represent document components (passage and section titles) extracted from the top-$1000$ initially retrieved documents, while edges encode 8 distinct relation types, grouped into two broad classes: intra-document and inter-document relations. Intra-document relations capture passage ordering, hierarchical structure, and internal citation, whereas inter-document relations model external citation between nodes. The inverse of these four relations is also considered in the graph construction phase. Inspired by ColBERT \cite{khattab2020colbert}, the authors propose two GAT-based ranking models: a merged interaction model (MiM) and a late interaction model (LiM), which differ in how they represent passage content. MiM simultaneously leverages the concepts and contexts of each passage to estimate the relevance score for a given query. In this setting, GAT computes the in-context representation of a passage $E_p$ from its content, its neighbors, and the neighboring graph structure. Subsequently, the query-passage relevance is computed as follows:

\begin{equation}
\mathrm{sim}(E_q, E_p) = \sum_{i \in [1, \lvert E_q \rvert]} \max_{j \in [1, \lvert E_p \rvert]} E_{q_i} \cdot E_{p_j}^{\top}
\label{eq:ND_1}
\end{equation}

where $E_q$ and $E_p$ denote the multiple representation embeddings of the query $q$ and passage $p$, respectively. In contrast, LiM models passage content and passage context separately during relevance estimation. While a ColBERT-based text encoder facilitates content embedding, a modified GAT, the context-only graph attention network (CGAT), is employed for context embedding, considering only a passage's contextual information by removing all its outgoing edges. The query-content relevance and the query-context relevance are then calculated using Eq. \ref{eq:ND_1}. The final query-passage relevance score is obtained as a weighted sum of these two similarities, given by Eq. \ref{eq:ND_2}. 

\begin{equation}
\mathrm{relevance}(q,p) = (1 - \lambda)\,\mathrm{sim}(E_q, E_p) + \lambda\,\mathrm{sim}(E_q, E_{\mathrm{context}_p})
\label{eq:ND_2}
\end{equation}

Building on the transformer architecture \cite{vaswani2017attention}, pairwise re-ranking has achieved strong effectiveness. However, the high inference overhead introduced by the quadratic comparison on the document size diminishes its feasibility. To address this issue, \cite{gienapp2022sparse} explores the possibility of comparing a subset of pairs sampled from an entire document pair set. In this setting, the aggregation stage becomes especially important, since it must derive the final rankings from pairwise preference probabilities and ensure reliable re-ranking of the sampled documents. One of the proposed strategies is a graph-based aggregation method inspired by PageRank \cite{page1999pagerank}. In this approach, the sampled document subset $C$ is represented as a directed graph $D_k$, where nodes correspond to documents and edges encode preference probabilities. For a document pair $(d_i, d_j)$, the preference probability, $p_{ij}$ indicates the likelihood that $d_i$ should be ranked above $d_j$. Final re-ranking is then produced through a PageRank-style aggregation procedure adapted for weighted graphs \cite{mihalcea2004textrank} as represented in Eq. \ref{eq:M9_1}. 

\begin{equation}
s_i = \gamma \cdot \frac{1}{\lvert D_k \rvert} + (1 - \gamma) \cdot \sum_{(d_j, d_i) \in C} \frac{p_{ji}}{\sum_{l \in [1,k]} p_{jl}} \cdot s_j
\label{eq:M9_1}
\end{equation}

Pairwise ranking prompting (PRP) has emerged as a promising zero-shot re-ranking approach based on LLMs. Despite their significant contribution to zero-shot re-ranking, concurrent PRP methods do not account for the uncertainty associated with labels; instead, they directly output labels for pairwise comparisons, leading to suboptimal performance \cite{qin2024large}. To overcome this limitation, \cite{luo2024prp} proposes a PRP-graph together with a new scoring unit that utilizes the degree of uncertainty from the output probabilities. The PRP unit facilitates pairwise comparison from a dual perspective as represented by Eq. \ref{eq:M5_1}

\begin{equation}
\begin{gathered}
s_{j \rightarrow i} = \mathrm{softmax}\!\left( \mathrm{LLM}\!\left( u(q, d_i, d_j) \right) \right) \\
s_{i \rightarrow j} = \mathrm{softmax}\!\left( \mathrm{LLM}\!\left( u(q, d_j, d_i) \right) \right)
\end{gathered}
\label{eq:M5_1}
\end{equation}

Here, $s_{j \rightarrow i}$ denotes the probability that document $i$ is more relevant to the query than document $j$, and $s_{i \rightarrow j}$ is the other way around. Built on the PRP unit, the PRP-graph ranks the top-$N$ documents $\mathcal{D}$ returned by the initial retriever through two main stages: ranking graph construction and ranking graph aggregation. The graph construction method is inspired by the Swiss-system tournament’s dynamic pairing and point accumulation methods \cite{csato2013ranking}. It iteratively constructs a document graph by conducting pairwise comparisons across $\mathcal{D}$ in $R$ rounds. The documents ranking scores are initialized as $\mathbf{\mathcal{S}}^{0} = \left[ 1,\; 1 - \frac{1}{N},\; 1 - \frac{2}{N},\; \ldots,\; \frac{1}{N} \right]$. In the $r$-th round, documents are traversed in order of their ranks in $\mathbf{\mathcal{S}}^{r-1}$ and are paired with the closest subsequent documents such that this pair hasn’t been compared in any previous round, including the current round. Next, the vertex weights  $S_{i}^{r}$ for each document pair ($d_i$, $d_j$) are updated following Eq. \ref{eq:M5_2}, which act as the ranking scores for the corresponding documents in the $r$-th round.

\begin{equation}
\begin{gathered}
S_{i}^{r} = S_{i}^{r-1} + s_{j \rightarrow i} \times \frac{S_{j}^{r-1}}{r} \\
S_{j}^{r} = S_{j}^{r-1} + s_{i \rightarrow j} \times \frac{S_{i}^{r-1}}{r}
\end{gathered}
\label{eq:M5_2}
\end{equation}

Here, the fraction in the equation denotes the difficulty level associated with the other document. The final ranking graph $\mathcal{G}$ is obtained after $R$ rounds, where directed edges between document vertices are weighted by the comparison scores $s_{j \rightarrow i}$ and $s_{i \rightarrow j}$. Following this, the graph aggregation phase revises the re-ranking scores by incorporating edge weights using a modified PageRank algorithm, as formulated in Eq. \ref{eq:M5_3}. Here, $s(i)$ denotes the relevance score of document $d_i$ at node $i$, $df$ is the damping factor, and $N$ denotes the total number of vertices in $\mathcal{G}$. The set $In(i)$ contains the vertices pointing to $i$, while $Out(i)$ contains the vertices to which $i$ points. Nodes are initialized with BM25 scores. The re-ranking scores are iteratively updated until convergence, determined by a threshold $\delta$.

\begin{equation}
s(i) = df \times \left[ \sum_{j \in \mathrm{In}(i)} \frac{s(j)}{\sum_{k \in \mathrm{Out}(j)} w_{jk}} \times w_{ji} \right] + \frac{1 - df}{N}
\label{eq:M5_3}
\end{equation}

\subsection{Question Answering (QA)}
Question answering is a widely studied NLP task that aims to automatically answer natural language queries. Graph-based re-ranking methods complement the QA pipeline by providing the answer generator with the most relevant candidates. This section reviews graph-based re-ranking approaches in various QA tasks, including open-domain question answering, knowledge graph question answering, temporal knowledge graph question answering, and multi-hop question answering. ODQA aims to answer queries using large, unstructured corpora. KGQA focuses on answering a query $q$ by leveraging knowledge graph triples $<e_s, r, e_t>$, which represent source entities, relations, and tail entities. TKGQA extends KGQA by incorporating temporal facts. Multi-hop QA generates an answer by reasoning over multiple pieces of evidence.

\subsubsection{Open-Domain Question Answering (ODQA)}
Fusion-in-Decoder (FiD), which combines dense passage retriever with a generative reader, is a representative architecture for ODQA \cite{izacard2021leveraging}. However, its inherent independence assumption regarding retrieved passages often constrains re-ranking performance and can lead to suboptimal answer generation. To address this limitation, \cite{yu2022kg} proposes KG enhanced FiD (KG-FiD), which incorporates GNNs to exploit the inter-passage relationships encoded in a knowledge graph. KG-FiD employs a two-stage hierarchical pruning strategy. In the first stage, the model re-ranks the $N_0$ passages returned by the initial retriever, retaining the top $N_1$ passages for further processing. In the second stage, the re-ranker takes the top $N_2$ passages from $N_1$ as input and jointly performs re-ranking and answer generation. In both phases, passage graphs $\mathcal{G}$ are constructed where nodes represent passages, initialized with DPR embeddings \cite{karpukhin2020dense}, and edges encode topological relationships derived from a knowledge graph. For the initial re-ranking, node states in $\mathcal{G}_0$ are refined through an $L_g$-layer GAT, as depicted in Eq. \ref{eq:M12_1}. The refined node representations are subsequently utilized to compute query-passage relevance scores, defined as $s_i^{stage-1} = Q^TE_i^{L_g}$, where $Q$ denotes the query embedding derived from the DPR retriever. Based on these initial scores, the top-$N_1$ passages are propagated to the second-stage re-ranker. Similar to the first stage, the second-stage re-ranker constructs a passage graph $\mathcal{G}_1 \subset \mathcal{G}_0$, where nodes are initialized with the first-token embeddings extracted from the final layer of the FiD encoder, as defined in Eq. \ref{eq:M12_2}.

\begin{equation}
E_i^{(l)} = h\!\left( E_i^{(l-1)},\; \{ E_j^{(l-1)} \}_{(i,j) \in \mathcal{G}_0} \right)
\label{eq:M12_1}
\end{equation}

\begin{equation}
Z_i^{(0)} = P_i^{(L)}(0) \in \mathbb{R}^{D},\; \text{where } P_i^{(L)} = \mathrm{T5\text{-}Encoder}_{L}\!\left( P_i^{(L-1)} \right) \in \mathbb{R}^{T_p \times H}
\label{eq:M12_2}
\end{equation}

To improve computational efficiency, an alternative variant initializes nodes using token embeddings extracted from an intermediate FiD encoder layer $(1 \le L_1 < L)$, defined as $Z_i^{(0)} = P_i^{(L_1)}(0)$. A GAT subsequently refines these representations $Z^{(L_g)} = GAT(Z^{(0)}, \mathcal{G}_1')$. The refined node embeddings are then used to compute query-passage relevance scores $s_i^{stage-2} = W^TZ_i^{L_g}$, where $W$ is a learnable parameter. Finally, the T5 decoder generates the target answer by utilizing the top-$N_2$ ranked passages. 
 
\subsubsection{Knowledge Graph Question Answering (KGQA)}
\cite{sun2024efficient} aims to extend the success of the neural retrieve-and-generate framework for question answering to KGQA by verbalizing structured KG triples and converting them into unstructured text. To mitigate the limited relevance of individually verbalized triples, the proposed approach groups verbalized triples that share a common subject entity into a single document. This design enables the re-ranker to exploit contextual information through triples that share the same subject entity during re-ranking. Triple verbalization follows a well-established template strategy \cite{agarwal2021knowledge}. Given a query $q$, the first stage of the KGQA pipeline employs BM25 and DPR \cite{karpukhin2020dense} to retrieve $N$ relevant documents, generated by verbalizing grouped triples. Subsequently, each triple $T_j \in D_i$ within a document is re-ranked by leveraging its contextual information $C_j$, defined as the concatenation of all remaining triples ($T_{1…j-1}+T_{j+1…n}$) in the same document. The re-ranking model, based on ELECTRA-large \cite{clark2020electra}, is trained as a binary classifier. To enable effective supervision, the authors introduce a triple-level labeling strategy: triples containing both the topic entity and the gold answer are labeled positive; if no such triple exists, a triple containing only the gold answer is labeled positive. The resulting top-ranked triples serve as the input for the answer generation module.

A prevalent approach to KGQA leverages the synergy between LLMs and KGs, in which a subgraph is heuristically extracted around topic entities and their k-hop neighbors to support LLM-generated answers. Although this strategy has shown strong performance, its reliance on heuristically constructed subgraphs often limits effectiveness, as these subgraphs tend to include substantial irrelevant information. To address this limitation, \cite{zhang2024question} proposes a question-guided knowledge graph re-scoring method (Q-KGR), which refines the extracted subgraph by eliminating noisy pathways through an edge re-scoring mechanism that evaluates triplet relevance based on semantic similarity to the input question. Specifically, for each subject-object node pair and the question $q$, embeddings $(e_s, e_o, e_q)$ are produced using a frozen PLM. The question embedding $e_q$ is computed by averaging the entity embeddings corresponding to $q$. A bilinear layer is subsequently applied to estimate a relevance score for each edge, as defined below:

\begin{equation}
\eta = \mathrm{Normalize}\!\left( \mathrm{Bilinear}\!\left( [e_s, e_o],\, e_q \right) \right)
\label{eq:M55_1}
\end{equation}

Normalization is performed using either the Gumbel-max or Gumbel-softmax technique \cite{jang2016categorical}. The resulting relevance score is treated as a latent variable and is optimized end-to-end with the QA objective, given the absence of a gold standard annotation for an ideal subgraph. The re-scored knowledge graph is further refined through a GAT, and the resulting representations are fed into a customized transformer to generate the final answer.

\subsubsection{Temporal Knowledge Graph Question Answering (TKGQA)}
Motivated by the strong performance of LLMs in KGQA, \cite{qian2024timer4} proposes TimeR\textsuperscript{4}, a retrieve-rewrite-retrieve-rerank framework designed to address temporal questions in temporal knowledge graph question answering. A temporal knowledge graph (TKG) is defined as a directed graph $G = \{E,P, T ,F\}$, where entities $E$ constitute the vertices, predicates $P$ associated with timestamps $T$ form the edges, and relations are represented as a set of temporal quadruples $F = \{ (s, p, o, t) \} \subseteq E \times P \times E \times T$, with $s$, $p$, $o$, and $t$ denoting the subject, predicate, object, and timestamp, respectively. Despite their strengths, LLMs face two major challenges in TKGQA: i) hallucination caused by implicit temporal cues in questions, and ii) insufficient temporal knowledge. To address the former, TimeR\textsuperscript{4} incorporates a retrieve-rewrite module that leverages TKGs to transform implicit temporal information into explicit representations. To enhance the temporal reasoning, the framework further introduces a retrieve-rerank module that captures semantic similarity between the question and candidate while enforcing temporal constraints. Specifically, a time-aware retrieval component first extracts relevant facts from a temporal knowledge store (TKS) that satisfy both semantic relevance and temporal consistency. The TKS contains time-aware facts generated by a fine-tuned language model, as defined in Eq. \ref{eq:M56_1}.

\begin{equation}
\mathrm{TKS} = \left\{ E_t \mid E_t = LM_t\!\left( S(s,p,o,t) \right),\; (s,p,o,t) \in \mathcal{G} \right\}
\label{eq:M56_1}
\end{equation}




A re-ranking mechanism is subsequently applied to filter candidate facts based on temporal relevance, prioritizing time-consistent information over irrelevant evidence. Let $t_q$ denote the timestamp extracted from the rewritten question $q^*$. The re-ranker computes the time difference between $t_q$ and each candidate timestamp $t \in F$ to discard temporally inconsistent candidates and estimate question-candidate relevance $\phi(q, t)$. The relevance score is obtained by normalizing valid time differences using a time filtering function $\phi_t(t_q, t)$, as defined in Eq. \ref{eq:M56_3}. Finally, the rewritten question and the re-ranked temporal facts are fed into a fine-tuned LLM to generate the answer. 

\begin{equation}
\begin{gathered}
\phi(q,t) = \mu \cdot \phi_{\mathrm{TKS}}(E_{q^{*}}, E_{t}) + (1 - \mu) \cdot \phi_{t}(t_q, t) \\
\phi_{t}(t_q, t) = \begin{cases} 1 - \dfrac{|t_q - t|}{\max(t_q - t)}, & \text{if } (t_q - t) > 0 \\ -100, & \text{otherwise} \end{cases}
\end{gathered}
\label{eq:M56_3}
\end{equation}


\subsubsection{Multi-hop Question Answering}
In the open-domain paradigm, answering multi-hop questions remains a challenging yet critical task, as it requires reasoning across multiple documents, many of which may exhibit little to no lexical overlap with the query. To tackle multi-hop QA, numerous prior studies adopt iterative document retrieval strategies \cite{nie2019revealing, feldman2019multi, das2019multi}. Building on this line of work, \cite{asai2019learning} proposes a graph-based recurrent retrieval framework for multi-hop open-domain question answering, which employs a recurrent neural network (RNN) to iteratively select candidate paragraphs conditioned on those retrieved in previous steps. The framework relies on a strong interplay between two components, a recurrent retriever and a multi-task reader, operating over the entire Wikipedia graph $G$. The process begins by populating an initial candidate set $C_1$ with top-ranked paragraphs obtained using TF-IDF. During each iteration $t$, the system identifies the next paragraph in the reasoning chain by scoring the neighbors of the current nodes in $G$ with a BERT-based relevance model. The iterative process terminates upon encountering a special end-of-evidence ([EOE]) symbol in the candidate list. The top-$B$ reasoning paths $E$, identified via Beam search, are subsequently passed to the reader module. The reader module first re-ranks each reasoning path $E' \in E$ using [CLS] token representation produced by BERT, as depicted in Eq. \ref{eq:M28_1}, where $w_n \in \mathbb{R}^D$ represents the weight vector. The resulting best reasoning path $E_{best}$ is then utilized to generate the answer span $S_{read}$, as defined in \ref{eq:M28_2}. In this formulation, $p_i^{start}$ and $P_j^{end}$ are the corresponding probability of i\textsuperscript{th} and j\textsuperscript{th} token be the start and end position from $E_{best} = \arg\max_{E \in \mathbf{E}} P(E \mid q)$. 

\begin{equation}
P(E \mid q) = \sigma\!\left( {w}_{n} \cdot {u}_{E} \right)\;\text{s.t.}\; {u}_{E} = \mathrm{BERT}_{\mathrm{[CLS]}}(q, E) \in \mathbb{R}^{D}
\label{eq:M28_1}
\end{equation}

\begin{equation}
S_{\mathrm{read}} = \arg\max_{i,j,\; i \le j} P_{i}^{\mathrm{start}} P_{j}^{\mathrm{end}}
\label{eq:M28_2}
\end{equation}

However, the traditional iterative retrieval approach introduces noise into the reader module by retrieving many relevant but non-supporting documents. Furthermore, existing methods typically assess query-document relevance independently of other candidates, often leading to suboptimal performance due to strong lexical similarity with non-supporting content. To overcome these limitations, \cite{zhang2021answering} proposes iterative document re-ranking (IDR), which employs a document graph $G$ to model inter-document relationships, enabling iterative re-ranking and filtering of retrieved documents. In $G$, two documents are connected if they share entities extracted from the corresponding query-document pairs. Following an initial retrieval step, IDR updates the query using text spans extracted from the re-ranked documents and retrieves additional documents based on the updated query. The newly retrieved documents are merged into the existing document pool and re-ranked using the graph-based module, while the least relevant documents are pruned at each iteration. The reader module adaptively halts the process once a satisfactory answer is identified. The graph-based re-ranking component contains four main stages: contextual encoding, graph attention, multi-document fusion, and document filtering. The process begins by encoding each query-document pair ($q$, $d_k$) by feeding their aggregated tokens ($[\mathrm{CLS}]\, q_{1} \ldots q_{|q|}\, [\mathrm{SEP}]\, t_{1}^{(k)} \ldots t_{|d_k|}^{(k)}\, [\mathrm{SEP}]$) through a pre-trained language model. The resulting representations from all documents are concatenated to form a matrix $v \in \mathbb{R}^{L \lvert \mathcal{D} \rvert \times h}$, where $L$ is the maximum  input length, $\lvert\mathcal{D}\rvert$ denotes the number of documents retrieved by the sparse retriever, and $h$ is the embedding dimension. A document graph is then constructed, with connections between documents determined by shared entities extracted via NER. An entity representation $e_i$ for each shared entity $E_i$ is obtained by pooling its token embeddings from $v$ through a combination of max-pooling and mean-pooling. Query-relevant entities are further emphasized by applying a dynamic softmax function \cite{qiu2019dynamically}, which produces masked entity representations $\mathbf{g}_i$ weighted by their relevance to the query. These representations are subsequently refined using a GAT with message passing, as formulated in Eq. \ref{eq:M1_1}.

\begin{equation}
\begin{gathered}
\mathbf{h}_{i}^{(t)} = {W}_{i}\mathbf{g}_{i}^{(t-1)} + {b}_{i} \\
\mathbf{g}_{i}^{(t)} = \mathrm{ReLU}\!\left( \sum_{j \in \mathcal{N}(i)} \alpha_{i,j}^{(t)} \mathbf{h}_{j}^{(t)} \right)
\end{gathered}
\label{eq:M1_1}
\end{equation}

Here, $\mathcal{N}(i)$ denotes the neighborhood of entity $E_i$, and $\alpha_{i,j}^{(t)}$ is the entity-entity attention, as defined in Eq. \ref{eq:M1_2}. In the multi-document fusion stage, non-entity token representations are updated by concatenating the refined masked embedding of each entity token with its corresponding token embeddings from $v$ and projecting back to the embedding dimension $h$, producing $\hat{t}_{j}^{(i)} = \mathbf{W}_{3}\!\left[ t_{j}^{(i)} ; \mathbf{g}_{i}^{(T)} \right]$, where $W_3 \in \mathbb{R}^{h \times 3h}$. The updated token representations replace their corresponding vectors in $v$, yielding the updated representation $\hat{v}$, over which a transformer encoder fuses contextual information across all tokens to produce  $\tilde{v} \in \mathbb{R}^{L \lvert \mathcal{D} \rvert \times h}$. The [CLS] token from this fused representation is then passed to a binary classifier to compute the query-document relevance score. 

\begin{equation}
\begin{gathered}
\alpha_{i,j}^{(t)} = \frac{\exp\!\left( s_{i,j}^{(t)} \right)}{\sum_{k} \exp\!\left( s_{i,k}^{(t)} \right)},\; \text{where }
s_{i,j}^{(t)} = \mathrm{LeakyReLU}\!\left( {W}_{2}\!\left[ \mathbf{h}_{i}^{(t)} ; \mathbf{h}_{j}^{(t)} \right] \right)
\end{gathered}
\label{eq:M1_2}
\end{equation}

\subsection{Retrieval-Augmented Generation (RAG)}
RAG augments text generation by retrieving information from external sources to complement the parametric knowledge of LLMs. Within this framework, the re-ranker plays a pivotal role in surfacing the most relevant factual content for generation. To this end, G-RAG \cite{dong2024don} explicitly models inter-document relationships through a document graph, enabling the identification of salient yet less obvious documents relevant to a given question. The re-ranking process begins with a query $q$ and top-$100$ documents $\{ p_1, p_2, …, p_n \}$ retrieved by an initial retriever, DPR \cite{karpukhin2020dense}. An abstract meaning representation (AMR) graph $G_{qp} = \{ V, E \}$ is subsequently constructed for each query-document pair $(q, p)$ using AMRBART \cite{bai2022graph}. AMR is a semantic representation language that captures sentence meaning through rooted directed graphs \cite{langkilde1998generation, banarescu2013abstract}. The resulting AMR graphs $\{ G_{qp1}, G_{qp2}, …, G_{qpn} \}$ are merged into an undirected document graph $\mathcal{G}_q = \{ \mathcal{V}, \mathcal{E} \}$, where each node $v_i \in \mathcal{V}$ corresponds to a document $p_i$, and an edge between $v_i, v_j \in \mathcal{V}$ is established whenever their respective AMR graphs $G_{qpi}$ and $G_{qpj}$ share common AMR nodes. To capture both structural and semantic information, a PLM encoder produces a fused document representation $x_i = Encode(concat(p_i, a_i))$, where $a_i$ denotes the AMR-derived information. However, directly concatenating document text with full AMR graphs incurs significant computational overhead \cite{wang2023exploiting}. To mitigate this, G-RAG filters AMR information through the shortest single source paths (SSSPs), which identify the shortest paths from the question node to all other nodes, representing the AMR semantics as a sequence of concepts $a_i$ along these paths. Two-dimensional edge features are subsequently incorporated into the document graph $\mathcal{G}_q$, defined as follows:

\begin{equation}
\hat{E}_{ij} =
\begin{cases}
0, & \text{no connection between } G_{q p_i} \text{ and } G_{q p_j} \\
\hat{E}_{ij1}, & \text{\# common nodes between } G_{q p_i} \text{ and } G_{q p_j} \\
\hat{E}_{ij2}, & \text{\# common edges between } G_{q p_i} \text{ and } G_{q p_j}
\end{cases}
\label{eq:NA_1}
\end{equation}

where $\hat{E}_{ijk}$ denotes $k$-th dimension of the edge-feature vector associated with nodes $i$ and $j$. The node representations for each vertex $x_v \in \mathbb{R}^d$ are then refined by a GNN according to the update rule in Eq. \ref{eq:NA_2}. The re-ranking model subsequently computes query-document similarity using Eq. \ref{eq:NA_3}, where $y$ denotes the query representation generated by the same encoder used to produce the node representations.

\begin{equation}
x_{v}^{\ell} = g\!\left( x_{v}^{\ell-1},\; \bigcup_{u \in \mathcal{N}(v)} f\!\left( x_{u}^{\ell-1}, e_{uv}^{\ell-1} \right) \right)
\label{eq:NA_2}
\end{equation}

\begin{equation}
s_i = \mathbf{y}^{\top} \mathbf{x}_{v_i}^{L}
\label{eq:NA_3}
\end{equation}

\cite{christmann2024rag} presents QUASAR, a RAG framework that integrates structured sources, including structured tables and knowledge graphs, alongside unstructured text. These heterogeneous contents are verbalized and indexed for subsequent use. The process begins with an initial retriever fetching the top-$1000$ candidate evidence given a query. Before passing these candidates to the answer generator, a GNN-based re-ranker filters the evidence set to a smaller, highly relevant subset (e.g., to-$30$ or top-$10$). During re-ranking, a bipartite word graph is constructed in which the evidence documents and their constituent entities are represented as nodes, and an edge models the relationship between an evidence node and an entity node whenever the entity is present in the evidence. Evidence and entity nodes are jointly scored using a GNN within a multi-task learning framework \cite{christmann2023explainable}, and the top-$k'$ $(k’ << k)$ candidates are subsequently forwarded to the answer generator to produce the final answer. 

\cite{dampanaboina2025diffusion} proposes Diffusion-Aided RAG, a graph-based diffusion re-ranking approach integrated into the RAG pipeline to guide generation using highly relevant reference passages. The proposed framework follows a two-stage pipeline in which a dense retriever first retrieves the top-$50$ candidate passage chunks based on inner-product similarity. These passages are then modeled as nodes in a graph to capture the inter-passage semantic relationships, followed by a graph-based diffusion re-ranking process that scores candidates using refined node representations. At the outset of diffusion-based re-ranking, a weighted undirected graph $G  = (V, E)$ is constructed from the top-$n$ candidate passage chunks, where the node set $V$ corresponds to passage chunks and the edge set $E$ encodes pairwise cosine similarities between them. A personalized PageRank-based \cite{page1999pagerank} diffusion algorithm is then applied to iteratively compute the re-ranking scores $\pi \in \mathbb{R}^n$, as defined in Eq. \ref{eq:M41_1}. 

\begin{equation}
\boldsymbol{\pi} = \alpha A^{\top} \boldsymbol{\pi} + (1 - \alpha)\mathbf{p}
\label{eq:M41_1}
\end{equation}

Here, $A$ denotes the row-normalized adjacency matrix of $G$, $\alpha$ is the dampening factor, and $\mathbf{p}$ is the personalization vector that injects the initial relevance scores from dense retrieval into the re-ranking process. This process assigns higher weights to the top-ranked candidates from the initial retriever, as defined in Eq. \ref{eq:M41_2}. Consequently, the relevant passage chunks within the same cluster are promoted while noisy or irrelevant candidates are suppressed, thereby yielding improved ranking quality. 

\begin{equation}
p_i = \frac{s_i}{\sum_{j=1}^{n} s_j},\; i = 1, \ldots, n,\; \mathbf{p} \neq 0,\; \sum_{i=1}^{n} p_i = 1
\label{eq:M41_2}
\end{equation}

\subsection{Auxiliary \& Specialized Task}
This final category of the taxonomy comprises four subcategories: \textit{Answer Sentence Selection}, \textit{Semantic Parsing}, \textit{Entity Re-ranking}, and \textit{Term Re-ranking}. Answer sentence selection is closely related to document re-ranking, with AS2 operating at a fine-grained level, whereas document re-ranking operates at a coarse-grained level. While AS2 aims to directly identify the answer itself, document re-ranking seeks to identify content that is likely to contain the answer. Semantic parsing is a classical NLP task that translates natural language into structured logical forms. Term extraction, in contrast, focuses on identifying domain-relevant terms from unstructured textual corpora.

\subsubsection{Answer Sentence Selection}
\cite{iyer2023question} proposes EQAG-GNN, a graph-based framework that formulates AS2 as a node classification problem by jointly modeling question and candidate answers within a GNN. EQAG-GNN further exploits information from semantically similar questions and their corresponding answers, incorporating additional contextual cues and their correctness signals to guide relevance estimation for target question-answer pairs. For each target question, EQAG-GNN constructs a small-scale Effective QA Graph (EQAG), where each node represents a question-answer sentence pair $(q, a)$. Edges in the graph encode three types of semantic dependencies: question-answer (QA), question-question (QQ), and answer-answer (AA). QA dependency models the relationship between query and candidate answers by connecting the top K\textsubscript{intra} candidates ranked by TANDA \cite{garg2020tanda}. QQ dependencies are established by identifying the top K\textsubscript{rows} semantically similar questions from the training data using a  RoBERTa model trained on Quora, which are then linked to the target question. AA captures similar answers based on the semantic similarity of their corresponding questions. Following graph construction, a two-layer GCN is employed to estimate the relevance of each $(q, a)$ pair as a node classification task. Each node $v_i = (q, a)$ is initialized using TANDA's relevance score, which is refined using GCN as follows:

\begin{equation}
h_i^{(l+1)} = \sigma\!\left( W_l^{\top} \left( \sum_{j \in N_i \cup \{i\}} \frac{e_{j,i}}{\sqrt{m_j m_i}}\, h_j^{(l)} \right) \right)
\label{eq:M59_1}
\end{equation}

where $h_i$ represents the embedding of node $v_i$ at layer $l \in [0,2]$, while $N_i$ denotes the set of neighboring nodes of $v_i$. The matrix $W$ is a learnable weight matrix. The terms $m_i$ and $m_j$ corresponds to entities of degree matrix, where $m_i = 1 + \sum_{j \in N_i} e_{j,i}$. The model is optimized using binary-cross entropy loss to achieve the desired binary node classification objective, as depicted in Eq. \ref{eq:M59_2}, where $y_i$ and $\hat{y}_i$ denote the true label and model’s prediction, respectively. The resulting refined scores are used to re-rank the candidate answers. 

\begin{equation}
\mathrm{BCE} = -\frac{1}{N} \sum_{i=0}^{n} \left( y_i \log(\hat{y}_i) + (1 - y_i)\log(1 - \hat{y}_i) \right)
\label{eq:M59_2}
\end{equation}

To overcome the limitation of traditional AS2 models that ignore the contextual relationships among candidates, \cite{tian2020capturing} introduces the multi-perspective graph encoder (MPGE) that explicitly captures inter-candidate dependencies by modeling their interactions through a series of complementary viewpoints. MPGE constructs several candidate graphs, each capturing a distinct relational perspective, to generate corresponding contextual representations that are subsequently aggregated into a unified embedding. The graphs employed in MPGE are broadly categorized into static graphs and dynamic graphs. In both categories, nodes correspond to candidate sentences, and edge definitions vary by the graph type. Based on different heuristics, MPGE constructs three types of static graphs: the \textit{entity graph}, the \textit{distance graph}, and the \textit{similarity graph}. In the \textit{entity graph}, an edge $A_{ij}^{\mathrm{ent}}$ between two sentences $S_i$ and $S_j$ is defined based on the co-occurrence of shared entities. Motivated by the observation that sentence proximity often correlates with relevance, the \textit{distance graph} models the edge between $S_i$ and $S_j$ using a distance-based score derived from a Gaussian distribution, as formulated in Eq. \ref{eq:M61_2}.


\begin{equation}
A_{ij}^{\mathrm{dist}} = \frac{1}{\sigma \sqrt{2\pi}} \exp\!\left( -\frac{(j - i)^2}{2\sigma^2} \right)
\label{eq:M61_2}
\end{equation}

The \textit{similarity graph} defines the edge $A_{ij}^{\mathrm{simi}}$ between sentences $S_i$ and $S_j$ using the cosine similarity of their BERT-based contextual representation. Since static graphs rely on heuristic rules and are therefore limited in capturing instance-specific relationships, MPGE further introduces a dynamic graph construction method based on a self-attention mechanism. Under this setup, the edge between $S_i$ and $S_j$ is computed according to Eq. \ref{eq:M61_3}, where $\sigma$ denotes an activation function and $w_s \in \mathbb{R}^{d \times d}$ is a learnable weight matrix. A GCN is then applied to update the initial sentence representation $H^0$ by leveraging the four adjacency matrices A\textsuperscript{ent}, A\textsuperscript{dist}, A\textsuperscript{simi}, and A\textsuperscript{dyn}. This process yields in refined sentence representations $H_{\mathrm{ent}}^{L},\; H_{\mathrm{dist}}^{L},\; H_{\mathrm{simi}}^{L},\; H_{\mathrm{dyn}}^{L} \in \mathbb{R}^{N \times d}$, as depicted in Eq. \ref{eq:M61_4}, where $H^t \in \mathbb{R}^{N \times d}$ represents the input node embeddings at layer $t$, $N$ represents the total number of nodes in the graph, $d$ is the embedding dimension, and $\tilde{D}_{ii}$ denotes the layer specific normalization matrix. The refined representations, together with the original sentence embeddings, are subsequently fed into a bidirectional GRU to generate a fused representation H\textsuperscript{mpge}. Finally, this fused representation is used to score candidate answers using either a linear projection or a bilinear matching function. 

\begin{equation}
A_{ij}^{\mathrm{dyn}} = \frac{\exp(\alpha_{ij})}{\sum_{j'} \exp(\alpha_{ij'})},\; \text{where }
\alpha_{ij} = \sigma\!\left( {w}_{s}{h}_{i}^{0} \right)^{\top} \sigma\!\left( {w}_{s}{h}_{j}^{0} \right)
\label{eq:M61_3}
\end{equation}

\begin{equation}
H^{(t+1)} = \sigma\!\left( \tilde{D}^{-\frac{1}{2}} {A} \tilde{D}^{-\frac{1}{2}} H^{(t)} W^{(t)} \right)
\label{eq:M61_4}
\end{equation}

A common paradigm in AS2 is to construct the transformer input by concatenating contextual information with the query and its candidate answers. Although the self-attention mechanism of PLMs makes this strategy appear effective, it often yields suboptimal performance in AS2, as it fails to sufficiently filter noisy contextual sentences and capture fine-grained question-candidate alignment. To overcome this limitation, \cite{van2023question} introduces CASSIE, a framework that improves candidate scoring in AS2 by explicitly modeling dependencies between the question and answer contexts. The process begins by generating a  unified contextual representation of the question, candidate answers, and their neighboring sentences via a PLM. A query-context alignment module based on optimal transport \cite{monge1781memoire, cuturi2013sinkhorn} is then applied to align question tokens with candidate sentences, enabling more precise question-candidate word alignment beyond standard transformer self-attention. Subsequently, dependencies between the candidate answer and its surrounding sentences are modeled using a fully-connected graph. Let $[r1, r2, r3]$ denote the representations of answer candidates $s_1 \equiv c$, its preceding sentence $s_2 \equiv s_{prev}$, and its succeeding sentence $s_3 \equiv s_{next}$, respectively. The candidate sentences are represented as vertices in a fully-connected graph $G = (V, E)$, where $V = \{ s_i \mid 1 \le i \le 3 \}$. Each edge in $E$ encodes the pairwise dependencies $\alpha_{ij} \in (0, 1)$ between sentences $s_i$ and $s_j$, whose weight is computed by a feed-forward neural network operating on their semantic representations $(r_i, r_j)$ and transportation cost to the query $(d_{qsi}, d_{qsj})$, as formulated in Eq. \ref{eq:M60_1}. Here, $\odot$ denotes element-wise multiplication, "$;$" denotes concatenation, and FFF\textsubscript{DEP} is the feed-forward neural network. A GCN \cite{kipf2016semi} then refines the sentence representations using these learned edge weights, as defined in Eq. \ref{eq:M60_2}, where $W$ and $b$ denote trainable parameters and $h_i^0$ is initialized with $r_i$. Next, the relevance score $p_c \in (0, 1)$ of each candidate answer is computed by passing its refined representation $h_1$ through a feed-forward network followed by a sigmoid activation, i.e., $p_c = FFN_{AS2}(h1)$. The overall model is jointly optimized using cross-entropy loss and MI optimization loss \cite{belghazi2018mutual, devon2018learning}.

\begin{equation}
\alpha_{ij} = \frac{\exp(u_{ij})}{\sum_{j'=1}^{K} \exp(u_{ij'})},\; \text{where }
u_{ij} = \mathrm{FFN}_{\mathrm{DEP}}\!\left( [\, r_i \odot r_j \,;\, d_{qs_i} \,;\, d_{qs_j} \,] \right)
\label{eq:M60_1}
\end{equation}

\begin{equation}
h_i^{l} = \mathrm{ReLU}\!\left( \sum_{j=1}^{K} \alpha_{ij} W^{l} h_{j}^{\,l-1} + b^{l} \right)
\label{eq:M60_2}
\end{equation}

\subsubsection{Semantic Parsing}
To address the challenges of candidate selection in semantic parsing, \cite{shao2020graph} introduces GTCV, a graph-based transformer architecture that improves the ranking of logical outputs. Existing ranking models often fail to capture the structural characteristics of logical forms \cite{wen2015stochastic, einolghozati2019improving}. Concretely, vanilla transformer models rely on all-to-all token interactions, which disregard the tree structure inherent in logical forms. To address this limitation, GTCV augments the transformer multi-head attention mechanism with a graph-aware structural information derived from logical forms. The semantic parsing pipeline begins by generating a set of candidate logical forms for an input question $Q$. A BERT-base encoder then derives a contextual representation $h_0$ for each logical form $L_i$ using the input format [CLS] $Q$ [SEP] $L_i$ [SEP]. To incorporate structural dependencies, an adjacency matrix $M$ is constructed from each logical form and used to mask the transformer's multi-head attention, enabling structure-aware interactions. Multiple graph-aware transformer layers are then aggregated via average pooling to obtain a structure-aware representation $Z$. This representation is concatenated with $h_0$ to form $g  = [Z;h_0]$, which is passed through a sigmoid layer to produce a matching score $m_i$. Based on the assumption that correct logical forms tend to share subtrees with other high-quality candidates, GTCV further incorporates a cross-candidate verification score $v_i$ into the final relevance estimation $p_i$, as defined in Eq. \ref{eq:M36_1}. Here, $\alpha \in [0,1]$ is a balancing parameter, and $v_i$ is computed from normalized pairwise similarities among candidate logical forms.

\begin{equation}
p_i = (1 - \alpha)\, m_i + \alpha\, v_i
\label{eq:M36_1}
\end{equation}

\subsubsection{Entity Re-ranking}
Unlike most information retrieval approaches, which leverage entities to enhance text retrieval, \cite{dietz2019ent} reverses the paradigm by using textual information to facilitate entity retrieval. The authors introduce ENT Rank, a method for identifying one or more entities relevant to a given topical query (e.g., “Zika fever”). Whereas  prior approaches typically assume all entity connections are equally informative with respect to the query, ENT Rank explicitly models the relevance of neighboring entities by incorporating their associated textual context. To this end, relations are represented as a hypergraph, where nodes correspond to entities and hyperedges capture shared textual context. The hypergraph is subsequently transformed into a directed multi-graph, in which a directed edge $(e_i, t_k, e_j)$ represents a connection between entities ($e_i$ and $e_j$) mediated through the textual context $t_k$. Entity relevance is ultimately estimated using a traditional LTR framework that combines entity-specific features with aggregated relevance signals derived from neighboring entities and their shared contexts.

\subsubsection{Term Re-ranking}
Existing term extraction methods fall into two categories: statistical and graph-based methods. Statistical approaches rely heavily on term-frequency, which often leads to suboptimal performance by overlooking highly relevant but infrequent terms. Graph-based approaches \cite{brin1998anatomy, mihalcea2004textrank, rospocher2012corpus} address this limitation by modeling term importance through relational structures rather than raw frequency. Several studies further enhance term extraction by incorporating external knowledge (e.g., ontology and thesaurus) into the extraction process \cite{vivaldi2010finding, gazendam2010thesaurus, cambriasentic}. However, these methods are fundamentally limited by the assumption that such resources are universally available across domains. To overcome this limitation, \cite{khan2016term} proposes Term Ranker, a graph-based re-ranking framework for term extraction. The proposed method first extracts candidate terms using statistical methods, then constructs a refined similarity graph by leveraging domain-relevant knowledge derived from patents, and finally re-ranks candidate terms using TextRank \cite{mihalcea2004textrank}, a graph-based ranking algorithm that predates modern graph neural networks.
\section{Experimental Settings}
\label{sec:experimental_settings}

This section presents the experimental setups employed across the 52 surveyed studies, covering datasets, evaluation metrics, learning paradigms, their corresponding objective functions, and code availability. Table \ref{tab:experimental_settings} provides a summarized overview of these aspects.

{\footnotesize
\newcolumntype{L}[1]{>{\raggedright\arraybackslash}p{#1}}
\setlength{\LTcapwidth}{\textwidth}
\begin{longtable}{L{2cm} L{3cm} L{2.3cm} L{1.5cm} L{2.4cm} L{2.7cm}}
\caption{Experimental setup overview of 52 selected graph-based re-ranking methods. Studies without official model names are represented by their corresponding references. The content enclosed in parentheses after \textit{Custom Dataset(s)} indicates the sources used to build the dataset(s).  The full form of the evaluation metrics is provided in the following discussion. Each model is color-coded according to its primary taxonomy class: \textcolor{cust_red}{Question Answering (QA)}, \textcolor{cust_purple}{General Information Retrieval}, \textcolor{cust_pink}{RAG}, and \textcolor{cust_cyan}{Auxiliary \& Specialized Tasks}}\label{tab:experimental_settings}\\
\hline
\textbf{Model} & \textbf{Dataset} & \textbf{Evaluation Metric} & \textbf{Learning Paradigm} & \textbf{Training Objective} & \textbf{Open-source Code} \\
\hline
\endfirsthead

\hline
\textbf{Model} & \textbf{Dataset} & \textbf{Evaluation Metric} & \textbf{Learning Paradigm} & \textbf{Training Objective} & \textbf{Open-source Code} \\
\hline
\endhead

\hline
\endfoot

\hline
\endlastfoot

\textcolor{cust_red}{KG-FiD}\cite{yu2022kg} & $\bullet$ Natural Questions (NQ)\newline $\bullet$ TriviaQA & No Ranking Evaluation \newline  & Supervised learning & Cross-entropy loss & Not Available \\
\hline
\textcolor{cust_red}{\cite{sun2024efficient}} & $\bullet$ FreebaseQA\newline $\bullet$ WebQSP & Hit@k & Supervised learning & Binary classification objective & Not Available \\
\hline
\textcolor{cust_red}{Q-KGR}\cite{zhang2024question} & $\bullet$ OpenBookQA\newline $\bullet$ ARC\newline $\bullet$ Riddle\newline $\bullet$ PIQA & No Ranking Evaluation & Supervised learning & End-to-end QA objective & \url{https://github.com/EchoDreamer/Q-KGR} \\
\hline
\textcolor{cust_red}{TimeR\textsuperscript{4}}\cite{qian2024timer4} & $\bullet$ MULTITQ\newline $\bullet$ TimeQuestions & No Ranking Evaluation & Ad-hoc & - & \url{https://github.com/qianxinying/TimeR4} \\
\hline
\textcolor{cust_red}{\cite{asai2019learning}} & $\bullet$ HotpotQA\newline $\bullet$ SQuAD Open\newline $\bullet$ NQ Open & Answer Recall, Paragraph Recall, Paragraph EM & Supervised learning & Binary cross-entropy loss & \url{https://github.com/AkariAsai/learning\_to\_retrieve\_reasoning\_paths} \\
\hline
\textcolor{cust_red}{IDRQA}\cite{zhang2021answering} & $\bullet$ HotpotQA\newline $\bullet$ NQ Open\newline $\bullet$ SQuAD Open & Paragraph EM & Supervised learning & Binary classification objective & Not Available \\
\hline
\textcolor{cust_purple}{\cite{bendersky2008re}} & $\bullet$ TREC AP\newline $\bullet$ TREC WSJ\newline $\bullet$ TREC8 & P, MRR & Ad-hoc & - & Not Available \\
\hline
\textcolor{cust_purple}{GBRM}\cite{deng2009effective} & $\bullet$ Custom Dataset \newline(DBLP XML records) & P, MAP, Bpref & Unsupervised learning & Joint regularized cost function & Not Available \\
\hline
\textcolor{cust_purple}{PsgAidRank}\cite{krikon2010utilizing} & $\bullet$ TREC AP\newline $\bullet$ TREC WSJ\newline $\bullet$ TREC8\newline $\bullet$ FR\newline $\bullet$ WT10G & P & Ad-hoc & - & Not Available \\
\hline
\textcolor{cust_purple}{\cite{kurland2010pagerank}} & $\bullet$ TREC AP\newline $\bullet$ TREC WSJ\newline $\bullet$ TREC8 & P, MRR & Ad-hoc & - & Not Available \\
\hline
\textcolor{cust_purple}{TRM \& TDM}\cite{veningston2014information} & $\bullet$ OHSUMED & MRR, MAP, P, R, F-Measure & Ad-hoc & - & Not Available \\
\hline
\textcolor{cust_purple}{EsdRank}\cite{xiong2015esdrank} & $\bullet$ ClueWeb09\newline $\bullet$ ClueWeb12\newline $\bullet$ OHSUMED  & ERR, nDCG, MAP & Supervised learning & Maximize log-likelihood via Expectation-Maximization (EM) & Not Available \\
\hline
\textcolor{cust_purple}{GAR}\cite{macavaney2022adaptive} & $\bullet$ TREC DL19, DL20\newline $\bullet$ MSMARCO dev (small) & nDCG, MAP, R, RR & Ad-hoc & - & \url{https://github.com/terrierteam/pyterrier\_adaptive/tree/main} \\
\hline
\textcolor{cust_purple}{QDG}\cite{frayling2024effective} & $\bullet$ TREC DL19, DL20 & nDCG, MAP, R & Ad-hoc & - & Not Available \\
\hline
\textcolor{cust_purple}{\cite{macavaney2022adaptiveagent}} & $\bullet$ TREC DL19, DL20 & nDCG, R & Ad-hoc & - & Not Available \\
\hline
\textcolor{cust_purple}{\cite{jaenich2024fairness}} & $\bullet$ TREC 2021 Fair Ranking Track\newline $\bullet$ TREC 2022 Fair Ranking Track & nDCG, AWRF, FairRel & Ad-hoc & - & Not Available \\
\hline
\textcolor{cust_purple}{SlideGAR}\cite{rathee2025guiding} & $\bullet$ MSMARCO Passage Ranking\newline $\bullet$ TREC DL19, DL20, DL21, DL22 & nDCG, R & Ad-hoc & - & \url{https://github.com/Mandeep-Rathee/llmgar} \\
\hline
\textcolor{cust_purple}{Quam}\cite{rathee2025quam} & $\bullet$ TREC DL19, DL20 & nDCG, R & Supervised fine-tuning & Binary cross-entropy loss & \url{https://github.com/Mandeep-Rathee/quam} \\
\hline
\textcolor{cust_purple}{ORE}\cite{rathee2025breaking} & $\bullet$ MSMARCO passage\newline $\bullet$ TREC DL19, DL20, DL21, DL22 & nDCG, R & Online learning & Squared error & \url{https://github.com/elixir-research-group/ORE} \\
\hline
\textcolor{cust_purple}{LADR}\cite{kulkarni2023lexically} & $\bullet$ TREC DL19, DL20\newline $\bullet$ MSMARCO Dev (Small) & nDCG, R, RR & Ad-hoc & - & \url{https://github.com/Georgetown-IR-Lab/ladr} \\
\hline
\textcolor{cust_purple}{\cite{xiong2016bag}} & $\bullet$ ClueWeb09\newline $\bullet$ ClueWeb12 & ERR, nDCG & Ad-hoc & - & Not Available \\
\hline
\textcolor{cust_purple}{AttR-Duet}\cite{xiong2017word} & $\bullet$ ClueWeb09-B\newline $\bullet$ ClueWeb12-B13 & nDCG, ERR & Supervised learning & Pairwise hinge loss & Not Available \\
\hline
\textcolor{cust_purple}{EDRM}\cite{liu2018entity} & $\bullet$ Query log from Sogou (Chinese search engine) & nDCG, MRR & Supervised learning & Pairwise loss & \url{https://github.com/thunlp/EntityDuetNeuralRanking} \\
\hline
\textcolor{cust_purple}{GAT-reRanker}\cite{vollmersdocument} & $\bullet$ MSMARCO Document Ranking & MRR, MAP & Supervised learning & Not specified & Not Available \\
\hline
\textcolor{cust_purple}{KGPR}\cite{fang2023kgpr} & $\bullet$ MSMARCO Passage Ranking\newline $\bullet$ TREC DL19, DL20\newline $\bullet$ TREC DL Hard & MRR, nDCG & Supervised fine-tuning & Cross-entropy loss & \url{https://github.com/jyfang6/KGPR} \\
\hline
\textcolor{cust_purple}{KERM}\cite{dong2022incorporating} & $\bullet$ MSMARCO Passage Ranking\newline $\bullet$ TREC DL19\newline $\bullet$ Ohsumed query set & MRR, MAP & Pre-training \& Supervised fine-Tuning & Masked Language Model (MLM) \& Sentence Relation Prediction (SRP); Cross-entropy loss & \url{https://github.com/CSQianDong/KERM} \\
\hline
\textcolor{cust_purple}{QDER}\cite{chatterjee2025qder} & $\bullet$ TREC Robust 2004\newline $\bullet$ TREC CAR 2017\newline $\bullet$ TREC News 2021\newline $\bullet$ TREC Core 2018\newline $\bullet$ CODEC & P, nDCG, MRR, MAP & Supervised fine-tuning & Cross-entropy loss & \url{https://github.com/shubham526/SIGIR2025-QDER} \\
\hline
\textcolor{cust_purple}{DREQ}\cite{chatterjee2024dreq} & $\bullet$ TREC Robust 2004\newline $\bullet$ TREC News 2021\newline $\bullet$ TREC Core 2018\newline $\bullet$ CODEC & P, nDCG, MAP & Supervised fine-tuning & Binary cross-entropy loss & \url{https://github.com/shubham526/ECIR2024-DREQ} \\
\hline
\textcolor{cust_purple}{GraphMonoT5}\cite{gupta2024empowering} & $\bullet$ BioASQ8B\newline $\bullet$ TREC-COVID\newline $\bullet$ HotPotQA & R, nDCG, MAP & Supervised learning & Log-likelihood (max.), Mutual Information (min.) & Not Available \\
\hline
\textcolor{cust_purple}{GRAPH-JPDRMM}\cite{pappas2020aueb} & $\bullet$ BIOASQ 8 & MAP & Supervised learning & Hinge loss \& Binary cross-entropy loss & Not Available \\
\hline
\textcolor{cust_purple}{KEGNR}\cite{liu2025knowledge} & $\bullet$ TREC PM 2017, 2018, 2019, 2020 & P, nDCG & Supervised learning & Cross-entropy loss & Not Available \\
\hline
\textcolor{cust_purple}{KG-Rank}\cite{yang2024kg} & $\bullet$ LiveQA\newline $\bullet$ ExpertQA-Bio\newline $\bullet$ ExpertQA-Med\newline $\bullet$ MedQA & No Ranking Evaluation & Ad-hoc & - & \url{https://github.com/ruiyang-medinfo/KG-Rank} \\
\hline
\textcolor{cust_purple}{\cite{kohail2017matching}} & $\bullet$ Custom Dataset (German news articles) & P & Supervised learning & Not specified & Not Available \\
\hline
\textcolor{cust_purple}{PassageRank}\cite{reed2020faster} & $\bullet$ TREC DL20 & nDCG, MAP, MRR & Supervised fine-tuning & Passage classification objective & \url{https://github.com/kylereed96/trec-dl-2020} (Error Code: 404) \\
\hline
\textcolor{cust_purple}{\cite{sarwar2021graph}} & $\bullet$ Cranfield\newline $\bullet$ WebAp\newline $\bullet$ Ohsumed & MRR, P, MAP & Ad-hoc & - & Not Available \\
\hline
\textcolor{cust_purple}{GRMM}\cite{zhang2021graph} & $\bullet$ Robust04\newline $\bullet$ ClueWeb09-B & nDCG, P & Supervised learning & Pairwise hinge loss & \url{https://github.com/CRIPAC-DIG/GRMM} \\
\hline
\textcolor{cust_purple}{LGRe}\cite{dong2021latent} & $\bullet$ Robust04\newline $\bullet$ WebTrack2009-12 & P, nDCG & Supervised learning & Pairwise ranking loss, Triangle distance loss & \url{https://github.com/CSQianDong/LGRe} \\
\hline
\textcolor{cust_purple}{DGRe}\cite{dong2022disentangled} & $\bullet$ Robust04\newline $\bullet$ WebTrack2009-12 & P, nDCG & Supervised learning & Pairwise ranking loss, Triangle distance loss, Decomposition loss & \url{https://github.com/DQ0408/DGRe} (Error Code: 404) \\
\hline
\textcolor{cust_purple}{PGT}\cite{yu2021pgt} & $\bullet$ MSMARCO Passage Ranking\newline $\bullet$ TREC DL19 & nDCG, MAP, R & Supervised learning & Cross-entropy loss & Not Available \\
\hline
\textcolor{cust_purple}{GNRR}\cite{di2024graph} & $\bullet$ MSMARCO passage\newline $\bullet$ TREC DL19, DL20\newline $\bullet$ TREC DL Hard & AP, P, nDCG, RR & Supervised learning & Pairwise loss & Not Available \\
\hline
\textcolor{cust_purple}{MiM \& LiM}\cite{albarede2022passage} & $\bullet$ CLEF-IP2013 & PRES, MAP, R, PREC(D) & Supervised learning & Pairwise softmax cross-entropy loss & Not Available \\
\hline
\textcolor{cust_purple}{\cite{gienapp2022sparse}} & $\bullet$ ClueWeb09\newline $\bullet$ ClueWeb12\newline $\bullet$ MSMARCO & nDCG & Ad-hoc & - & \url{https://github.com/webis-de/ICTIR-22} \\
\hline
\textcolor{cust_purple}{PRP-Graph}\cite{luo2024prp} & $\bullet$ BEIR Benchmark\newline $\bullet$ TREC DL19, DL20 & nDCG & Zero-shot & - & \url{https://github.com/Memelank/PRP-Graph} \\
\hline
\textcolor{cust_pink}{G-RAG}\cite{dong2024don} & $\bullet$ NQ \newline $\bullet$  Trivia QA (TQA) & MRR, TMHits, MTRR, MHits   & Supervised learning & Pairwise ranking loss & Not Available \\
\hline
\textcolor{cust_pink}{QUASAR}\cite{christmann2024rag} & $\bullet$ CompMix\newline $\bullet$ CRAG\newline $\bullet$ TimeQuestions & MRR, P & Weak supervision & Not specified & \url{https://github.com/PhilippChr/QUASAR} \\
\hline
\textcolor{cust_pink}{Diffusion-Aided RAG}\cite{dampanaboina2025diffusion} & $\bullet$ OVGU QA dataset & nDCG, MRR, R & Ad-hoc & - & \url{https://github.com/sai0499/Diffusion-Page-Reranking-aided-Dense-Retrieval-for-RAG-} \\
\hline
\textcolor{cust_cyan}{EQAG-GNN}\cite{iyer2023question} & $\bullet$ WikiQA\newline $\bullet$ TREC-QA\newline $\bullet$ WQA & P, MAP, MRR & Supervised learning & Binary cross-entropy loss & Not Available \\
\hline
\textcolor{cust_cyan}{MPGE}\cite{tian2020capturing} & $\bullet$ WikiQA\newline $\bullet$ SQuAD & MAP, MRR \newline  & Supervised learning & Cross entropy loss & Not Available \\
\hline
\textcolor{cust_cyan}{CASSIE}\cite{van2023question} & $\bullet$ WikiQA\newline $\bullet$ WDRASS & P, MAP, MRR & Supervised learning & Binary cross-entropy loss \& Mutual information (MI) loss & Not Available \\
\hline
\textcolor{cust_cyan}{GTCV}\cite{shao2020graph} & $\bullet$ ATIS\newline $\bullet$ JOBS\newline $\bullet$ TOP & No Ranking Evaluation & Supervised learning & Binary cross-entropy loss & Not Available \\
\hline
\textcolor{cust_cyan}{ENT Rank}\cite{dietz2019ent} & $\bullet$ TREC CAR\newline $\bullet$ DBpedia-Entity v2 & R-Precision, MAP, nDCG & Supervised learning & Maximize Mean Average Precision (MAP) & \url{https://github.com/TREMA-UNH/ENT-rank} \\
\hline
\textcolor{cust_cyan}{Term Ranker}\cite{khan2016term} & $\bullet$ Custom Datasets (Domain-specific text corpora, Patent corpora)  & P & Ad-hoc & - & Not Available \\
\hline

\end{longtable}
}

\subsection{Dataset}
The datasets used in the reviewed studies can be broadly categorized into four groups: core IR datasets, QA datasets, domain-specific datasets, and specialized datasets. In this discussion, we primarily focus on IR datasets, as graph-based re-ranking is predominantly an information retrieval task. The statistics of the IR datasets most commonly used in graph-based re-ranking studies are presented in Table \ref{tab:Dataset-table}. Here, \textit{\#Corpus} denotes the total number of documents contained in a dataset, while \textit{\#Queries} denotes the number of queries available for training or evaluation. Each dataset provides query-document relevance judgments, indicating the extent to which a document is related to a query. Binary relevance indicates whether a query is relevant to a document, whereas multi-level relevance enables a more fine-grained assessment of query-document relevance. For example, TREC DL19 provides four levels of relevance, where $0$, $1$, $2$, and $3$ denote irrelevant, relevant, highly relevant, and perfectly relevant, respectively. 

As shown in Table \ref{tab:Dataset-table}, datasets from Text REtrieval Conference (TREC), an annual conference organized by the National Institute of Standards and Technology (NIST) to advance the IR field, account for the majority of datasets used in the literature. Early IR models heavily relied on classical TREC datasets such as TREC8, TREC AP, and TREC WSJ \cite{bendersky2008re, kurland2010pagerank}. More recently, datasets such as TREC Deep Learning (DL), TREC Robust04, TREC News, TREC CORE, and TREC COVID have become widely used by researchers. The Microsoft machine reading comprehension (MSMARCO) \cite{bajaj2016ms} is another widely adopted dataset, created by sampling user queries from the Bing search engine. The TREC DL19 \cite{craswell2020overview} and DL20 \cite{craswell2021overview} inherit their document corpus from the MSMARCO v1 dataset, while MSMARCO v2 provides the corpus for TREC DL21 and DL22 \cite{craswell2025overview}. 

Ohsumed\cite{hersh1994ohsumed} is a well-known clinically oriented IR dataset derived from MEDLINE\footnote{\url{https://www.nlm.nih.gov/medline/medline_overview.html}}. ClueWeb09 and its successor, ClueWeb12, are large-scale datasets crawled from the web to support the TREC web track and related research. The complex document and entity collection (CODEC) \cite{mackie2022codec} is a relatively recent dataset designed for ranking complex documents and entities. Among the datasets listed in Table \ref{tab:Dataset-table}, only MSMARCO supports large-scale supervised training or fine-tuning of re-ranking models, primarily due to its substantial number of queries and extensive document corpus. 

\begin{table}[h]
\caption{Statistics of widely used IR datasets in the reviewed literature. \textbf{\#Corpus} and \textbf{\#Queries} denote the total number of documents and queries, respectively. \textbf{Relevance Level} indicates the degree of query-document relevance. \textbf{Usage} specifies whether a dataset is used for model training or evaluation.}
\centering
\small
\begin{tabular}{l l l l l}
\hline
\textbf{Dataset} & \textbf{\#Corpus} & \textbf{\#Queries} & \textbf{Relevance Level} & \textbf{Usage}\\
\hline
MSMARCO Passage Ranking \cite{bajaj2016ms} & 8.8M & 809k & Binary & Train \\
\hline
MSMARCO Passage Ranking v2 & 138M & 277K & Binary & Train \\
\hline
MSMARCO Dev(small) & 8.8M & 6,980 & Binary & Test \\
\hline
TREC DL19 (judged) \cite{craswell2020overview} & 8.8M & 43 & 4-level & Test \\
\hline
TREC DL20 (judged) \cite{craswell2021overview} & 8.8M & 54 & 4-level & Test  \\
\hline
TREC DL21 (judged) & 138M & 53 & 4-level & Test \\
\hline
TREC DL22 (judged) \cite{craswell2025overview} & 138M & 76 & 4-level & Test  \\
\hline
TREC DL Hard \cite{mackie2021deep} & 8.8M & 50 & 4-level & Test  \\
\hline
TREC Robust04 \cite{voorhees2005overview} & 528K & 250 & 3-level & Test  \\
\hline
TREC News 2021\tablefootnote{\url{https://trec.nist.gov/data/news2021.html}} \cite{soboroff2021overview} & 728,626 & 51 & 5-level & Test  \\
\hline
TREC CORE 2018\tablefootnote{\url{https://github.com/trec-core/2018}} & 595K & 50 & 3-level & Test  \\
\hline
TREC COVID \cite{thakur2021beir} & 171,332 & 50 & 3-level & Test \\
\hline
CODEC \cite{mackie2022codec} & 729,824 & 42 & 4-level & Test  \\
\hline
Ohsumed \cite{hersh1994ohsumed} & 348,566 & 106 & 3-level & Test  \\
\hline
ClueWeb09\tablefootnote{\url{https://lemurproject.org/clueweb09/}} & 1B & 200 & 3-level/6-level & Test  \\
\hline
ClueWeb12\tablefootnote{\url{https://lemurproject.org/clueweb12/}} & 733M & 100 & 6-level & Test  \\
\hline
TREC8\tablefootnote{\url{https://ir-datasets.com/disks45.html}} & 528K & 50 & Binary & Test  \\
\hline
TREC AP \cite{bendersky2008re, kurland2010pagerank} & 242,918 & 99 & Binary & -  \\
\hline
TREC WSJ \cite{bendersky2008re, kurland2010pagerank} & 173,252 & 50 & Binary & -  \\
\hline
\end{tabular}
\label{tab:Dataset-table}
\end{table}

\subsection{Evaluation Metric}
Evaluation metrics are essential to assess the effectiveness of ranking models and to allow fair comparison between studies. In total, 18 different evaluation metrics are employed in the literature reviewed in this survey, with certain metrics being more widely adopted within the IR community. The following provides a detailed description of the evaluation metrics.  

\textit{Precision (P)} indicates the fraction of retrieved documents that are relevant to a given query, while \textit{recall (R)} reflects the fraction of all relevant documents that have been retrieved. The \textit{F-measure} balances precision and recall by computing their harmonic mean as depicted in Eq. \ref{eq:F1}. At the passage level, \textit{PREC(D)} infers the relevance of a passage from that of its containing document. Complementing this, \textit{R-precision (Rprec)} is defined as the precision at a cutoff R equal to the total number of relevant documents. One of the widely used IR metrics, the \textit{mean average precision (MAP)}, is calculated as the mean of \textit{average precision (AP)} scores across a set of queries, where AP corresponds to the mean of the precision computed at the ranking positions of relevant documents. Formally, MAP is defined in Eq. \ref{eq:MAP}. 

\begin{equation}
\label{eq:F1}
\text{F-measure} = 2 \times \frac{\text{Precision} \times \text{Recall}}{\text{Precision} + \text{Recall}}
\end{equation}

\begin{equation}
\label{eq:MAP}
\text{MAP} = \frac{1}{|Q|} \sum_{q=1}^{|Q|} \text{Average Precision}_q
\end{equation}

For a given query, the \textit{reciprocal rank (RR)} is the inverse of the rank position of the first relevant result. The \textit{mean reciprocal rank (MRR)} extends this by averaging RR across multiple queries, as defined in Eq. \ref{eq:MRR}, where ${\text{rank}_q}$ indicates the position of the first relevant document for query $q$. Building on this, \textit{expected reciprocal rank (ERR)} \cite{chapelle2009expected} models user behavior such as scanning a ranked list of retrieved documents from top to bottom and stopping once a document satisfies the information need. It is formulated in Eq. \ref{eq:ERR}, where $n$ denotes the ranked documents, and $R_i$ denotes the probability that the $i$-th document satisfies the user's information need. 

\begin{equation}
\label{eq:MRR}
\text{MRR} = \frac{1}{|Q|} \sum_{q=1}^{|Q|} \frac{1}{\text{rank}_q}
\end{equation}
 
\begin{equation}
\label{eq:ERR}
\text{ERR} = \sum_{r=1}^{n} \frac{1}{r}
\prod_{i=1}^{r-1} (1 - R_i)\, R_r
\end{equation}

The \textit{normalized discounted cumulative gain (nDCG)} is computed by normalizing the \textit{discounted cumulative gain (DCG)} with respect to the Ideal DCG (IDCG). Using relevance labels, DCG quantifies the usefulness of the resulting documents while applying a logarithmic discount based on their positions in the ranked list, thereby emphasizing higher-ranked results. The nDCG at rank $k$ is defined in Eq. \ref{eq:nDCG}, where $rel_i$ defines the relevance label of the document at position $i$.

\begin{equation}
\label{eq:nDCG}
\text{nDCG@}k = \frac{\text{DCG@}k}{\text{IDCG@}k},\; \text{where } \text{DCG@}k = \sum_{i=1}^{k} \frac{rel_i}{\log_2(i + 1)}
\end{equation}

\textit{Hits@k} measures the proportion of queries for which at least one correct answer appears within the top-$k$ ranked results. The \textit{mean hits@k (MHits@k)}, on the other hand, measures the percentage of all positive documents that are retrieved in the top-$k$ ranked results, as depicted in Eq. \ref{eq:MHits} following G-RAG \cite{dong2024don}. In this formulation, $\mathbb{I}$ denotes the indicator function. Furthermore, \cite{dong2024don} introduces \textit{mean tied reciprocal ranking (MTRR)} and \textit{tied mean hits@k (TMHit@k)} to explicitly account for the ties in the ranking scores. 

\begin{equation}
\label{eq:MHits}
\text{MHits@}k = \frac{1}{|Q|} \sum_{q \in Q} 
\left(
\frac{1}{|P^{+}|} \sum_{p \in P^{+}} \mathbb{I}(r_p \leq k)
\right)
\end{equation}

\textit{Attention weighted ranked fairness (AWRF)} \cite{sapiezynski2019quantifying} evaluates the fairness of the ranked results by comparing the actual exposure allocated to different groups to a target or ideal distribution. Building on this, \textit{FairRel} combines AWRF and nDCG to achieve a balance between fairness and ranking quality \cite{jaenich2024fairness}. Moving to domain-specific evaluation, \textit{PRES@K (Patent Retrieval Evaluation Score)}, as adopted in \cite{albarede2022passage}, assesses the effectiveness of a ranking system by measuring how high it places relevant documents on the ranked list. At a finer granularity, \textit{Paragraph exact match (EM)} evaluates whether the ground truth evidence paragraph appears within the retrieved set. Finally, \textit{Binary preference (Bpref)} \cite{buckley2004retrieval} evaluates a system's ability to rank relevant candidates ahead of irrelevant ones, as depicted in Eq. \ref{eq:bpref}. In this formulation, $R$ denotes the number of relevant candidates in the judged set, and $n$ refers to the first irrelevant candidate $R$. 

\begin{equation}
\label{eq:bpref}
\text{bpref} = \frac{1}{R} \sum_{r=1}^{N} 
\left(
1 - \frac{\#n \text{ ranked higher than } r}{R}
\right)
\end{equation}

\subsection{Learning Paradigm}
Figure \ref{fig:learning_method} presents an overview of the learning paradigms adopted in the selected studies. These statistics pertain only to the re-ranking module's training, although a study may include additional modules within its overall architecture. Among the 52 studies, 48.08\% employs supervised learning to train their re-ranking models, making it the most prevalent strategy. The second most common approach is Ad-hoc models, used in 32.69\% of the studies, followed by Supervised Fine-tuning, which accounts for 11.54\%. The remaining learning paradigms, zero-shot learning, unsupervised learning, online learning, and weak supervision, are each represented by a single study.    

\begin{figure}[htbp]
    \centering
	\includegraphics[width=\columnwidth, height=0.25\textheight, keepaspectratio]{learning_method.pdf}
	\caption{Influence of learning paradigms in graph-based re-ranking methods}
	\label{fig:learning_method}
\end{figure}

\subsection{Training Objective}
Studies adopting a training-oriented learning paradigm rely on loss functions to guide model optimization. As summarized in Table \ref{tab:experimental_settings}, two dominant loss functions account for 67.65\% of the 34 studies employing training-oriented learning paradigms: pairwise ranking loss, specifically the pairwise hinge loss, and cross-entropy loss. Pairwise ranking loss is a broad category of loss functions that measures the relative ordering between pairs of candidate documents by assigning higher scores to the more relevant candidate. Pairwise hinge loss is a margin-based pairwise ranking loss that enforces a minimum score difference between candidate pairs, as defined in Eq. \ref{eq:hinge}. In this formulation, $m$ represents the margin, $s_i$ and $s_j$  are the query-document relevance scores for documents $d_i$ and $d_j$, respectively. 

\begin{equation}
\label{eq:hinge}
\mathcal{L}_{hinge}(d_i, d_j) = \max \left( 0,\; m - (s_i - s_j) \right)
\end{equation}

Cross-entropy loss, a standard objective in classification problems, quantifies the discrepancy between two probability distributions. In information retrieval, the binary cross-entropy loss facilitates learning for many ranking models, as shown in Eq. \ref{eq:CE}. Some studies further extend the cross-entropy loss by applying softmax over the query-candidate relevance score, as represented in Eq. \ref{eq:CE_2}. 

\begin{equation}
\label{eq:CE}
\mathcal{L}_{\text{CE}} = -\left( y \log p + (1 - y)\log(1 - p) \right)
\end{equation}

\begin{equation}
\label{eq:CE_2}
\mathcal{L}_{\text{CE}} = -\log \left( \frac{e^{s_i}}{e^{s_i} + e^{s_j}} \right)
\end{equation}

\subsection{Open-source Code}
Among the reviewed papers, only 23 (44.23\%) provide open-source implementation hosted on GitHub. Of this, 21 repositories remain publicly accessible at the time of the analysis. Consequently, reproducing results for most studies requires implementation from scratch. The links to available open-source repositories are listed in Table \ref{tab:experimental_settings}.
\section{Findings \& Discussion}
\label{sec:findings}

Graph structures and graph learning techniques hold significant potential for enhancing re-ranking frameworks. In the preceding sections, we examined graph-based re-ranking techniques across a wide range of tasks and different levels of graph granularity. In this section, we analyze the key findings, provide insights into graph-based re-ranking methods, and outline their limitations and promising directions for future research. 

\subsection{Key Findings}
Through an extensive analysis of the reviewed literature, we have identified several graph-related characteristics that play an important role in the ranking process. The LTR paradigms and graph-related characteristics of the surveyed studies are summarized in Table \ref{tab:model_comparison}.

{\footnotesize
\newcolumntype{L}[1]{>{\raggedright\arraybackslash}p{#1}}
\setlength{\LTcapwidth}{\textwidth}
\begin{longtable}{p{3.2cm} p{1.4cm} p{1.8cm} p{1.4cm} p{1.8cm} L{1.8cm} p{1.9cm}}
\caption{Comparison of the reviewed literature in terms of \textbf{LTR Paradigm} and graph-related characteristics. \textbf{Node Granularity} represents the content encoded by graph nodes. \textbf{KG Integration} indicates the use of knowledge graphs in the re-ranking module. \textbf{Graph-refined Features} denotes whether node features are updated using graph learning techniques (e.g., GNNs, PageRank). \textbf{Graph-based Re-ranking} indicates the direct use of graph learning techniques in the ranking process. \textbf{Inter-document Relation} indicates whether semantic or structural relationships among candidate documents are utilized. \textit{Doc, Q,} and \textit{A} stands for \textit{Document, Query,} and \textit{Answer}, and are occasionally used as shorthand for space efficiency. Each model is color-coded according to its primary taxonomy class: \textcolor{cust_red}{Question Answering (QA)}, \textcolor{cust_purple}{General Information Retrieval}, \textcolor{cust_pink}{RAG}, and \textcolor{cust_cyan}{Auxiliary \& Specialized Tasks}}\label{tab:model_comparison}\\
\hline
\textbf{Model} & \textbf{LTR\newline Paradigm} & \textbf{Node\newline Granularity} & \textbf{KG\newline Integration} & \textbf{Graph-refined Features} & \textbf{Graph-based Re-ranking} & \textbf{Inter-document Relation} \\
\hline
\endfirsthead

\hline
\textbf{Model} & \textbf{LTR\newline Paradigm} & \textbf{Node\newline Granularity} & \textbf{KG\newline Integration} & \textbf{Graph-refined Features} & \textbf{Graph-based Re-ranking} & \textbf{Inter-document Relation} \\
\hline
\endhead

\hline
\endfoot

\hline
\endlastfoot

\textcolor{cust_red}{KG-FiD}\cite{yu2022kg} & Listwise & Document & \textcolor{darkgreen}{\checkmark} & \textcolor{darkgreen}{\checkmark} & \textcolor{darkred}{\ding{55}} & \textcolor{darkgreen}{\checkmark}  \\
\hline
\textcolor{cust_red}{\cite{sun2024efficient}} & Pointwise & - & \textcolor{darkgreen}{\checkmark} & \textcolor{darkred}{\ding{55}} & \textcolor{darkred}{\ding{55}} & \textcolor{darkred}{\ding{55}} \\
\hline
\textcolor{cust_red}{Q-KGR}\cite{zhang2024question} & Pointwise & KG Entity & \textcolor{darkgreen}{\checkmark} & \textcolor{darkred}{\ding{55}} & \textcolor{darkred}{\ding{55}} & \textcolor{darkred}{\ding{55}}  \\
\hline
\textcolor{cust_red}{TimeR\textsuperscript{4}}\cite{qian2024timer4} & Pointwise & - & \textcolor{darkgreen}{\checkmark} & \textcolor{darkred}{\ding{55}} & \textcolor{darkred}{\ding{55}} & \textcolor{darkred}{\ding{55}}  \\
\hline
\textcolor{cust_red}{\cite{asai2019learning}} & Pointwise & Paragraph & \textcolor{darkred}{\ding{55}} & \textcolor{darkred}{\ding{55}} & \textcolor{darkred}{\ding{55}} & \textcolor{darkred}{\ding{55}} \\
\hline
\textcolor{cust_red}{IDRQA}\cite{zhang2021answering} & Pointwise & Entity & \textcolor{darkred}{\ding{55}} & \textcolor{darkgreen}{\checkmark} & \textcolor{darkred}{\ding{55}} & \textcolor{darkgreen}{\checkmark}  \\
\hline
\textcolor{cust_purple}{\cite{bendersky2008re}} & Pointwise & Doc, Segment & \textcolor{darkred}{\ding{55}} & \textcolor{darkred}{\ding{55}} & \textcolor{darkred}{\ding{55}} & \textcolor{darkgreen}{\checkmark} \\
\hline
\textcolor{cust_purple}{GBRM}\cite{deng2009effective} & Listwise & Document & \textcolor{darkred}{\ding{55}} & \textcolor{darkred}{\ding{55}} & \textcolor{darkgreen}{\checkmark} & \textcolor{darkgreen}{\checkmark}  \\
\hline
\textcolor{cust_purple}{PsgAidRank}\cite{krikon2010utilizing} & Pointwise & Doc, Segment & \textcolor{darkred}{\ding{55}} & \textcolor{darkred}{\ding{55}} & \textcolor{darkgreen}{\checkmark} & \textcolor{darkgreen}{\checkmark}  \\
\hline
\textcolor{cust_purple}{\cite{kurland2010pagerank}} & Listwise & Document & \textcolor{darkred}{\ding{55}} & \textcolor{darkred}{\ding{55}} & \textcolor{darkgreen}{\checkmark} & \textcolor{darkgreen}{\checkmark} \\
\hline
\textcolor{cust_purple}{TRM \& TDM}\cite{veningston2014information} & Pointwise & Term & \textcolor{darkred}{\ding{55}} & \textcolor{darkred}{\ding{55}}  & \textcolor{darkgreen}{\checkmark} & \textcolor{darkgreen}{\checkmark}  \\
\hline
\textcolor{cust_purple}{EsdRank}\cite{xiong2015esdrank} & Listwise & - & \textcolor{darkgreen}{\checkmark} & \textcolor{darkred}{\ding{55}} & \textcolor{darkred}{\ding{55}} & \textcolor{darkred}{\ding{55}}  \\
\hline
\textcolor{cust_purple}{GAR}\cite{macavaney2022adaptive} & Pointwise & Document & \textcolor{darkred}{\ding{55}} & \textcolor{darkred}{\ding{55}} & \textcolor{darkred}{\ding{55}} & \textcolor{darkred}{\ding{55}} \\
\hline
\textcolor{cust_purple}{QDG}\cite{frayling2024effective} & Pointwise & Doc, Query & \textcolor{darkred}{\ding{55}} & \textcolor{darkred}{\ding{55}} & \textcolor{darkred}{\ding{55}} & \textcolor{darkred}{\ding{55}}  \\
\hline
\textcolor{cust_purple}{\cite{macavaney2022adaptiveagent}} & Pointwise & Document &  \textcolor{darkred}{\ding{55}} &  \textcolor{darkred}{\ding{55}} &  \textcolor{darkred}{\ding{55}} & \textcolor{darkred}{\ding{55}}  \\
\hline
\textcolor{cust_purple}{\cite{jaenich2024fairness}} & Pointwise & Document & \textcolor{darkred}{\ding{55}} & \textcolor{darkred}{\ding{55}} & \textcolor{darkred}{\ding{55}} & \textcolor{darkred}{\ding{55}} \\
\hline
\textcolor{cust_purple}{SlideGAR}\cite{rathee2025guiding} & Listwise & Document & \textcolor{darkred}{\ding{55}} & \textcolor{darkred}{\ding{55}} & \textcolor{darkred}{\ding{55}} & \textcolor{darkgreen}{\checkmark}  \\
\hline
\textcolor{cust_purple}{Quam}\cite{rathee2025quam} & Pointwise & Document & \textcolor{darkred}{\ding{55}} & \textcolor{darkred}{\ding{55}} & \textcolor{darkred}{\ding{55}} & \textcolor{darkred}{\ding{55}}  \\
\hline
\textcolor{cust_purple}{ORE}\cite{rathee2025breaking} & Pointwise & Document & \textcolor{darkred}{\ding{55}} & \textcolor{darkred}{\ding{55}} & \textcolor{darkred}{\ding{55}} & \textcolor{darkgreen}{\checkmark}  \\
\hline
\textcolor{cust_purple}{LADR}\cite{kulkarni2023lexically} & Pointwise & Document & \textcolor{darkred}{\ding{55}} & \textcolor{darkred}{\ding{55}} & \textcolor{darkred}{\ding{55}} & \textcolor{darkred}{\ding{55}}  \\
\hline
\textcolor{cust_purple}{\cite{xiong2016bag}} & Pointwise & - & \textcolor{darkgreen}{\checkmark} & \textcolor{darkred}{\ding{55}} & \textcolor{darkred}{\ding{55}} & \textcolor{darkred}{\ding{55}}  \\
\hline
\textcolor{cust_purple}{AttR-Duet}\cite{xiong2017word} & Pairwise & - & \textcolor{darkgreen}{\checkmark} & \textcolor{darkred}{\ding{55}} & \textcolor{darkred}{\ding{55}} & \textcolor{darkred}{\ding{55}} \\
\hline
\textcolor{cust_purple}{EDRM}\cite{liu2018entity} & Pairwise & - & \textcolor{darkgreen}{\checkmark} & \textcolor{darkred}{\ding{55}} & \textcolor{darkred}{\ding{55}} & \textcolor{darkred}{\ding{55}} \\
\hline
\textcolor{cust_purple}{GAT-reRanker}\cite{vollmersdocument} & Pointwise & KG Entity & \textcolor{darkgreen}{\checkmark} & \textcolor{darkgreen}{\checkmark} & \textcolor{darkgreen}{\checkmark} & \textcolor{darkgreen}{\checkmark}  \\
\hline
\textcolor{cust_purple}{KGPR}\cite{fang2023kgpr} & Pointwise & KG Entity & \textcolor{darkgreen}{\checkmark} & \textcolor{darkred}{\ding{55}} & \textcolor{darkred}{\ding{55}} & \textcolor{darkred}{\ding{55}}  \\
\hline
\textcolor{cust_purple}{KERM}\cite{dong2022incorporating} & Pointwise & KG Entity & \textcolor{darkgreen}{\checkmark} & \textcolor{darkgreen}{\checkmark} & \textcolor{darkred}{\ding{55}} & \textcolor{darkred}{\ding{55}}  \\
\hline
\textcolor{cust_purple}{QDER}\cite{chatterjee2025qder} & Pointwise & - & \textcolor{darkgreen}{\checkmark} & \textcolor{darkred}{\ding{55}} & \textcolor{darkred}{\ding{55}} & \textcolor{darkred}{\ding{55}}  \\
\hline
\textcolor{cust_purple}{DREQ}\cite{chatterjee2024dreq} & Pointwise & - & \textcolor{darkgreen}{\checkmark} & \textcolor{darkred}{\ding{55}} & \textcolor{darkred}{\ding{55}} & \textcolor{darkred}{\ding{55}}  \\
\hline
\textcolor{cust_purple}{GraphMonoT5}\cite{gupta2024empowering} & Pointwise & Entity & \textcolor{darkgreen}{\checkmark} & \textcolor{darkgreen}{\checkmark} & \textcolor{darkred}{\ding{55}} & \textcolor{darkred}{\ding{55}}  \\
\hline
\textcolor{cust_purple}{GRAPH-JPDRMM}\cite{pappas2020aueb} & Pairwise & Entity & \textcolor{darkred}{\ding{55}} & \textcolor{darkred}{\ding{55}} & \textcolor{darkred}{\ding{55}} & \textcolor{darkred}{\ding{55}} \\
\hline
\textcolor{cust_purple}{KEGNR}\cite{liu2025knowledge} & Pointwise & Term, Sentence & \textcolor{darkgreen}{\checkmark} & \textcolor{darkgreen}{\checkmark} & \textcolor{darkred}{\ding{55}} & \textcolor{darkred}{\ding{55}}  \\
\hline
\textcolor{cust_purple}{KG-Rank}\cite{yang2024kg} & Pointwise & - & \textcolor{darkgreen}{\checkmark} & \textcolor{darkred}{\ding{55}} & \textcolor{darkred}{\ding{55}} & \textcolor{darkred}{\ding{55}}  \\
\hline
\textcolor{cust_purple}{\cite{kohail2017matching}} & Pointwise & Word & \textcolor{darkred}{\ding{55}} & \textcolor{darkred}{\ding{55}} & \textcolor{darkred}{\ding{55}} & \textcolor{darkred}{\ding{55}}  \\
\hline
\textcolor{cust_purple}{PassageRank} \cite{reed2020faster} & Pointwise & Segment & \textcolor{darkred}{\ding{55}} & \textcolor{darkred}{\ding{55}} & \textcolor{darkgreen}{\checkmark} & \textcolor{darkred}{\ding{55}} \\
\hline
\textcolor{cust_purple}{\cite{sarwar2021graph}} & Pointwise & Segment & \textcolor{darkred}{\ding{55}} & \textcolor{darkred}{\ding{55}} & \textcolor{darkred}{\ding{55}} & \textcolor{darkred}{\ding{55}}  \\
\hline
\textcolor{cust_purple}{GRMM\cite{zhang2021graph}} & Pairwise & Word & \textcolor{darkred}{\ding{55}} & \textcolor{darkgreen}{\checkmark} & \textcolor{darkred}{\ding{55}} & \textcolor{darkred}{\ding{55}} \\
\hline
\textcolor{cust_purple}{LGRe\cite{dong2021latent}} & Pairwise & Word & \textcolor{darkred}{\ding{55}} & \textcolor{darkgreen}{\checkmark} & \textcolor{darkred}{\ding{55}} & \textcolor{darkred}{\ding{55}}  \\
\hline
\textcolor{cust_purple}{DGRe\cite{dong2022disentangled}} & Pairwise & Word & \textcolor{darkred}{\ding{55}} & \textcolor{darkgreen}{\checkmark} & \textcolor{darkred}{\ding{55}} & \textcolor{darkred}{\ding{55}} \\
\hline
\textcolor{cust_purple}{PGT\cite{yu2021pgt}} & Pointwise & Doc, Query & \textcolor{darkred}{\ding{55}} & \textcolor{darkgreen}{\checkmark} & \textcolor{darkred}{\ding{55}} & \textcolor{darkgreen}{\checkmark}  \\
\hline
\textcolor{cust_purple}{GNRR\cite{di2024graph}} & Pairwise & Document & \textcolor{darkred}{\ding{55}} & \textcolor{darkgreen}{\checkmark} & \textcolor{darkred}{\ding{55}} & \textcolor{darkgreen}{\checkmark}  \\
\hline
\textcolor{cust_purple}{MiM \& LiM}\cite{albarede2022passage} & Pairwise & Segment & \textcolor{darkred}{\ding{55}} & \textcolor{darkgreen}{\checkmark} & \textcolor{darkgreen}{\checkmark} & \textcolor{darkgreen}{\checkmark} \\
\hline
\textcolor{cust_purple}{\cite{gienapp2022sparse}} & Pairwise & Document & \textcolor{darkred}{\ding{55}} & \textcolor{darkred}{\ding{55}} & \textcolor{darkgreen}{\checkmark} & \textcolor{darkgreen}{\checkmark}  \\
\hline
\textcolor{cust_purple}{PRP-Graph\cite{luo2024prp}} & Pairwise & Document & \textcolor{darkred}{\ding{55}} & \textcolor{darkred}{\ding{55}} & \textcolor{darkgreen}{\checkmark} & \textcolor{darkgreen}{\checkmark}  \\
\hline
\textcolor{cust_pink}{G-RAG}\cite{dong2024don} & Pairwise & Document & \textcolor{darkred}{\ding{55}} & \textcolor{darkgreen}{\checkmark} & \textcolor{darkgreen}{\checkmark} & \textcolor{darkgreen}{\checkmark}  \\
\hline
\textcolor{cust_pink}{QUASAR}\cite{christmann2024rag} & Pointwise & Doc, KG Entity & \textcolor{darkgreen}{\checkmark} & \textcolor{darkgreen}{\checkmark} & \textcolor{darkgreen}{\checkmark} & \textcolor{darkred}{\ding{55}}  \\
\hline
\textcolor{cust_pink}{Diffusion-Aided RAG}\cite{dampanaboina2025diffusion} & Listwise & Passage Chunks & \textcolor{darkred}{\ding{55}} & \textcolor{darkred}{\ding{55}} & \textcolor{darkgreen}{\checkmark} & \textcolor{darkgreen}{\checkmark}  \\
\hline
\textcolor{cust_cyan}{EQAG-GNN}\cite{iyer2023question} & Pointwise & (Q, A) Sentence & \textcolor{darkred}{\ding{55}} & \textcolor{darkgreen}{\checkmark} & \textcolor{darkgreen}{\checkmark} & \textcolor{darkgreen}{\checkmark} \\
\hline
\textcolor{cust_cyan}{MPGE}\cite{tian2020capturing} & Pointwise & Sentence & \textcolor{darkred}{\ding{55}} & \textcolor{darkgreen}{\checkmark} & \textcolor{darkgreen}{\checkmark} & \textcolor{darkgreen}{\checkmark} \\
\hline
\textcolor{cust_cyan}{CASSIE}\cite{van2023question} & Pointwise & Sentence & \textcolor{darkred}{\ding{55}} & \textcolor{darkgreen}{\checkmark} & \textcolor{darkred}{\ding{55}} & \textcolor{darkred}{\ding{55}}  \\
\hline
\textcolor{cust_cyan}{GTCV}\cite{shao2020graph} & Listwise & - & \textcolor{darkred}{\ding{55}} & \textcolor{darkred}{\ding{55}} & \textcolor{darkred}{\ding{55}} & \textcolor{darkred}{\ding{55}}  \\
\hline
\textcolor{cust_cyan}{ENT Rank}\cite{dietz2019ent} & Pointwise & KG Entity & \textcolor{darkgreen}{\checkmark} & \textcolor{darkred}{\ding{55}} & \textcolor{darkred}{\ding{55}} & \textcolor{darkred}{\ding{55}}  \\
\hline
\textcolor{cust_cyan}{Term Ranker}\cite{khan2016term} & - & Term & \textcolor{darkred}{\ding{55}} & \textcolor{darkred}{\ding{55}} & \textcolor{darkgreen}{\checkmark} & \textcolor{darkred}{\ding{55}}  \\
\hline
\end{longtable}
}


\subsubsection{LTR Paradigm}
As illustrated in the taxonomy (Figure \ref{fig:taxonomy}) and the timeline (Figure \ref{fig:timeline}), graph-based re-ranking is primarily applied in general IR tasks. Nevertheless, its contribution also extends to QA and numerous specialized tasks. Across these settings, graph-based ranking models typically follow one of three LTR paradigms: pointwise, pairwise, and listwise. Traditionally, these paradigms describe characteristics of the loss function used during supervised training. More recently, they have also been used to define a model's ranking mechanism during inference, particularly for unsupervised or zero-shot settings, such as in LLM-based zero-shot re-rankers. The majority of the models adopt a pointwise LTR paradigm, whereas only a few methods adopt pairwise or listwise approaches.

\subsubsection{Node Granularity}
Approximately 80\% of the surveyed studies construct graphs that directly or indirectly support the re-ranking process. These graphs primarily differ in the content of their nodes, while edges typically model semantic similarity or relational dependencies. Common node types include documents, segments, named entities, and KG entities, as illustrated in the \textit{Node Granularity} column. According to standard IR conventions, the term "document" refers to a textual unit at varying levels of granularity, ranging from full documents to passages and even sentences. Consequently, segments denote a finer-grained textual unit relative to the corresponding document definition within a given work. For example, when a document retains its conventional meaning, then segments represent passages. Conversely, when a document is used to denote passages, segments correspond to sentences. Most graphs are homogeneous, containing a single node type, as shown in the \textit{Node Granularity} column. In contrast, several studies employ heterogeneous or multi-type nodes. QDG \cite{frayling2024effective} constructs a bipartite graph between unique query nodes and document nodes. PGT \cite{yu2021pgt} incorporates two types of nodes within its graph structure: (query, document) node and feedback document nodes. \cite{bendersky2008re} creates a bipartite graph between documents and their constituent passages. QUASAR \cite{christmann2024rag} employs a bipartite graph between evidence documents and their associated entities. KEGNR \cite{liu2025knowledge} constructs a heterogeneous knowledge-query-doc graph consisting of three types of nodes: query term nodes, document term nodes, and sentence nodes. EQAG-GNN \cite{iyer2023question} builds a small-scale graph, where each node represents a question-answer (Q, A) sentence pair. 

\subsubsection{Knowledge Graph Integration}
Knowledge graphs constitute another critical component in many graph-based re-ranking frameworks, introducing external knowledge to guide the re-ranking process. GAT-reRanker\cite{vollmersdocument} incorporates external knowledge by constructing a query-document sub-graph, where entities associated with queries and documents are modeled as nodes. QDER \cite{chatterjee2025qder} represents documents as a combination of token and entity embeddings, while DREQ \cite{chatterjee2024dreq} leverages KG information to amplify query-relevant entities in document representations while suppressing irrelevant ones. KGPR\cite{fang2023kgpr} exploits a knowledge subgraph extracted from Freebase \cite{bollacker2008freebase} to enhance cross-encoder re-ranking. KERM \cite{dong2022incorporating} first prunes noise from a global KG before constructing a bipartite meta-graph for each query-passage pair of entities extracted from the refined KG. Similarly, Q-KGR \cite{zhang2024question} refines a knowledge subgraph through an edge re-scoring mechanism to eliminate noise for improved KGQA. TimeR\textsuperscript{4} \cite{qian2024timer4} extends this line of work to temporal settings, leveraging TKGs to convert implicit temporal information into explicit representations, thereby improving temporal KGQA tasks.

Several approaches rely on knowledge graphs to introduce structural relational signals into the ranking pipeline. KG-FiD \cite{yu2022kg} leverages topological relationships within a KG to establish node connections, thereby introducing relational information into the re-ranking process.  In QUASAR\cite{christmann2024rag}, KG entities are used as nodes in a bipartite graph to incorporate structured knowledge into an RAG framework. Earlier approaches also explored incorporating KG information to improve their ranking models: EsdRank \cite{xiong2015esdrank} incorporates rich external information, such as entities and relational structures, to complement query and corpus signals; EDRM \cite{liu2018entity} introduces knowledge graph semantics into an interaction-based neural network; and \cite{xiong2016bag} replaces the traditional bag-of-words (BoW) representation with bag-of-entities (BoE). In a similar vein, AttR-Duet\cite{xiong2017word} complements word-based representations with entity-based representations derived from a BoE approach using automatic entity linking \cite{xiong2016bag}. External knowledge also plays an important role in domain-specific and other specialized tasks. KG-Rank \cite{yang2024kg} uses external knowledge graphs to improve factual reasoning in medical QA. Similarly, GraphMonoT5 \cite{gupta2024empowering} incorporates external knowledge from a KG and integrates it with MonoT5 \cite{nogueira2020document} to improve the re-ranking of biomedical documents. 

Table \ref{tab:KG-table} summarizes the statistics of the knowledge graphs utilized in the surveyed literature. The statistics reflect the most up-to-date information available at the time of writing. Freebase  \cite{bollacker2008freebase}, a large-scale community-driven knowledge base, was first introduced by Metaweb in 2007. It was later acquired and maintained by Google before being discontinued in 2016. DBpedia is a large-scale multilingual knowledge graph founded in 2007 to store structured information from the Wikimedia projects, particularly Wikipedia. Inspired by DBpedia, \cite{xu2017cn} introduced CN-DBpedia, a knowledge graph dedicated to the Chinese language, generated and continuously updated through knowledge extracted from Chinese online encyclopedias. Wikidata is another widely recognized large-scale multilingual knowledge base operated by the Wikimedia Foundation. ConceptNet \cite{liu2004conceptnet} was initially released as a common sense knowledge base and later evolved into a multilingual knowledge graph in ConceptNet 5.5 \cite{speer2017conceptnet}. UMLS (Unified Medical Language System) is a biomedical knowledge graph developed by the U.S. National Library of Medicine to capture the relationships among medical terminologies. ICEWS \cite{o2010crisis}, the Integrated Crisis Early Warning System, provides information on numerous political events, along with associated timestamps. Using this resource, \cite{garcia2018learning} created the ICEWS05-15 temporal knowledge graph. 

Although we categorize all these resources as knowledge graphs, some of their official documentation may use different terminology. For instance, Freebase is described as a knowledge base, while UMLS is framed as a knowledge source of biomedical vocabularies. However, in the context of re-ranking research, they are typically treated as knowledge graphs because they provide graph-structured representations.

\begin{table}[h]
\caption{Statistics of the knowledge graphs used in the reviewed literature. \textit{Entity} represents real-world objects. \textit{Triple} is a fundamental unit of a KG, represented as \textit{(Head Entity, Relation, Tail Entity)}. \textit{Relation} denotes the connection between head and tail entities. \textit{Statement} extends a triple with additional qualifiers or metadata. \textit{K}, \textit{M}, and \textit{B} denote thousand, million, and billion, respectively.}
\centering
\small
\begin{tabular}{l p{4.2cm} l}
\hline
\textbf{Knowledge Graph} & \textbf{Statistics} & \textbf{Literature} \\
\hline
Freebase \cite{bollacker2008freebase} & Entities: 48M \cite{pellissier2016freebase}\newline Triples: 1.9B\tablefootnote{\url{https://developers.google.com/freebase}} & KGPR \cite{fang2023kgpr}, AttR-Duet \cite{xiong2017word}, \cite{sun2024efficient}, EsdRank \cite{xiong2015esdrank}, \cite{xiong2016bag}\\
\hline
Wikidata \cite{vrandevcic2014wikidata} & Entities: 120.6M\tablefootnote{\url{https://www.wikidata.org/wiki/Wikidata:Statistics}}\newline Statements: 1.65B \tablefootnote{\url{https://en.wikipedia.org/wiki/Wikidata}} & GAT-reRanker \cite{vollmersdocument}, KG-FiD \cite{yu2022kg}, QUASAR \cite{christmann2024rag} \\
\hline
ConceptNet \cite{speer2017conceptnet} & Nodes: Over 8M \newline Statements: 34M\tablefootnote{\url{https://github.com/commonsense/conceptnet5/wiki/FAQ}} & KERM \cite{dong2022incorporating}, GraphMonoT5 \cite{gupta2024empowering}, Q-KGR \cite{zhang2024question} \\
\hline
UMLS \cite{bodenreider2004unified} & Concepts: 3.48M\tablefootnote{\url{https://www.nlm.nih.gov/research/umls/knowledge_sources/metathesaurus/release/statistics.html}} & GraphMonoT5 \cite{gupta2024empowering}, KEGNR \cite{liu2025knowledge}, KG-Rank \cite{yang2024kg} \\
\hline
DBpedia \cite{auer2007dbpedia} & Entities: 33M, Triples: 1.32B\tablefootnote{\url{https://community.openlinksw.com/t/announcing-dbpedia-release-2025-06/5994}} & QDER \cite{chatterjee2025qder}, DREQ \cite{chatterjee2024dreq} \\
\hline
CN-DBpedia \cite{xu2017cn} & Entities: 10.3M\newline Relations: 88.45M & EDRM \cite{liu2018entity} \\
\hline
ICEWS05-15 \cite{garcia2018learning}  & Entities: 10K, Relations: 251 & TimeR\textsuperscript{4} \cite{qian2024timer4} \\
\hline
\end{tabular}
\label{tab:KG-table}
\end{table}

\subsubsection{Graph-refined Features}
Structural and relational information among documents plays a crucial role in re-ranking, enabling the model to assess the relative importance of documents when estimating query-document relevance. It also enriches document representations with information from similar documents. Graph data structures naturally model such structural relationships through their networked architecture. However, this information cannot be directly extracted from the graph itself and must instead be processed using graph learning models, particularly graph neural networks. GNNs capture structural relationships by updating node representations via message passing. This process typically begins with initializing node representation, after which a GNN iteratively updates them by aggregating information from neighboring nodes. As a result, each node representation gradually incorporates information from its surrounding nodes. After $l$ layers of message passing, a node can capture information from its $l$-hop neighbors, enabling a model to learn both local and global structural information. This additional signal can significantly enhance candidate document re-ranking. The refinement of node features is not limited to graph learning techniques. For example, PGT \cite{yu2021pgt} employs the Transformer-XH architecture to update node representations. The \textit{Graph-refined Features} column in Table \ref{tab:model_comparison} indicates whether a method uses graph learning techniques to update its node features. 

\subsubsection{Graph-based Re-ranking}
A notable aspect of graphs in re-ranking methods is their direct integration into the ranking model. We define graph-based re-ranking as pipelines in which the graph component directly contributes to the final relevance score with minimal intermediate processing. For instance, approaches that introduce a separate neural network to compute query-document relevance from graph-refined representations are not considered a graph-based re-ranking method. Notably, only 28.84\% of the surveyed works adopt such direct graph-based re-ranking strategies, highlighting a clear opportunity for future exploration. Table \ref{tab:GNN-table} summarizes the graph-based models utilized in the surveyed literature for node representation updates and graph-based re-ranking.

Graph neural networks constitute the primary graph learning technique in the graph-based re-ranking architecture. Section \ref{subsec:GNN} provides a brief overview of GNNs. In addition to GNNs, graph-based centrality measures and link analysis algorithms, such as HITS, PageRank, and their variants, also play an important role in graph-based re-ranking. Although these approaches originated in the early web era, they remain key components in many modern re-ranking frameworks. Hyperlink-induced topic search (HITS) \cite{kleinberg1999authoritative}, a link analysis algorithm, ranks web pages using hub and authority scores. This concept stems from early web structures, in which hubs serving as large directories hosted many authoritative web pages. In modern re-ranking frameworks, the hub score reflects how well a node identifies high-quality information sources, while the authority score measures a node's impact as an information provider. PageRank \cite{page1999pagerank}, a foundational graph-based ranking algorithm that underpins Google search, ranks web pages based on the quality and quantity of incoming edges. Building on this, TextRank \cite{mihalcea2004textrank} adapts the PageRank algorithm for natural language processing tasks by modeling relationships between textual units. 

\begin{table}[h]
\caption{Overview of the graph neural networks used in the reviewed literature.}
\centering
\small
\begin{tabular}{l p{11cm}}
\hline
\textbf{Graph Models}  & \textbf{Literature} \\
\hline
GAT \cite{velivckovic2017graph} & GNRR \cite{di2024graph}, MiM \& LiM \cite{albarede2022passage}, GAT-reRanker \cite{vollmersdocument}, IDRQA \cite{zhang2021answering}, KG-FiD \cite{yu2022kg}  \\
\hline
GCN \cite{kipf2016semi} & G-RAG \cite{dong2024don}, GNRR \cite{di2024graph}, EQAG-GNN \cite{iyer2023question}, CASSIE \cite{van2023question}, MPGE \cite{tian2020capturing}  \\
\hline
R-GCN \cite{schlichtkrull2018modeling} & KEGNR \cite{liu2025knowledge}  \\
\hline
GGNN \cite{li2015gated} & LGRe \cite{dong2021latent}, DGRe \cite{dong2022disentangled}, GRMM \cite{zhang2021graph}  \\
\hline
GMN \cite{dong2022incorporating} & KERM \cite{dong2022incorporating} \\
\hline
GraphSAGE \cite{hamilton2017inductive} & GNRR \cite{di2024graph}  \\
\hline
GIN \cite{xu2018powerful} & GNRR \cite{di2024graph}  \\
\hline
SignedConv \cite{derr2018signed} & GNRR \cite{di2024graph}  \\
\hline
HITS \cite{kleinberg1999authoritative} &  \cite{bendersky2008re}  \\
\hline
PageRank \cite{page1999pagerank} & TRM \& TDM \cite{veningston2014information}, PRP-Graph \cite{luo2024prp}, \cite{gienapp2022sparse}, PsgAidRank \cite{krikon2010utilizing}, \cite{kurland2010pagerank},  GBRM \cite{deng2009effective}, \newline Diffusion-Aided RAG \cite{dampanaboina2025diffusion} \\
\hline
TextRank \cite{mihalcea2004textrank} & Term Ranker\cite{khan2016term}, PassageRank \cite{reed2020faster}\\
\hline
\end{tabular}
\label{tab:GNN-table}
\end{table}

\subsubsection{Inter-document Relation}
It complements graph-refined features when graph models relationships among candidate documents, in which case inter-document relationships are encoded within the updated node representations learned through GNN message passing. In addition, several graph-based centrality measures, such as HITS and PageRank, enable modeling inter-document relationships without explicitly updating node representations. For instance, \cite{bendersky2008re} captures inter-document relations using the HITS algorithm, while \cite{veningston2014information, luo2024prp, gienapp2022sparse, deng2009effective, krikon2010utilizing, kurland2010pagerank, dampanaboina2025diffusion} employ PageRank or its variants to incorporate inter-document relationships into the ranking process. Furthermore, although implicitly, LLM-based listwise re-rankers such as SlideGAR \cite{rathee2025guiding} can capture limited representations of inter-document relationships through their parametric modeling capabilities. Finally, ORE \cite{rathee2025breaking}, an adaptive re-ranking technique, captures inter-document relationships as an auxiliary signal by estimating the average relevance of a candidate's neighbors within a set of highly relevant documents.

\subsection{Limitations and Future Directions}
\noindent\textbf{Latency.} It is a key evaluation criterion of information retrieval and re-ranking systems, as it is closely related to user satisfaction and practical usability of retrieval systems. This consideration is particularly critical for re-ranking modules, as they are often integrated into more complex pipelines, such as a RAG or QA framework, where other components may already introduce additional latency. Therefore, IR systems must maintain a prompt response time to meet real-world requirements. However, as shown in Table \ref{tab:experimental_settings}, none of the graph-based re-ranking methods report latency as an evaluation metric. This is especially concerning given that graph-based methods are often associated with longer execution times. Consequently, we encourage future researchers to include latency as a standard evaluation metric. Moreover, future work should explicitly address the latency challenges associated with graph-based re-ranking methods to improve their practicality. 

\noindent\textbf{LTR Paradigm.} The performance of IR models is often influenced by the LTR paradigm they use. Each paradigm offers distinct advantages and limitations. However, most graph-based re-ranking methods in the surveyed literature rely on a pointwise approach, which is generally more efficient than other paradigms but suffers from independently estimating query-document relevance. As a result, it fails to utilize the comparative relation among the candidate documents. Therefore, future research should place greater emphasis on pairwise, listwise, and setwise approaches. In particular, the setwise paradigm can benefit from the effectiveness of listwise approaches while requiring lower latency than traditional pairwise methods. 

\noindent\textbf{Diversity of Node's Content.} As illustrated in Table \ref{tab:model_comparison}, most graph-based re-ranking methods use a single node type within the graph. While this design is effective for capturing the structural relationships among documents, incorporating multiple node types could enable modeling more nuanced relationships, particularly in longer documents. Another promising future direction is the development of query-dependent graph structures, where introducing an independent query node or composite query-document nodes may facilitate query-aware document representations, potentially leading to improved ranking performance. 

\noindent\textbf{Learning Paradigm.} Existing graph-based re-ranking methods predominantly rely on supervised and ad-hoc learning strategies, as shown in Figure \ref{fig:learning_method}. Therefore, we encourage future researchers to explore alternative learning paradigms, particularly LLM-based supervised fine-tuning and zero-shot approaches. Supervised fine-tuning enables LLMs to adapt their capabilities to task-specific scenarios, facilitating the development of more sophisticated re-ranking architectures. In contrast, zero-shot LLM-based approaches help address the common limitations of supervised learning, such as limited generalization and reliance on expensive labeled datasets, by leveraging LLMs' parametric knowledge through prompt engineering. Although general IR has seen substantial progress in these directions, its application within graph-based re-ranking remains relatively unexplored.

\noindent\textbf{Supervised Datasets.} Supervised LTR models require large amounts of query-document relevance data with associated relevance labels. MSMARCO serves as the primary source of such data for modern re-ranking architectures. It contains real-world user queries extracted from Bing search logs, along with a binary relevance label obtained through crowdsourcing. However, binary relevance may not adequately capture the full spectrum of query-document relationships, as a document can be partially or moderately relevant to a query. Classifying such cases strictly as relevant or irrelevant may therefore result in an underrepresentation of their true relevance. While we acknowledge the cost of constructing supervised LTR datasets, developing large-scale datasets with fine-grained relevance labels would help advance not only graph-based re-ranking but also the broader field of information retrieval.

\noindent\textbf{Advanced Graph-learning Techniques.} Most modern graph-based re-ranking methods are built on graph neural networks, as depicted in Table \ref{tab:GNN-table}. While GNNs are effective at capturing local neighborhood information, they often struggle to model long-range dependencies. In addition, they are known to suffer from over-smoothing and over-squashing. Over-smoothing refers to the convergence of node representations as the number of GNN layers increases, resulting in node embeddings that become indistinguishable. Over-squashing, on the other hand, arises from structural bottlenecks in the graph that cause information loss from distant nodes. We suggest that future work address these challenges in graph-based re-ranking, particularly by adapting more advanced graph-learning models such as graph transformers. Graph transformers can capture long-range dependencies via a global self-attention mechanism and have been shown to mitigate issues of over-smoothing and over-squashing. Another promising direction is the use of graph foundation models in re-ranking. Owing to their training on large-scale graph data, GFMs have the potential to improve both the performance and generalization capabilities of graph-based re-ranking models. 

\section{Conclusion}
\label{sec:conclusion}

In this survey, we provide a comprehensive review of graph-based re-ranking methods, examining their historical development, evolution, and emerging research trends. To organize existing approaches, we introduce a structured taxonomy that categorizes graph-based re-ranking models into four primary classes based on their application domains, which are further subdivided by methodological characteristics. From a methodological perspective, we analyze the role of graphs in the re-ranking process across multiple levels of granularity, including the use of external graph-structured knowledge, the utilization of corpus-level graphs, the direct application of graph neural networks for document re-ranking, and hybrid approaches that integrate multiple strategies. Additionally, we present a chronological timeline highlighting the evolution of graph-based re-ranking methods across the four taxonomy categories. We further review the experimental settings of representative studies, including datasets, evaluation metrics, learning paradigms, training objectives, and the availability of open-source implementations. Finally, we summarize key insights from the literature by highlighting important findings, identifying current limitations, and discussing promising directions for future research. We hope this survey will serve as a valuable resource for researchers and encourage further advances in graph-based re-ranking for next-generation retrieval systems.


\section*{Acknowledgment}
This research was supported by a U.S. Department of Defense (DoD) grant (Grant No. FA-8075-18-D-0004). Any opinions, findings, or conclusions presented in this work are solely those of the author(s) and do not necessarily represent those of DoD.

\section*{Declaration of generative AI and AI-assisted technologies in the manuscript preparation process}
During the preparation of this work, the authors used OpenAI's ChatGPT in order to refine the manuscript’s language. The original work was composed entirely by the authors, with AI assistance limited to moderate editing to improve clarity. Occasionally, ChatGPT was used to obtain a preliminary overview of the relevant literature before the authors' detailed review. After using this tool/service, the authors reviewed and edited the content as needed and take full responsibility for the content of the published article.

\bibliographystyle{elsarticle-num} 
\bibliography{references}

\end{document}